\documentclass[a4paper,12pt]{article}
\pdfoutput=1 

\usepackage{jheppub} 

\usepackage{tabularx,subcaption,fancyhdr,mathtools,physics}
\usepackage[makeroom]{cancel}
\usepackage{bm}
\usepackage{csquotes}
\usepackage[nottoc,notlot,notlof]{tocbibind}
\usepackage{float}
\usepackage{relsize}
\usepackage{hyperref}
\usepackage{color,latexsym, array,multirow,url,verbatim, enumerate,cancel}
\DeclareUnicodeCharacter{2212}{-}
\hypersetup{colorlinks=true,linkcolor=blue,citecolor=magenta,filecolor=magenta,
	urlcolor=cyan}
\DeclareUnicodeCharacter{2212}{-}
\usepackage{setspace}
\usepackage{colortbl}
\usepackage{cancel}
\usepackage{xcolor}
\usepackage{multicol}
\usepackage{graphicx}
\usepackage[normalem]{ulem}
\usepackage{textgreek}

\def \gl{{\widetilde g}}
\def \mgl{m_{\widetilde g}}
\def \msq{m_{\widetilde q}}
\def \st1{{\widetilde{t_1}}}
\def \mst1{m_{\widetilde{t_1}}}

\def \lspone{\widetilde\chi_1^0}
\def \mlspone{m_{\lspone}}
\def \lsptwo{\widetilde\chi_2^0}

\def\chonepm{\widetilde{\chi}_1^{\pm}}
\def\chonemp{\widetilde{\chi}_1^{\mp}}
\def\mchonepm{m_{\chonepm}}
\def\champ2{\widetilde{\chi}_2^{\mp}}

\def\staur{\widetilde{\tau}_R}

\def \met{\rm E{\!\!\!/}_T}

\def \lum{\mathcal{L}}
\def \ifb{\rm {fb}^{-1}}

\title{ \Large{ Searches for the BSM scenarios at the LHC using decision tree based machine learning algorithms: A comparative study and review of Random Forest, AdaBoost, XGBoost and LightGBM frameworks 
} }

\author[]{Arghya Choudhury,} 
\author[]{Arpita Mondal,}
\author{and Subhadeep Sarkar}

\affiliation[]{Department of Physics, Indian Institute of Technology Patna, Bihar - 801106, India}

\emailAdd{arghya@iitp.ac.in}
\emailAdd{arpita\_1921ph15@iitp.ac.in}
\emailAdd{subhadeep\_1921ph21@iitp.ac.in}

\abstract{
Machine learning algorithms are now being extensively used in our 
daily lives, spanning across diverse industries as well as academia. In 
the field of high energy physics (HEP), the most common and challenging task 
is separating a rare signal from a much larger background. 
The boosted decision tree (BDT) algorithm has been a cornerstone of the 
  high energy physics  for analyzing event triggering, particle 
  identification, jet tagging, object reconstruction, event  classification, 
  and other related tasks for quite some time.
   This article  presents a comprehensive overview of research conducted by both 
  HEP experimental and phenomenological groups that utilize decision tree
  algorithms in the context of the Standard Model and Supersymmetry (SUSY). 
  We also summarize the basic concept of machine learning and decision tree algorithm along with the working principle of \texttt{Random Forest},  
  \texttt{AdaBoost} and two gradient boosting frameworks, such as 
  \texttt{XGBoost}, and \texttt{LightGBM}. Using a case study of 
  electroweakino productions at the high luminosity LHC, we demonstrate 
  how these algorithms lead to improvement in the search sensitivity 
  compared to traditional cut-based methods in both compressed and 
  non-compressed   R-parity conserving SUSY scenarios. The effect of different 
  hyperparameters and their optimization, feature importance study using 
  SHapley values are also discussed in detail. }

\begin{document} 
\maketitle

\flushbottom

\section{Introduction}
\label{sec:intro}

Supersymmetry (SUSY)~\cite{Martin:1997ns,drees2004theory,baer2006weak} is one of the most compelling extensions of beyond the Standard Model (BSM) scenario, and the pursuit of supersymmetric partners of the SM particles (sparticles) remains as a primary objective at the Large Hadron Collider (LHC). Since the inception of the LHC, both the ATLAS and CMS collaborations have already conducted numerous searches to explore the SUSY particles utilizing the LHC Run-I and Run-II dataset~\cite{atlas_susy,cms_susy}. In the absence of any statistical deviations from the SM predictions, the LHC has set stringent lower bounds on the masses of particles. For example, in the R-parity conserving (RPC) SUSY\footnote{The RPC SUSY model offers a stable lightest supersymmetric particle (LSP), which can be a promising candidate for dark matter. 
The weakly interacting massive particle  (the most popular choice is the lightest neutralino $\lspone$) is able to evade detection, and this results in a distinct signature of significant missing energy.}  scenarios with relatively light neutralinos ($\lspone$), the LHC Run-II data has extended the lower bounds on the masses of gluino ($\mgl$), first two generations light squarks ($\msq$), the lightest stop ($\mst1$) and the lightest chargino ($\mchonepm$) upto $\sim$ 2.3 TeV, 1.6 TeV, 1.2 TeV and 1.2 TeV respectively~\cite{atlas_susy,cms_susy} depending on the branching ratios and the simplified models assumptions.        

It is important to note that the accumulated luminosity of Run-II data is approximately $\sim$ 140 $\ifb$, which is about 5\% of the luminosity of planned upgrade of high luminosity LHC (HL-LHC) run ($\lum = 3000 \ifb$). The LHC Collaboration initiated the Run-III operation in July 2022, and it will continue the operation until the planned three-year-long shutdown (LS3) in preparation for the high luminosity upgrade, which is scheduled to commence in 2026. The constraints on sparticle masses, derived from  LHC Run-I and Run-II data, are primarily derived within the framework of simplified SUSY scenarios, which involve specific assumptions regarding decay modes, branching ratios, etc. Also, the majority of the analyses have concentrated on prompt decay scenarios. However, there are several pockets of SUSY parameter spaces with light sparticles that remain less explored, e.g., quasi-degenerate SUSY scenarios where the next to LSP (NLSP) - LSP mass gap is very small~\cite{Bhattacherjee:2013wna,Dutta:2015exw,Chakraborti:2017dpu,Chowdhury:2016qnz,Dutta:2017jpe}. Such compressed SUSY models are also highly motivated in the context of Dark Matter (DM) relic density~\cite{KumarBarman:2020ylm, Barman:2022jdg, He:2023lgi,Chakraborti:2015mra,Chakraborti:2014gea,Chakraborti:2017dpu,Chowdhury:2016qnz,Bhattacharyya:2011se,Choudhury:2012tc,Baer:2021aax}, Muon (g-2) anomaly~\cite{Chakraborti:2022vds, He:2023lgi, Chakraborti:2015mra, Chakraborti:2014gea, Baer:2021aax, Athron:2021iuf, Endo:2021zal, Chakraborti:2021bmv, Choudhury:2017acn, Choudhury:2017fuu, Banerjee:2018eaf, Banerjee:2020zvi, Chakraborti:2021dli, Frank:2021nkq, Ali:2021kxa, Kowalska:2015zja, Chakrabortty:2015ika, Choudhury:2016lku,Cao:2022htd,Cao:2023juc}. It has been observed that conventional cut and count analyses are highly competitive for non-degenerate SUSY searches compared to machine learning (ML) analyses. However, ML-based analyses have better sensitivity in both degenerate and non-degenerate BSM physics searches.

The term ``machine learning" (ML)  was first proposed by A. Samuel in 1959 \cite{Samuel1988}. After six decades, ML algorithms are now being widely used in our daily lives and across various industries, e.g., filtering email, social media recommendation, cyber security, image analysis and disease detection in healthcare, data analysis in the finance sector, autonomous vehicles, marketing and advertising, etc. In the future, the applications of ML will continue to expand rapidly with the advancement of technology and the availability of more data. 
high energy physics (HEP) experimental and phenomenological analyses deal with large amount of data and it has been observed that the use of ML techniques leads to improvements in the data analyses of several HEP fields. For particle identification, event selection, object reconstruction, event classification, etc., the experimental high energy physics (HEP-Ex) collaborations have been using the conventional ML algorithms for more than three decades. Boosted decision tree (BDT) is one of the most popular algorithms which has been widely used by the HEP-Ex collaborations, e.g., the search of single top quark prediction by the CDF and D0 collaborations~\cite{CDF:2010eor,D0:2008wma,D0:2006ngk}, Higgs discovery at the LHC~\cite{CMS:2014afl}. For a long time, the HEP community has used BDT and other algorithms implemented in the Toolkit for Multivariate Data Analysis (TMVA) software package~\cite{TMVA:2007ngy} while XGBoost (Extreme Gradient Boosting) has gained immense popularity in recent years~\cite{Chen:2016btl}.   However, it should be noted that the deep neural networks (DNN) or deep learning (DL) techniques, which are based on multilayer NN,  are becoming more popular nowadays~\cite{doi:10.1142/S0217751X20020030,dnn,Bhat:2010zz,Guest:2018yhq,Bourilkov:2019yoi,Schwartz:2021ftp,Carleo:2019ptp,Shlomi:2020gdn,Abdughani:2019wuv,Hammad:2023sbd,Hammad:2024cae,Arganda:2024eub}.
There are several broad reviews in the literature on the applications of 
BDT~\cite{Bhat:2010zz,Cornell:2021gut,Coadou:2022nsh,Arganda:2024eub}, deep learning algorithm \cite{doi:10.1142/S0217751X20020030,dnn,Bhat:2010zz,Guest:2018yhq,Bourilkov:2019yoi,Schwartz:2021ftp,Carleo:2019ptp,Shlomi:2020gdn,Abdughani:2019wuv,Hammad:2023sbd,Hammad:2024cae,Arganda:2024eub} like convolutional neural networks (CNN), recurrent neural networks (RNN)  etc. in the context of high energy physics. 

The collider analyses in HEP typically involve searches of new physics signals and the precise measurements of the existing known SM processes. To achieve this, one has to look for faint signals from a large amount of background 
where the distribution of signal and backgrounds have significant overlap. Traditional cut-based analysis shows less sensitivity for significant overlap scenarios. In such cases, the ML algorithms are more powerful in discriminating signals from backgrounds. In this article, we will focus on the various kinds of decision tree (DT) based algorithms and study how these algorithms lead to improvement in the search sensitivity compared to cut-based methods considering a case study of electroweakino pair production at the HL-LHC. The structure of the article is as follows. In Sec.~\ref{sec:ml_outline}, we introduce the basic concepts of machine learning along with different kinds of metrics. In Sec.~\ref{sec:ml_hep}, we briefly summarize the major important analyses involving the use of ML algorithms with an emphasis on DT-based algorithms. The basic concepts of decision trees are discussed in Sec.~\ref{sec:decision_tree}. Also, we present a concise overview of Random Forest, AdaBoost, and extreme gradient boosting algorithms such as  XGBoost and LightGBM in this section. In Sec.~\ref{sec:case_study}, we investigate the improvement of search sensitivity by these four ML algorithms with comparison to cut-and-count analysis using a SUSY scenario with $\chonepm\lsptwo$ (wino-like) pair production. We also present the role of hyperparameters in different ML frameworks along with the feature importance study using Shapley values. Finally, we summarize the paper in Sec.~\ref{sec:summary}.



\section{Basic concepts of machine learning}
\label{sec:ml_outline}

Machine learning, as a subspace of artificial intelligence (AI), involves the cultivation of various models that can learn from diverse datasets and execute tasks without requiring explicit programming. 
To perform a machine learning (ML) analysis, it is essential to collect data, which comprises information about the desired output that we want the computer to learn. The ML algorithms can be broadly categorized into supervised and unsupervised learning \cite{inbook,hastie2009elements,carbonell1983overview}. 
In addition to these two, there is also a third category known as semi-supervised learning. In supervised learning, each input data point ($x_i$, i = 1,....n, for n input data samples) is accompanied by a target variable or output label ($y_i$) and the algorithm learns to extrapolate patterns from the provided training data in order to predict the output labels for unseen (testing) data points. 
In other words, the aim of the supervised learning is to acquire a mapping function $f$ from the input data to output label: $f : x_i \rightarrow y_i$ such that it can accurately predict $y_i$ for a new data sets (testing) which do not have output labels. 
 
Supervised learning algorithms like decision trees (DT), support vector machines (SVM), logistic regression (LR), neural networks (NN), etc, are commonly used for two major tasks: classification and regression. In the classification task, the output/target is discrete, categorical, and finite, while the output is continuous and infinite for the regression task.  
For the HEP problems, the simplest example of binary classification is the categorization of  
the signal and backgrounds for the sensitivity study and one example of a regression task is object tagging. In this paper, we will concentrate on supervised DT algorithms. On the other hand, for the unsupervised learning algorithms (e.g.,  k-means clustering, autoencoders, etc.) the data only contains input features 
without any target output variables/values and the algorithms explore the hidden 
structures or patterns to perform tasks like clustering, anomaly detection, etc. 
For a recent review of unsupervised machine learning in the context of particle physics, see Ref. \cite{Bardhan:2024zla}.

In high energy physics (HEP), each and every particle of an event has different four-momentum information along with different energy depositions in the detector. These attributes serve as the basis for constructing different input features for analysis so that the ML algorithm can distinguish the signal and background efficiently. The available dataset is commonly divided into two subsets: training data and testing data. The training data is used to train the model, while the testing data is used to assess the model's performance on unseen data. Sometimes, we use another set of independent data, known as validation data,  to fine-tune the hyperparameters, such as learning rate, regularization strength, or the number of layers in a neural network, etc. 

\begin{figure}[!htb]
\begin{center}
\includegraphics[scale=0.55]{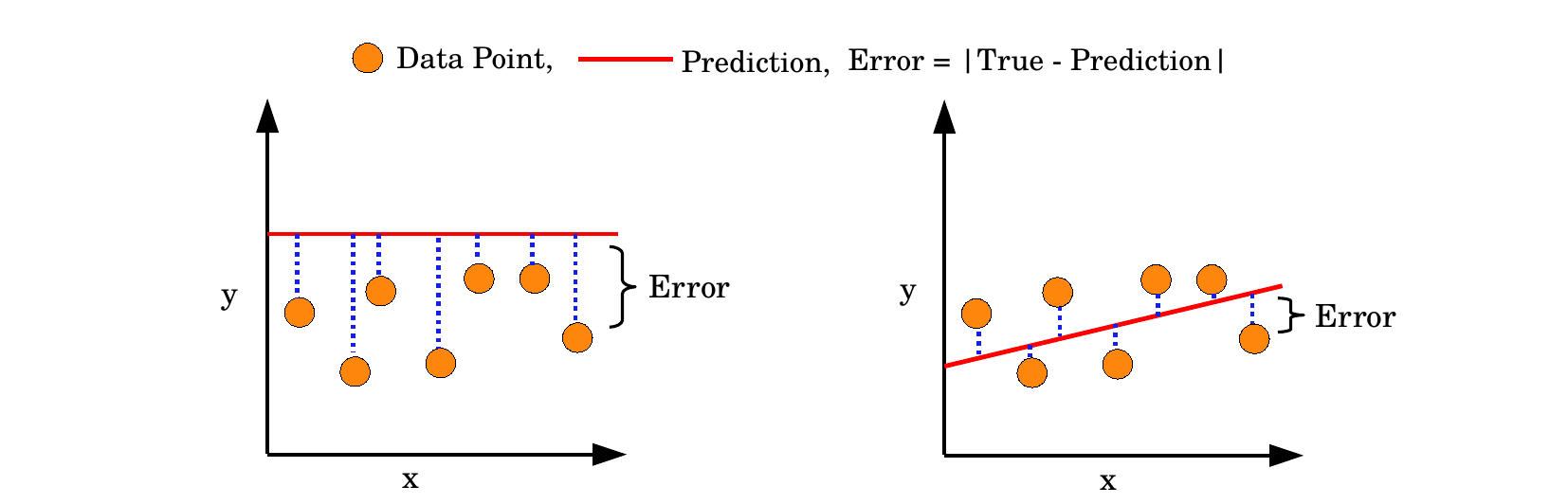}
\caption{Schematics to demonstrate optimization of mean squared error as a loss function ($L^{MSE}$). The data points are denoted as orange circles, and the prediction is denoted by the red line. The length of the blue dotted lines signifies the difference between the actual value and the predicted value. Left and right models correspond to high and low loss, respectively.}
\label{fig:loss_function}
\end{center}
\end{figure}
\vspace*{-1cm}
\subsection{Loss Function}
In supervised learning, the ML algorithm learns using a \textit{loss function}\footnote{It is also known as cost function or error function 
 and depends on the actual and predicted output.}. 
Using optimization techniques like Gradient Descent, the loss function is 
minimized during training by adjusting parameters, which are known as ``\textit{weights}". For example, 
the predicted output $y$ in a linear regression model is calculated as $y = f(x_i)  = w_i.x_i +b$, where b is the bias term. For decision trees, weights may indicate how much each feature contributes to the decision at each node of the tree.
The minimization of the \textit{loss function} specified that the model predicted values are close to the actual value/label. For different types of tasks or data, 
the choices of loss function vary. The commonly used function for regression 
tasks is the mean squared error (MSE) and is defined as \cite {hastie2009elements, article,bentejac2021comparative}
\begin{equation}
L^{MSE} = \frac{1}{n} \sum_{i =1}^{n} (\hat{y_i} -f(x_i))^2 = \frac{1}{n} \sum_{i =1}^{n} (\hat{y_i} -y_i)^2
\end{equation}
where $n$ is the number of samples in the dataset,  $y_i$ and $\hat{y_i}$ are 
the predicted and true target value of the $i^{th}$ sample. 
Another common loss function for regression tasks is mean absolute error and is defined as 
\begin{equation}
L^{MAE} =  \frac{1}{n} \sum_{i =1}^{n} |\hat{y_i} -y_i|
\end{equation}
For binary classification tasks the most commonly used loss function 
is Binary Cross-Entropy Loss or Log Loss, which is defined as: 
\begin{equation}
L^{BCE} = - \{\hat{y_i} \mathrm{log}(y_i) + (1 - \hat{y_i})(\mathrm{log}(1-y_i))\}
\end{equation}
where $y_i$ is the predicted probability and  $\hat{y_i}$ is the true binary label (0 or 1). Categorical cross-entropy loss
 ($\sum_{i =1}^{n} \hat{y_i}$  log(y)) is used in multi-class classification tasks.  
The gradient descent technique is used to find the local minima of loss function by an iterative approach. By the iterative method it finds the optimized parameters and improves the performance of the model. Let us consider a loss function $L$, which is differentiable in a neighborhood of a point $p$. $L(p)$ will decrease the fastest in the direction of the loss function ($-\nabla L(p)$). For two consecutive steps, say $n$ and $n+1$, one can write,
\begin{equation}
p_{n+1} = p_n - \eta \nabla L(p_n)
\end{equation}
where $\eta$ is the learning rate. It is worth noting that the choice of $\eta$ is crucial for function optimization. If $\eta$ is too big, it may not reach the minimum and can bounce back and forth.

\subsection{Overfitting and underfitting}
After training the dataset and adjusting the ``\textit{hyperparameters}", one proceeds to assess the model's ability to \textit{generalize} or to make accurate predictions for unseen testing data. However, the complexity of the trained model and the amount of training data can lead to two major issues: \textit{overfitting} and \textit{ underfitting}.
\begin{figure}[!htb]
\begin{center}
\includegraphics[scale=0.55]{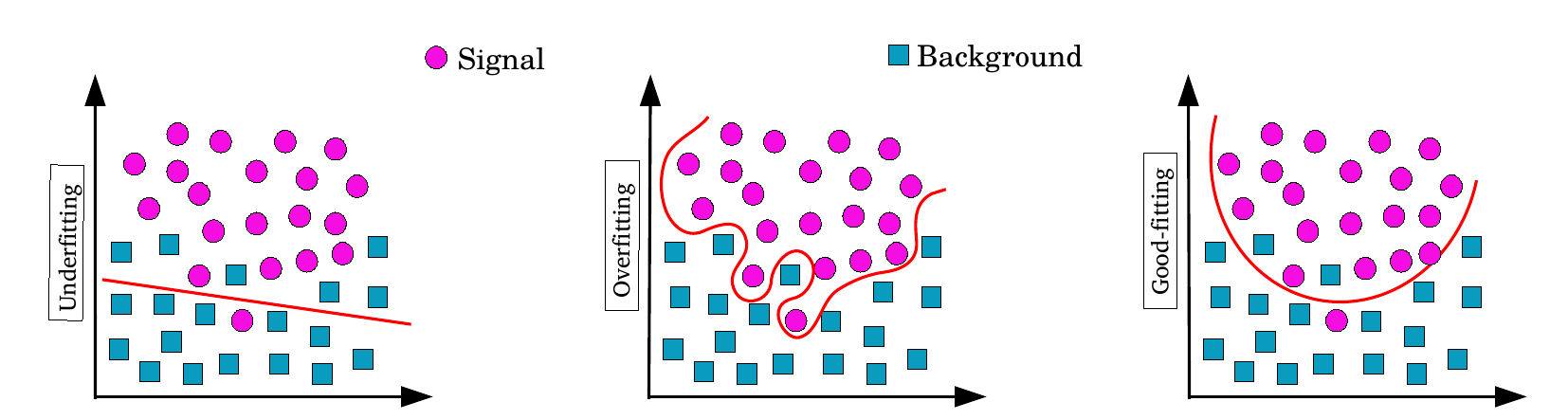}
\caption{A sample diagram to illustrate \textit{underfitting} (high bias, low variance), \textit{overfitting} (low bias, high variance) and good fitting (low bias, low variance) for a training dataset with signal (magenta points) and backgrounds (cyan points).} 
\label{fig:overfit}
\end{center}
\end{figure}
$Overfitting$ arises when a model becomes too complex and learns the training data 
too precisely. Then, the model captures noise in the data instead of the underlying patterns. As a result, the loss computed for the training data tends to approach zero, while it becomes significantly higher for unseen or new testing data. This lack of generalization leads to poor performance on the testing data. Bias represents the deviation of the predicted value from the actual value, while variance indicates the model's sensitivity to small fluctuations in the training data. Therefore, an \textit{ overfitted } model typically exhibits low bias and high variance. Conversely, if the model is too simple and fails to capture the underlying patterns in the training data, it performs poorly on both the training and testing data. This high bias and low variance scenario is known as \textit{ underfitting}. A good/robust model is a trade of both the parameters - bias and variance. In other words, the right balance between model complexity and the amount of training data is essential for building models that generalize well to unseen data. 
In Fig.~\ref{fig:overfit} we demonstrate an example of  
\textit{underfitting}, \textit{overfitting} and appropriate fitting by considering a dataset consisting of signal (circular magenta points) and background events (square cyan points). In the left panel, the model is too simple, underfitted, and has a very high bias. On the other hand, the lowest loss is obtained for the middle panel. However, the prediction of this overfitted model is not reliable for the new dataset. The right panel of  Fig.~\ref{fig:overfit}  represents a robust model that predicts accurately for untrained data.

\renewcommand{\arraystretch}{1.6}
\begin{table}[!htb]
\begin{center}
\begin{tabular}{||c|c||c||c||}
\cline{3-4}
\cline{3-4}
\multicolumn{2}{c||}{}& \multicolumn{2}{c||}{Predicted class}   \\ 
\cline{3-4}
\multicolumn{2}{c||}{} & Signal & Background \\
\hline\hline
  &  &  &   \\
{\rotatebox[origin=c]{90}{~Class}} & Signal & True Positive (TP) & False Negative (FN)  \\
 {\rotatebox[origin=c]{90}{True~}}& Background & False Positive (FP) & True Negative (TN)   \\
& & &  \\
\hline\hline
\end{tabular}
\caption{Graphical representation of confusion matrix for binary classification in a typical HEP event analysis. 
}
\label{tab:confusion_matrix}
\end{center}
\end{table}

\subsection{Measures  of classification performance}
After training, tuning the hyperparameter and testing the HEP data, 
our objective is to evaluate the effectiveness and accuracy of the 
machine learning models in solving a specific task, e.g.,  
classify or distinguish between signal and background events despite their similar 
signature. Performance metrics are quantitative indicators used to measure 
 the model's ability to make accurate predictions or classifications. By 
 evaluating performance metrics, we can compare different ML models 
 and optimize the hyperparameter tuning to improve the model's effectiveness. Sometimes, the highly skewed distribution of classes in the dataset leads to 
 \textit{class imbalance}, which affects the performance of the ML algorithm. 
Techniques like oversampling of the minority class or undersampling of the majority class, adjustment of class weights, cost-effective learning, etc., can be employed to address the class imbalance in ML \cite{Cornell:2021gut, weiss2003learning, Chawla2005, sulaiman, satio, LUQUE2019216, 10.1007/978-3-642-04962-0_53}. To evaluate the performance 
metrics, the algorithm checks the actual and predicted values of observations 
in the dataset by forming confusion matrix\footnote{It is also known as error matrix or classification table.}. 
When the signal/background  events are correctly classified as 
signal/background events, then those events are called True Positive (TP) or  
 True Negative (TN). On the other hand, False Positive (FP),   
 False Negative (FN)  are the number of events where the actual background/signal events are incorrectly classified as signal/background events. In the Table.~\ref{tab:confusion_matrix}, we present the confusion matrix graphically for binary classification. With these definitions, we can define the following 
 various performance metrics to evaluate the performance of the model:
 
 \vspace*{-0.3cm}

\begin{itemize}

\item \textbf{\textit{Sensitivity or Recall or True Positive Rate (TPR)}:}
Recall, also known as sensitivity or TPR,  is defined as the ratio of the number of true positives (correctly predicted signal events) to the total number of actual positives/signal events (TP + FN). Sensitivity is useful 
for minimizing the occurrence of false negatives. 
\begin{align}
\textrm{Recall~ =~TPR } = \frac{TP}{TP+FN} 
\end{align}

\item \textbf{\textit{precision}:} precision measures how many of the events predicted as signal events are actually signals. It is useful when the purpose is
to limit the number of false positives or incorrectly classified signal events. 
\begin{equation}
\textrm{Precision~  } = \frac{TP}{TP+FP} 
\end{equation}

\item \textbf{\textit{Accuracy}:}
Accuracy measures the overall correctness and is defined as the ratio of correct 
predictions to the total no of events/samples. 
\begin{equation}
\textrm{Accuracy~  } = \frac{TP+TN}{TP+TN+FP+FN} 
\end{equation}

\item \textbf{\textit{Specificity or True Negative Rate (TNR)}:}
Specificity is the ratio of true background to the total number of background events. 
\begin{equation}
\textrm{Specificity~  } = \frac{TN}{TN+FP} 
\end{equation}

\item \textbf{\textit{ROC curve and area under the curve (auc)}:}
The ROC (Receiver Operating Characteristic) 
curve\footnote{The ROC term originated in the context of electrical engineering 
during World War II, when electrical signals were used for the prediction of enemy objects.} is a graphical representation of TPR or recall against FPR or (1 - specificity) for various threshold settings. In HEP, the ROC curve relates the signal efficiency versus background efficiency/rejection plane as shown in Fig~\ref{fig:roc} and the curve illustrates the ability of a binary classifier to separate signal and background events. The ROC curve ends at (1,1) 
for a perfect classifier that accepts 100\% signal events and rejects 100\% background events.  
The area under the ROC curve is known as \textit{auc} metric and it varies from 0 to 1. For a perfect classifier, the \textit{auc} becomes one and random guessing 
leads to  \textit{auc} value = 0.5. Thus the \textit{auc} metric is a powerful tool for the evaluation of the overall ranking performance of a binary 
classifier~\cite{sulaiman}.

\item \textbf{\textit{F-score} metric:}  
F-score is a measure that combines recall and precision,  and it is basically 
the harmonic mean of them. F-score can be tuned via a real parameter ($\beta$) and the generic expression is given by: 
\begin{equation}
F_{\beta} = (1+\beta^2) \frac{precision\times recall}{(\beta^2\times precision) + recall}
\end{equation}
 For $\beta = 1$, it is known as $F_1$-score metric and expressed as 
$(2 \times precision\times recall)/(precision + recall)$. 
The F-score can be a better measure than the accuracy metric on imbalanced datasets. A high recall and precision rates indicate low FN and low FP rates and 
$F_1$ score can be useful for imbalanced HEP dataset where the signal events are very rare compared to backgrounds. 


\item \textbf{\textit{Approximate Median Significance (ams) score}:}
In the context high energy physics, the primary objective is to optimize the discovery significance. To estimate the discovery significance, the formulas $s/\sqrt{b}$ or $s/\sqrt{(s+b)}$  are commonly used\footnote{For discovery of a new particle the significance should be $\ge 5\sigma$ and for exclusion it should be
$\ge 2\sigma$  }, where s and b denote the numbers of signal and background events, respectively, that remain after the signal selection cuts. 
It may be noted that in a typical Poisson counting experiment, where $n$ events are observed, the Poisson distribution  often features a large mean value, 
$(s+b)$. However, the formula $s/\sqrt{b}$ is valid only for $b >> s$ and it 
overestimates the discovery significance when the background events are 
small \cite{Cowan2012DiscoverySF,Cowan:2010js}. For a Poisson counting experiment with negligible uncertainty,  the Asimov approximation for the median significance (ams score) is given by~\cite{Cowan2012DiscoverySF,Cowan:2010js}: 
\begin{equation}
\textrm{significance~(ams)} = \sqrt{2\left((s+b) \times ln(1+\frac{s}{b}) - s\right)}
\label{eq:ams1}
\end{equation}
For the scenarios with very small background events, $b$ is replaced by  
$(b + b_r)$, where $b_r$ is a regularization term typically set to stabilize 
the calculation. Expanding the logarithm in s/b, the Eq.\ref{eq:ams1} reduces to 
$\frac{s}{\sqrt{b}} (1+ \mathcal{O}(s/b))$. In the context of particle physics experiments, there is generally an uncertainty with the background and this   
uncertainty ($\Delta$) reduces the significance or the ams score. Due to the 
presence of the uncertainty, the Eq.\ref{eq:ams1} modifies as~\cite{Cowan2012DiscoverySF}
\begin{equation}
ams = \sqrt{2\left((s+b) ~
ln \left( \frac{(s+b)(b + \Delta^2 b^2)}{b^2+ (s+b)\Delta^2 b^2} \right)
- \frac{1}{\Delta^2} ~ln\left( 1 + \frac{ \Delta^2 b^2 s}{b(b+ \Delta^2 b^2)} \right) 
\right)}
\label{eq:ams2}
\end{equation}


\end{itemize} 

\section{Machine learning in High Energy Physics }

\label{sec:ml_hep}

In this section, we primarily review the studies that have utilized decision tree-based algorithms\footnote{We will also briefly mention the relevant works where algorithms other than DT have been used.} within the context of Standard Model (SM) and Beyond Standard Model (BSM) physics scenarios. A comprehensive list of references, grouped into a minimal number of topics, is regularly updated in the \textit{Living Review}~\citep{hepmllivingreview}, covering various categories like ML review works, classification and regression in supervised or unsupervised learning, generative models and more. 
%
\begin{figure}[!htb]
\begin{center}
\includegraphics[scale=0.25]{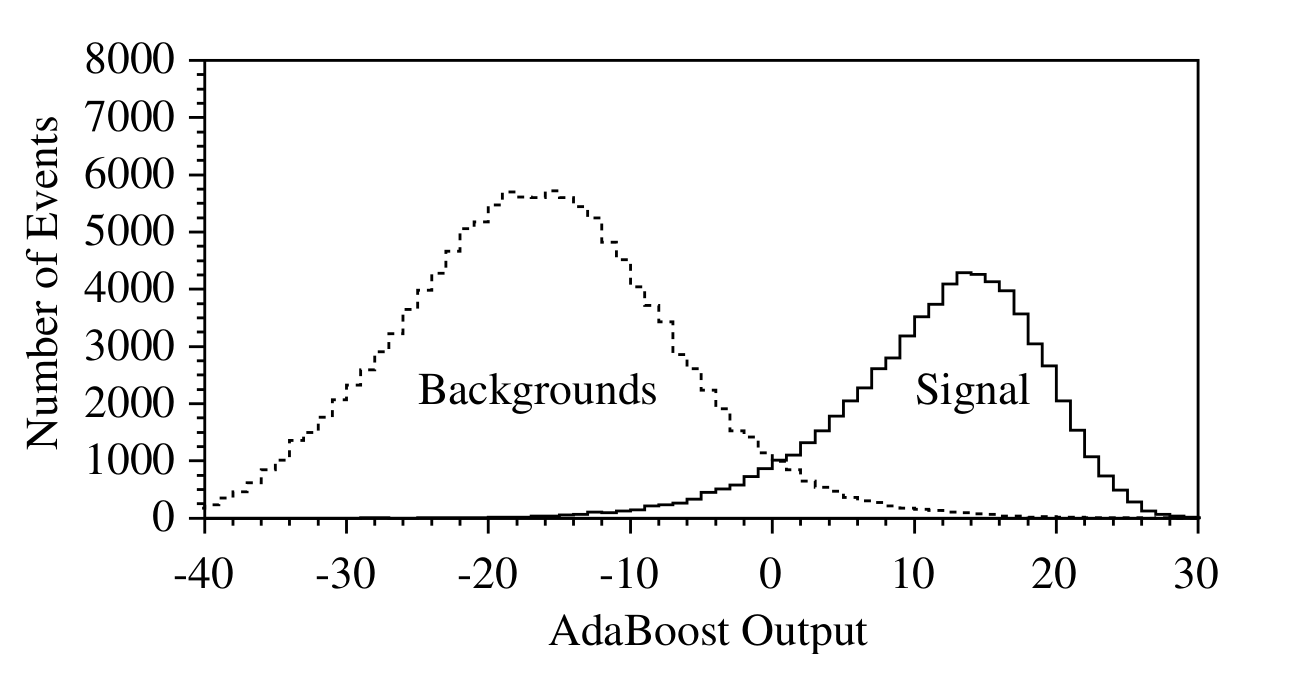}
\caption{AdaBoost output distributions for signal and background events used 
in MiniBoone experiment (adapted from Ref~\cite{ROE2005577}).}
\label{fig:mini} 
\end{center}
\end{figure}
Nearly two decades ago, the first notable use of a boosting algorithm and its performance comparison with the artificial neural network (ANN) and other algorithms was made in the  MiniBooNE experiment at Fermilab, which was designed to explore neutrino oscillations \cite{ROE2005577,YANG2005370}. 
It was observed that  particle identification (PID) with AdaBoost~\cite{FREUND1997119} 
the algorithm is better than the standard ANN PID technique or Random Forest~\cite {ROE2005577,YANG2005370}.  The separation between signal and background events from AdaBoost algorithms is shown in Fig.\ref{fig:mini} where the choice of parameters was:  $\beta$ = 0.5, the number of trees ($N_{tree}$) = 1000 and the number of leaves ($N_{leaves}$)  = 45. In the next few years, D$\cancel{0}$ and CDF experiments at Fermilab used the BDT algorithm along with other algorithms like neural network, matrix elements, etc. for the analyses of single top quark production from the Tevatron data \cite{D0:2006ngk, D0:2008wma, CDF:2010eor}. The performance of the BDT algorithm was slightly better than others and the combined results of different techniques led
to the initial evidence and subsequent observation of single top quark production~\cite{D0:2006ngk, D0:2008wma, CDF:2010eor}. The importance of BDT analysis for this observation has been discussed in a recent review~\cite{Coadou:2022nsh}. Over the past twenty years, the HEP community has been applying the BDT and other more advanced algorithms extensively for event triggering, event generation, parameter space exclusion/scan, jet identification/tagging, event classification and more. 

\noindent\underline{\textbf{Event Triggering:}} In particle physics experiments, the event triggering is very crucial for managing the enormous volume of data,  reducing the event rates for storage, conducting real-time analyses, enhancing the sensitivity of new physics searches, etc. The CMS collaboration has implemented 
BDT in the Endcap Muon Track Finder (EMTF) at the Level 1 (L1) trigger level for the LHC Run-II data collection~\cite{CMS:2018wav}. Also, after Phase-2 upgrade, the CMS experiment will use a dedicated 
BDT classifier at the HGCAL to achieve optimal signal efficiency while rejecting pileup-induced backgrounds~\cite{Zabi:2020gjd}. 
The high-level trigger (HLT) algorithms run in an online environment and they must be  
very fast. The LHCb collaboration has reoptimized the HLT using a bonsai BDT (BBDT) algorithm \cite{Gligorov:2012qt} within the Adaboost framework \cite{Likhomanenko:2015aba}. 
It may be noted that for exotic events with long-lived particles (LLPs) searches, the existing 
triggers are not suitable enough to select the displaced events at the HL-LHC.  In such cases, modern machine learning algorithms like lightweight graph autoencoder can be more promising \cite{Bhattacherjee:2023evs}.

\noindent\underline{\textbf{Event Simulation and Parameter space scanning:}}
Monte Carlo event generations, along with fast detector simulations, are becoming 
more and more computationally expensive as the size of the LHC data is increasing. 
Also, for the generation of events with multiple outgoing particles or the simulations of next-to-leading (NLO) order processes with a large number of additional jets at the LHC/HL-LHC, the evaluation of matrix elements becomes computationally intensive. A new machine learning algorithm based on gradient BDT (GBDT) has been proposed and tested for Monte Carlo integration in Ref.~\cite{Bendavid:2017zhk}.  
Most commonly, the exploration of the parameter space of a new physics model involves the evaluation of some complex likelihood function. Using popular 
approaches like frequentist and Bayesian statistics coupled with Markov Chain Monte Carlo (MCMC) methodology or  MultiNest \cite{Feroz:2007kg,Feroz:2008xx, Trotta:2008bp} algorithm based on Nested Sampling \cite{10.1111/j.1365-2966.2009.14548.x}, several phenomenological groups have analyzed and constrained the  SUSY parameter space \cite{Trotta:2008bp, GAMBIT:2017snp, GAMBIT:2017zdo,Choudhury:2023lbp}. 
A recent study has explored the effectiveness of Random Forest (RF) classifier, which is a decision tree based algorithm, in accurately predicting whether a particular SUSY model is excluded by LHC data or not \cite{Caron:2016hib}. 
For the Monte Carlo simulation and scanning or recasting of SUSY parameter space,  the HEP community has also extensively used more modern/advanced machine learning techniques 
like Multilayer Perceptron (MLP) \cite{Bridges:2010de, Buckley:2011kc} Bayesian Neural Network (BNN) \cite{Kronheim:2020vct}, Graph Neural Network \cite{Mullin:2019mmh}, Generative  Adversarial Network (GAN)
 \cite{DiSipio:2019imz,Butter:2019cae,Lin:2019htn,Musella:2018rdi}, active learning (AL) \cite{Ren:2017ymm, Caron:2019xkx} etc. These studies showed that the use of ML algorithms reduces CPU time and storage. For a recent review on the sampling of BSM parameter spaces subjected to available experimental data using machine learning algorithms, please see Ref.~\cite{Baruah:2024gwy}. 

\begin{figure}[!htb]
\begin{center}
\includegraphics[scale=0.22]{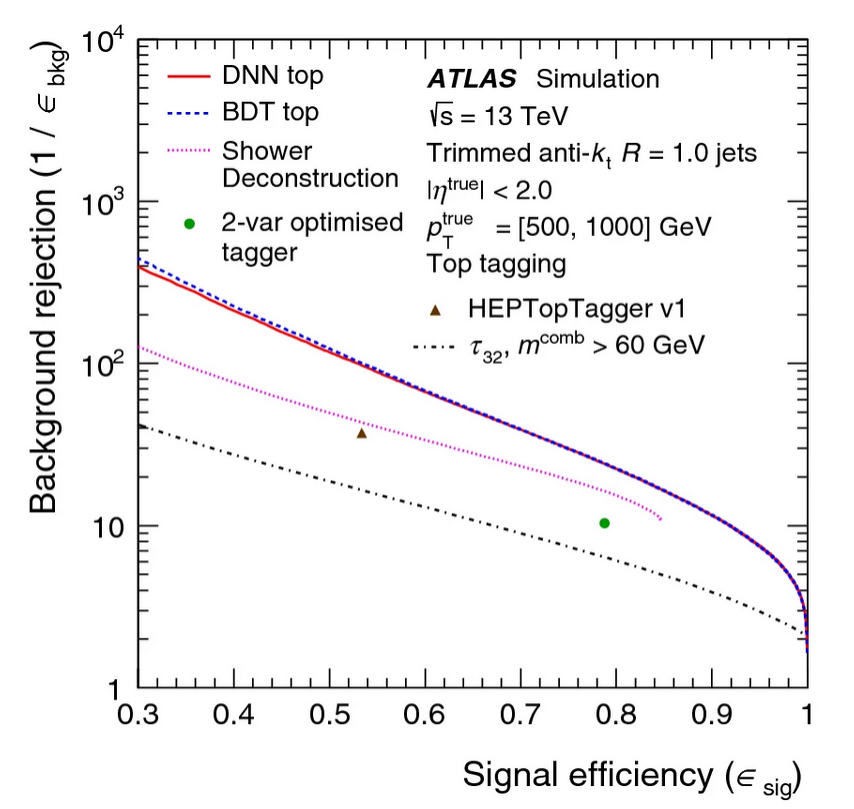}
\includegraphics[scale=0.22]{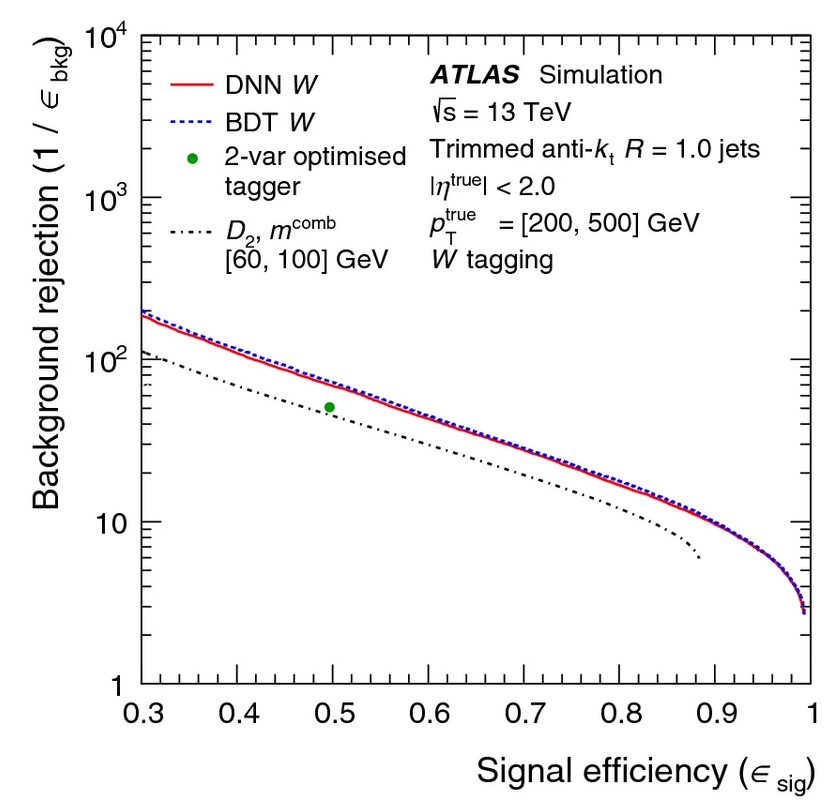}
\caption{The performance comparison (signal efficiency vs background rejection 
characterized in the form of a ROC curve) of the $W$-boson (left) and top-quark taggers are shown here as obtained by the ATLAS Collaboration 
\cite{ATLAS:2018wis} 
using LHC Run-II data with $\lum = 36 \ifb$. The performance of BDT and DNN algorithms is quite similar. }
\label{fig:w-top-tag} 
\end{center}
\end{figure}

\noindent\underline{\textbf{Jet tagging:}}
In collider experiments, the reconstruction and identification of hadronic jets are integral components of physics analyses. Jets originating from 
massive particles (e.g., bottom quark, top quark,  $W/Z/h$ bosons or supersymmetric partner of the top quark ($\mst1$)) are usually boosted and collinear. Jets emerging from b-quark are associated with long lifetime and secondary vertex and can be differentiated from other jets coming from light quarks. Motivated by computer vision techniques, \textit{jet images} was first introduced in Ref.~\cite{Cogan:2014oua} 
and then many groups have used modern deep neural network (DNN) architectures 
for jet tagging with  \textit{jet images} \cite{deOliveira:2015xxd, Baldi:2016fql, Komiske:2016rsd, Chakraborty:2019imr, ATLAS:2017dfg}. The CMS collaboration has developed DeepCSV~\cite{CMS:2017wtu, Bols:2020bkb}, DeepFlavour~ \cite{2020HeavyFI,deepflv} and DeepJet \cite{Bols:2020bkb} taggers for multiclass classification for light jets, gluon jets, c-jets and b-jets using a deep learning algorithm. In a recent work \cite{ATLAS:2019bwq}, the ATLAS Collaboration has developed two high-level b-tagging algorithms - MV2 and DL1 \cite{Krizhevsky2012ImageNetCW}. MV2 \cite{Krizhevsky2012ImageNetCW} algorithm is based on BDT
\footnote{This same algorithm has been used in another analysis \cite{ATLAS:2020hpj}, where the ATLAS collaboration has reported the first evidence of $t\bar t t\bar t$ production.}  and is trained within the TMVA framework \cite{TMVA:2007ngy}. The performance of b-jets tagging using the simulated $t\bar t$ events has been presented in Fig.1 in \cite{ATLAS:2019bwq} for BDT and deep feed-forward neural network algorithms. 
Although for jet images, tagging, and substructure studies, deep learning algorithms are the most efficient ones, the ATLAS collaboration recently showed that W-boson tagging and top quark tagging ML algorithms lead to 
a significant gain in efficiency compared to cut-based analysis, as presented 
in Fig.\ref{fig:w-top-tag}~\cite{ATLAS:2018wis}. This analysis has been performed using LHC Run-II data with  $\lum = 36 \ifb$. It may be noted that both the BDT and DNN-based algorithms perform similarly to each other for all signal efficiencies \cite{ATLAS:2018wis}.


\subsection{ Signal and background events classification}

As mentioned earlier, the most important goal of a collider experiment is the production and search of new particles by identifying rare (mostly) signal events from huge backgrounds. The last missing piece of the Standard Model, aka the Higgs boson, was discovered by the CMS and ATLAS collaboration in 2012 
\cite{ATLAS:2012yve,CMS:2012qbp}. The CMS Collaboration has used the BDT algorithm, implemented within the TMVA framework \cite{TMVA:2007ngy}, for the Higgs discovery analysis in Ref~\cite{CMS:2012qbp}. Using the full dataset collected in 2011 and 2012 from 7 \& 8 TeV LHC run, CMS has updated the $h \to \gamma \gamma$ analysis in Ref.~\cite{CMS:2014afl}, where BDT has been used extensively for several tasks such as -  photon identification, photon vertex reconstruction, signal-background classification with good diphoton mass resolution, classification of VBF, Vh, $t\bar t h$ tagged events etc. ATLAS Collaboration has also performed an analysis of 
$t\bar t h$ production, where the Higgs boson (h) decays to a $b \bar b$ pair, using Run-II data with $\lum = 36.1 \ifb$ \cite{ATLAS:2017fak}. In this analysis \textit{Classification BDT} has been trained to separate the signal ($t\bar t h$ ) 
from backgrounds and ATLAS has used \textit{Reconstruction BDT} to select the best combination of jet-parton to reconstruct the Higgs boson and top quark candidates \cite{ATLAS:2017fak}. In the next subsection, we will summarize the R-parity conserving (RPC) and R-parity violating (RPV)~\cite{Barbier:2004ez,Choudhury:2024ggy} searches using DT/BDT-based algorithms. The most distinct features between RPC and RPV SUSY scenarios are large missing energy from a stable LSP in the RPC case and higher lepton and jet multiplicity (arising from RPV couplings) in the latter case. Numerous experimental and phenomenological studies have explored ML algorithms to enhance the discovery reach and exclusion limit in contrast to traditional cut-based analyses. We outline some of the results below.


\subsubsection{Searches for  RPC SUSY scenarios using BDT}
Among different SUSY models, the RPC SUSY\footnote{RPC SUSY can provide the possible Dark Matter (DM) candidate as the LSP 
\cite{KumarBarman:2020ylm, Barman:2022jdg, He:2023lgi,Chakraborti:2015mra,Chakraborti:2014gea,Chakraborti:2017dpu,Chowdhury:2016qnz,Bhattacharyya:2011se,Choudhury:2012tc,Baer:2021aax} and also can explain the muon (g-2) excess \cite{Chakraborti:2022vds, He:2023lgi, Chakraborti:2015mra, Chakraborti:2014gea, Baer:2021aax, Athron:2021iuf, Endo:2021zal, Chakraborti:2021bmv, Choudhury:2017acn, Choudhury:2017fuu, Banerjee:2018eaf, Banerjee:2020zvi, Chakraborti:2021dli, Frank:2021nkq, Ali:2021kxa, Kowalska:2015zja, Chakrabortty:2015ika, Choudhury:2016lku}} is most widely studied by both the ATLAS and CMS collaborations~\cite{atlas_susy, cms_susy}. In several analyses, both collaborations have used the BDT algorithms to improve the sensitivity of sparticle 
searches. In this section, we will summarize the works where BDT algorithms have been used by the experimental and phenomenological groups in the context of RPC SUSY models.   

\noindent\underline{\textbf{Searches for first two generation squarks and gluinos:}}
The ATLAS collaboration has searched for squarks and gluinos in fully hadronic channel ($0l$ + jets + $\met$ final states) based on full Run-II dataset ($\lum =139 \ifb$) using three strategies: \texttt{multibin search, BDT search} and \texttt{model-independent search} \cite{ATLAS:2020syg}. Results were interpreted in various simplified scenarios where gluino can decay directly ($\gl \to q \bar q \lspone$) or via one step   ($\gl \to q \bar q^\prime \chonepm$) as shown in Fig. 1 of Ref.~\cite{ATLAS:2020syg}. Depending on the mass difference $(\Delta m(\gl, \lspone))$, 
ATLAS has separated the events into four categories for these gluino decay modes. Eight independent BDT were trained to obtain the optimal sensitivity  (see Table 7 of Ref.~\cite{ATLAS:2020syg}). Among the three search strategies, the 
\texttt{BDT search}  was the most robust tool, particularly effective in scenarios involving gluino decays via $ q \bar q^\prime \chonepm$, resulting in complex event topology with a large number of jets. The optimized BDT regions, chosen based on their BDT score, achieved the best sensitivity due to their ability to exploit the correlations between variables. In regions of the parameter space where the mass difference between the LSP - NLSP is small and approaches the kinematic limit, the \texttt{multi-bin search} is particularly effective and has excluded gluino and neutralino masses upto $\sim$ 900-1000 GeV (see Fig.\ref{fig:atlas_gluino}) 
from gluino pair production. For relatively light neutralino masses, the observed lower limit on the gluino mass reaches upto 2.3 (2.2) TeV for direct (one-step) gluino decay~\cite{ATLAS:2020syg} and is derived from the optimized BDT regions (see Fig.\ref{fig:atlas_gluino}). The same analysis has also excluded  $\msq$ upto 1.85 
TeV for massless lightest neutralino considering the pair production of mass degenerate first and second generation squarks.  

\begin{figure}[!htb]
\begin{center}
\includegraphics[width=0.488\textwidth]{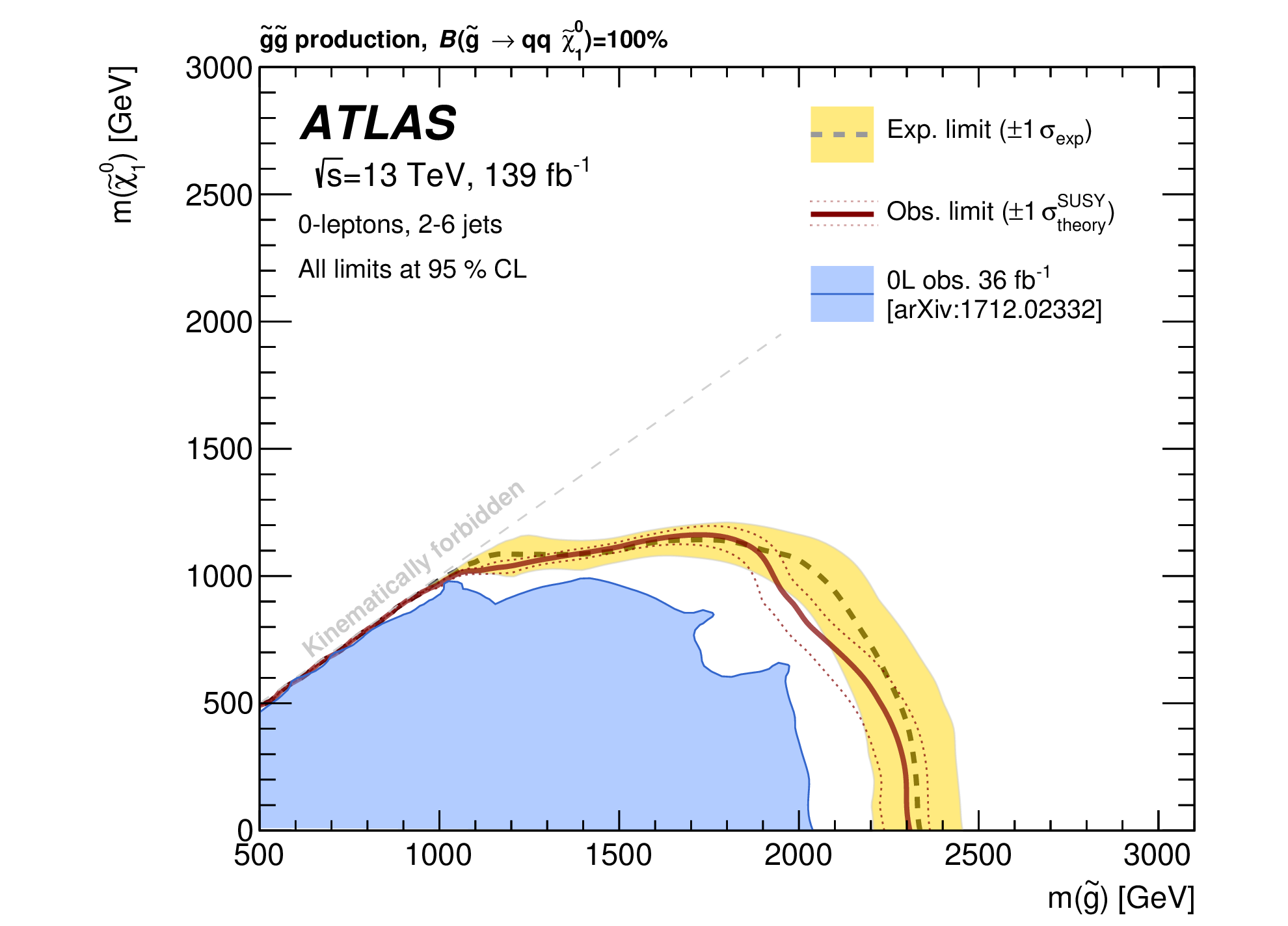}
\includegraphics[width=0.488\textwidth]{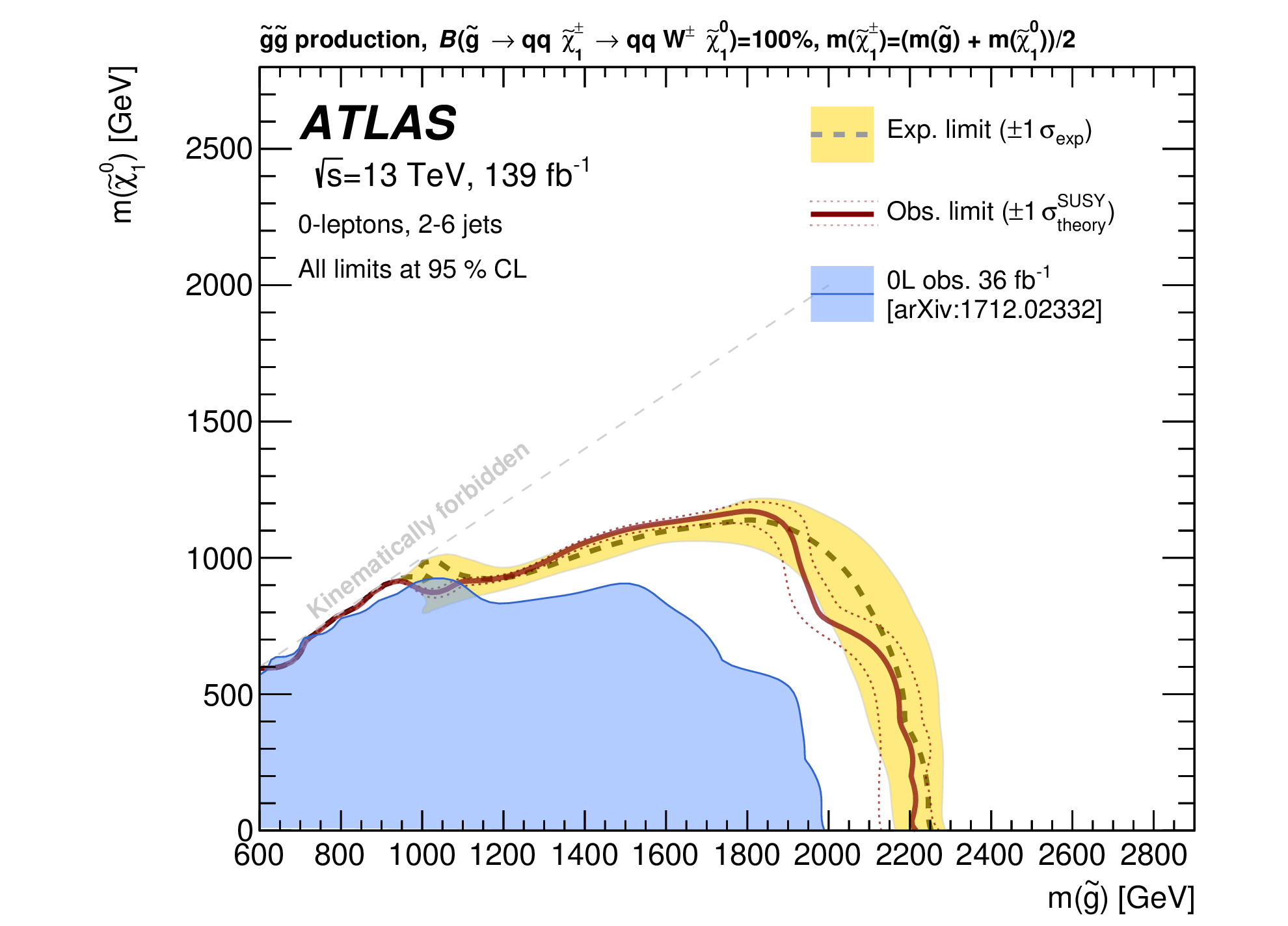}
\caption{ Exclusion limits in the $\mgl- \mlspone$ plane, obtained using \texttt{ multi-bin search, BDT search} and \texttt{model-independent search}, from gluino pair production  for direct decays ($\gl \to q \bar q \lspone$) in the left panel and for one-step   ($\gl \to q \bar q^\prime \chonepm$) decay in the right panel 
 \cite{ATLAS:2024fub}. 
  }
\label{fig:atlas_gluino} 
\end{center}
\end{figure}

\begin{figure}[!htb]
\begin{center}
\includegraphics[width=0.63\textwidth]{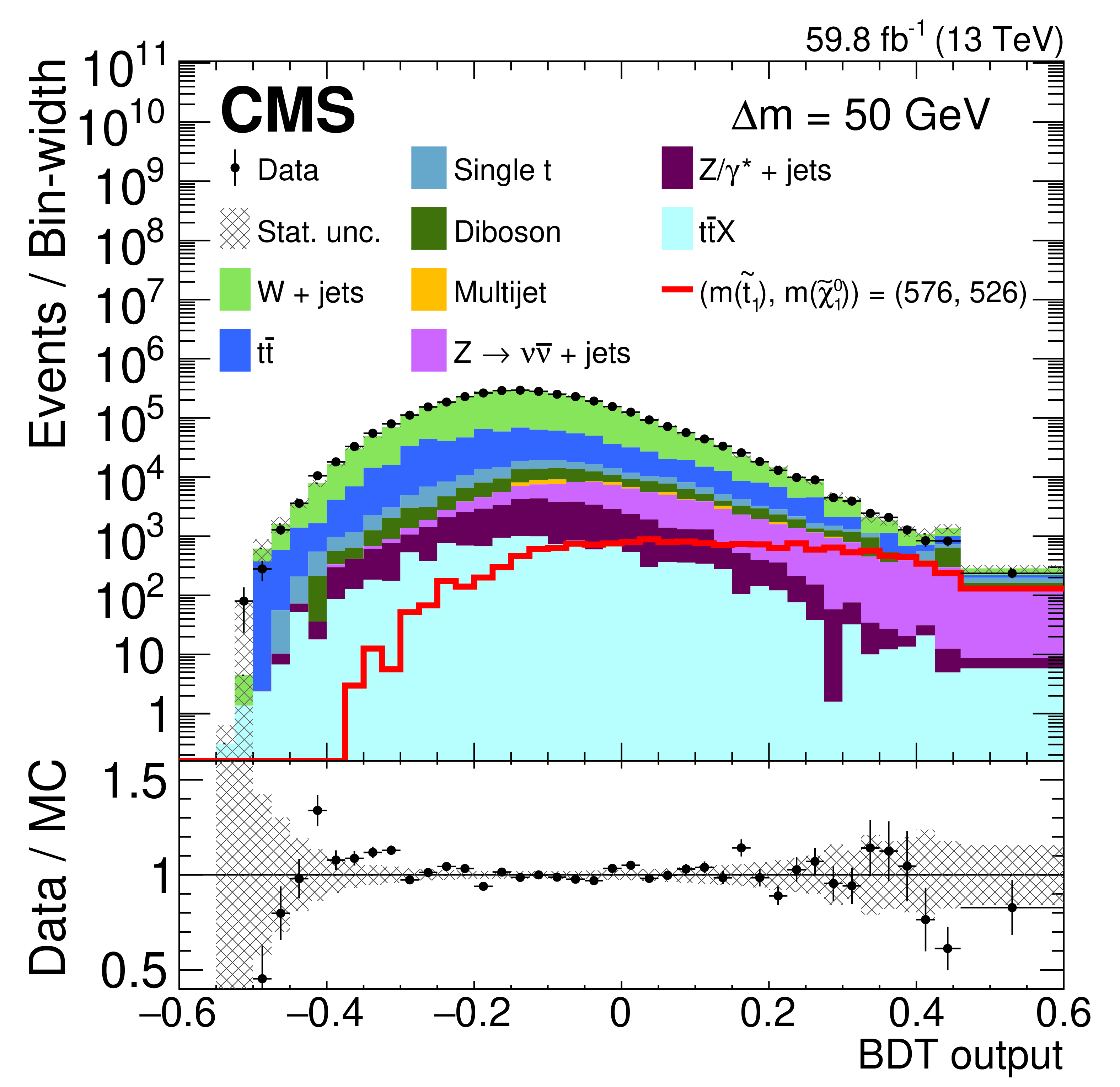}
\caption{ Left: Distribution of BDT output form 13 TeV CMS data with $\lum = 59.8 \ifb$ from stop pair production where $\Delta m (\st1 -\lspone) = 50$ GeV and stop decays via $b f \bar f^\prime \lspone$ mode \cite{CMS:2023ktc}. 
 } 
\label{fig:cms1} 
\end{center}
\end{figure}

\noindent\underline{\textbf{Searches for Stop:}}
The light stop scenario is theoretically well-motivated \cite{Balazs:2004bu} and also has a rich phenomenology. Both the ATLAS and  CMS collaborations have looked for the lightest stop squarks ($\st1$) through all possible decay modes with various final states. 
The possible decay modes of stop are $t \lspone$, $b \chonepm$, $bW\lspone$, $c \lspone$, $b f \bar f^\prime \lspone$. The last one (four body mode) may dominate in the compressed SUSY scenario where the other decay modes are kinematically forbidden and  $c \lspone$ is suppressed. In a recent study, the CMS collaboration 
has used the BDT algorithms \cite{TMVA:2007ngy} to optimize the separation between background and signal events~\cite{CMS:2023ktc}. The discriminating variables of signal and background processes have different correlations and CMS has performed the BDT  analysis for different values of $\Delta m (\st1 -\lspone)$. The distribution of the score of multivariate analysis (the BDT discriminator value or BDT output) is shown in Fig.\ref{fig:cms1} as obtained by CMS for $\Delta m (\st1 -\lspone) = 50$  GeV. This BDT search has excluded $\mst1$ upto 480 and 700 GeV for 
 $\Delta m$ = 10 and 80 GeV at 95$\%$ CL \cite{CMS:2023ktc}. 
 CMS had also implemented a  BDT multivariate approach to define the signal regions for the analysis of stop pair production with subsequent  $ \st1 \to t \lspone$, and  $ \st1 \to b \chonepm$ decays in Ref.~\cite{CMS:2013phu}. The comparison between the exclusion plots obtained by CMS \cite{CMS:2013phu} for  $ \st1 \to  b \chonepm$ decay using cut-based and BDT methods are presented in the left and right panels of 
 Fig.\ref{fig:cms2}.

\begin{figure}[!htb]
\begin{center}
\includegraphics[width=0.49503\textwidth]{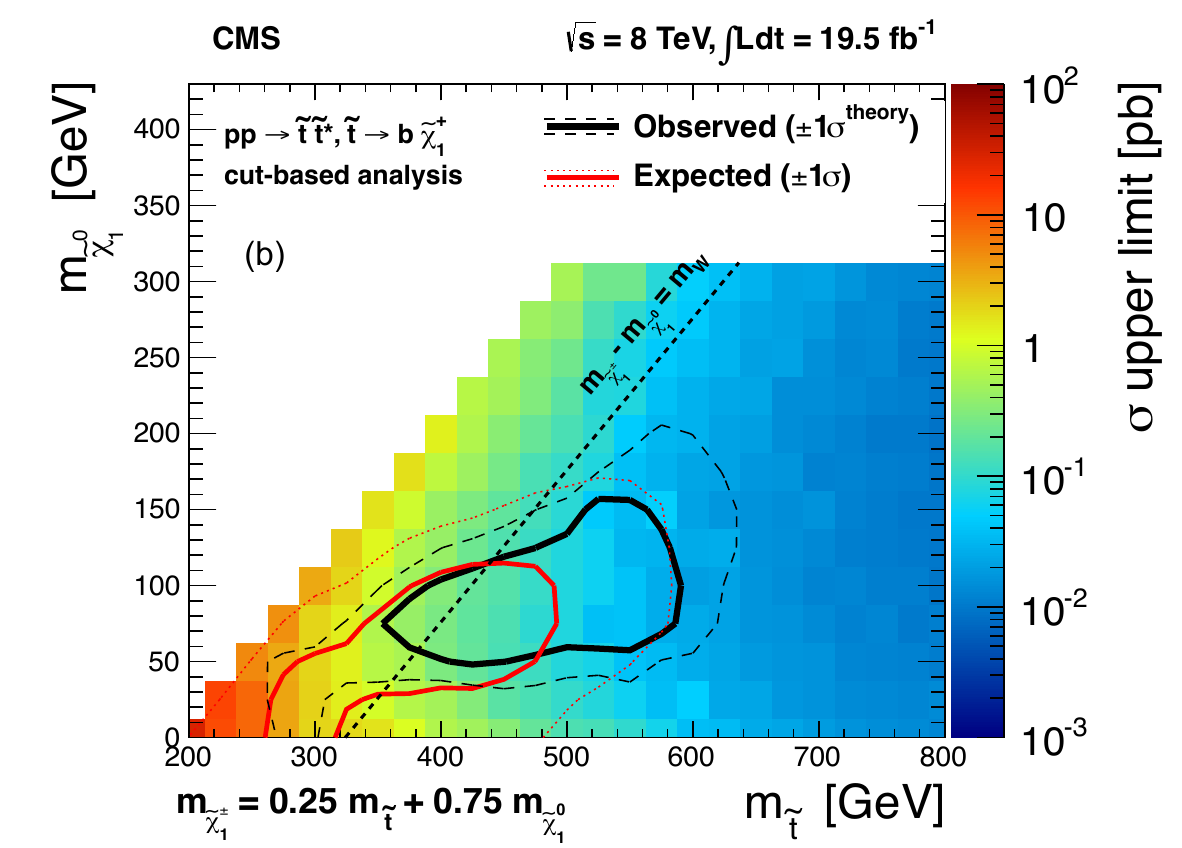}
\includegraphics[width=0.49503\textwidth]{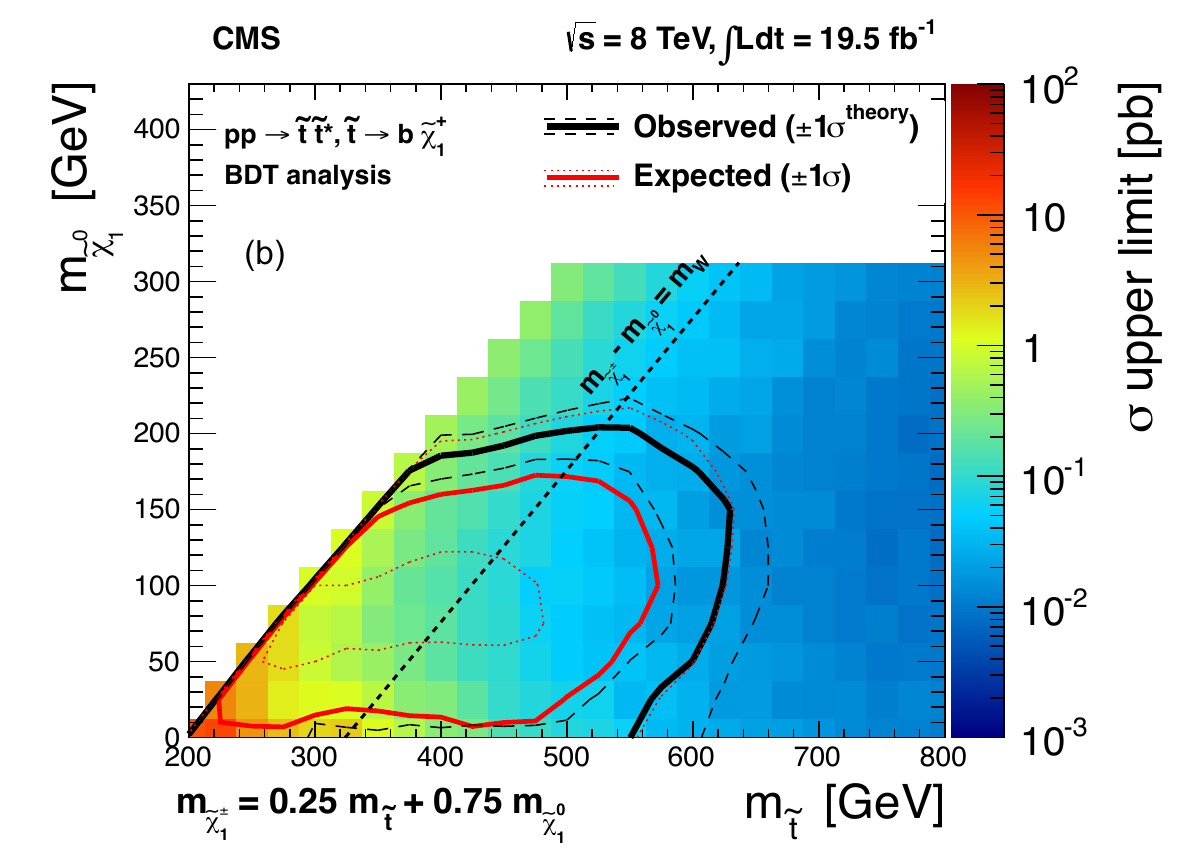}
\caption{ 
The exclusion plot in $\mst1 - \mlspone$ plane for 
$ \st1 \to b \chonepm$ decay from cut-based (left panel) and BDT method (right panel) respectively \cite{CMS:2013phu}. } 
  \label{fig:cms2} 
\end{center}
\end{figure}

A recent phenomenological work  \cite{Jorge:2021vpo} 
has expanded and improved the ATLAS analysis \cite{ATLAS:2020xzu} for 
$\st1 \st1$ pair production in the semileptonic channel where the stop squarks decay via 3 body mode $bW\lspone$. ATLAS has used a recurrent neural network (RNN) algorithm and excluded stop mass upto 710 GeV using 13 TeV LHC data with $\lum = 36.1 \ifb$~\cite{ATLAS:2020xzu}. The authors compared the performance of 
Logistic Regression, Random Forest, XGBoost and Neural Network algorithms 
in the Ref.~\cite{Jorge:2021vpo} at 13 TeV LHC with $\lum = 140 \ifb$. It has been shown that in such compressed scenarios, 
 XGBoost and Neural Network algorithms improve the signal significance 
 more efficiently compared to other algorithms as well as the cut-and-count approach.

\begin{figure}[!htb]
\begin{center}
\includegraphics[scale=0.15]{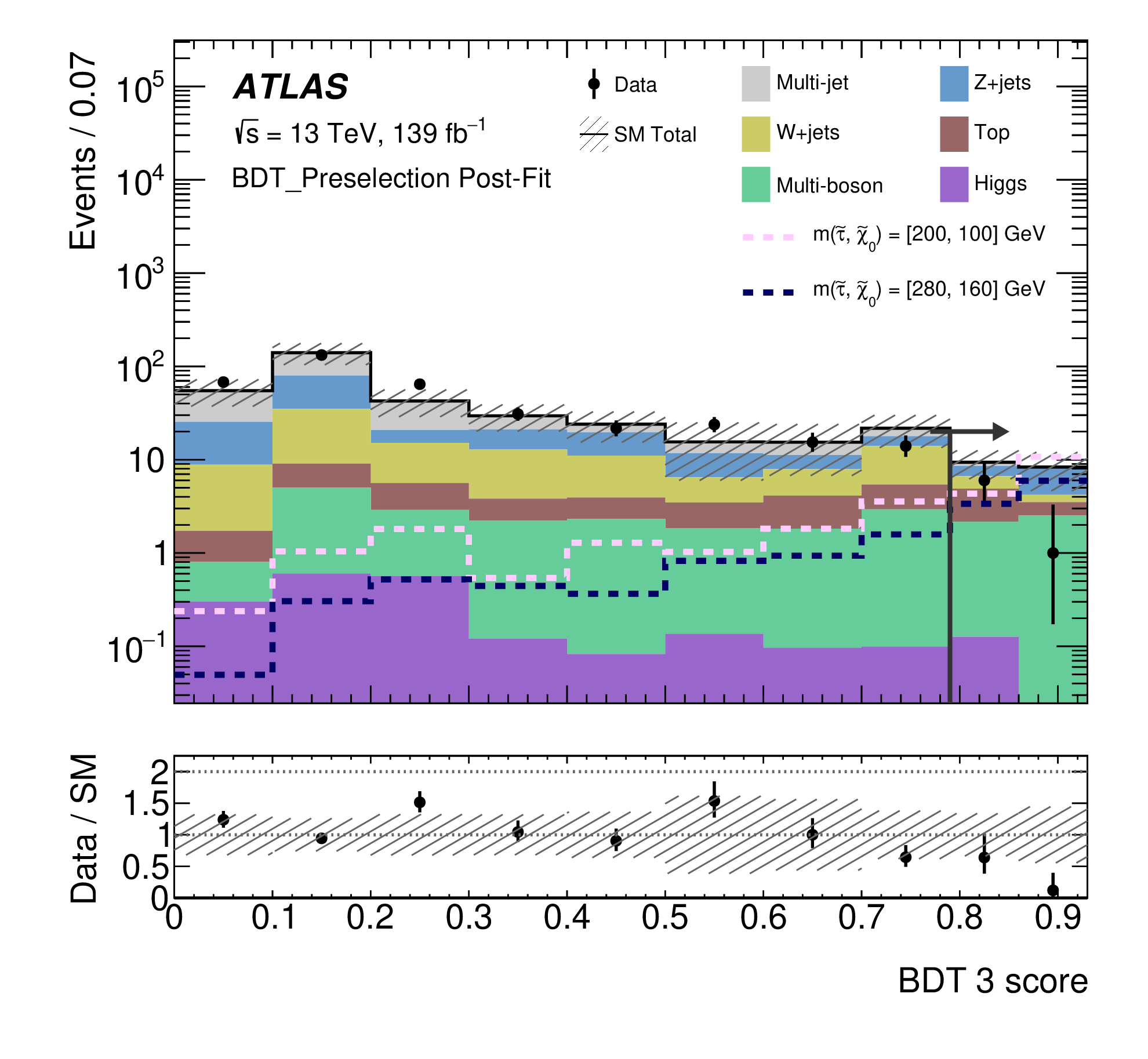}
\caption{ The post-fit BDT score distributions for the direct stau channel for BDT3 \cite{ATLAS:2024fub}. 
The black arrow represents the BDT score selection for the SR-BDT3. Two SUSY benchmark points are shown with a dashed line. For more details, see Ref. \cite{ATLAS:2024fub}. 
  }
\label{fig:atlas2} 
\end{center}
\end{figure}

\noindent\underline{\textbf{Searches of Electroweakinos and sleptons:}}
Very recently, for the first time, the ATLAS Collaboration has reported the sensitivity of $\staur$ only RPC SUSY scenarios with $2\tau + \met$ final states using the LHC Run-II dataset with $\lum = 139 \ifb$ \cite{ATLAS:2024fub}. In this analysis, other scenarios with $\chonepm \chonemp$ and $\chonepm \lsptwo$ pair productions also have been considered. The LightGBM package has been used  
to train multiple BDTs on the sensitivity of stau-LSP phase space. The ATLAS experiment has excluded $\staur$ masses upto 350 GeV by improving the signal background separation through the use of BDT~\cite{ATLAS:2024fub}. The training was done for four  BDT and for illustration purpose, we present the post-fit BDT score distribution for the signal region BDT3~\cite{ATLAS:2024fub} in Fig.\ref{fig:atlas2}
In another recent study, the ATLAS collaboration has looked for the chargino pair production via  2l + $\met$ final states where the mass gap between the charginos and lightest neutralinos are close to W boson mass~\cite{ATLAS:2022hbt}. For this analysis, ATLAS has performed a multiclass GBDT classification using LightGBM~\cite{NIPS2017_6449f44a}, and signal regions are defined according to BDT scores (see Table 3 of Ref.~\cite{ATLAS:2022hbt}). The observed and expected numbers of events are presented in Fig.~\ref{fig:atlas1} along with the significance. 
In the absence of any significant excess, ATLAS has excluded $\mchonepm$ up to 140 
GeV at 95\% CL for $\Delta m (\chonepm -\lspone) \sim 100$ GeV \cite{ATLAS:2022hbt}. 

\begin{figure}[!htb]
\begin{center}
\includegraphics[width=0.7\textwidth]{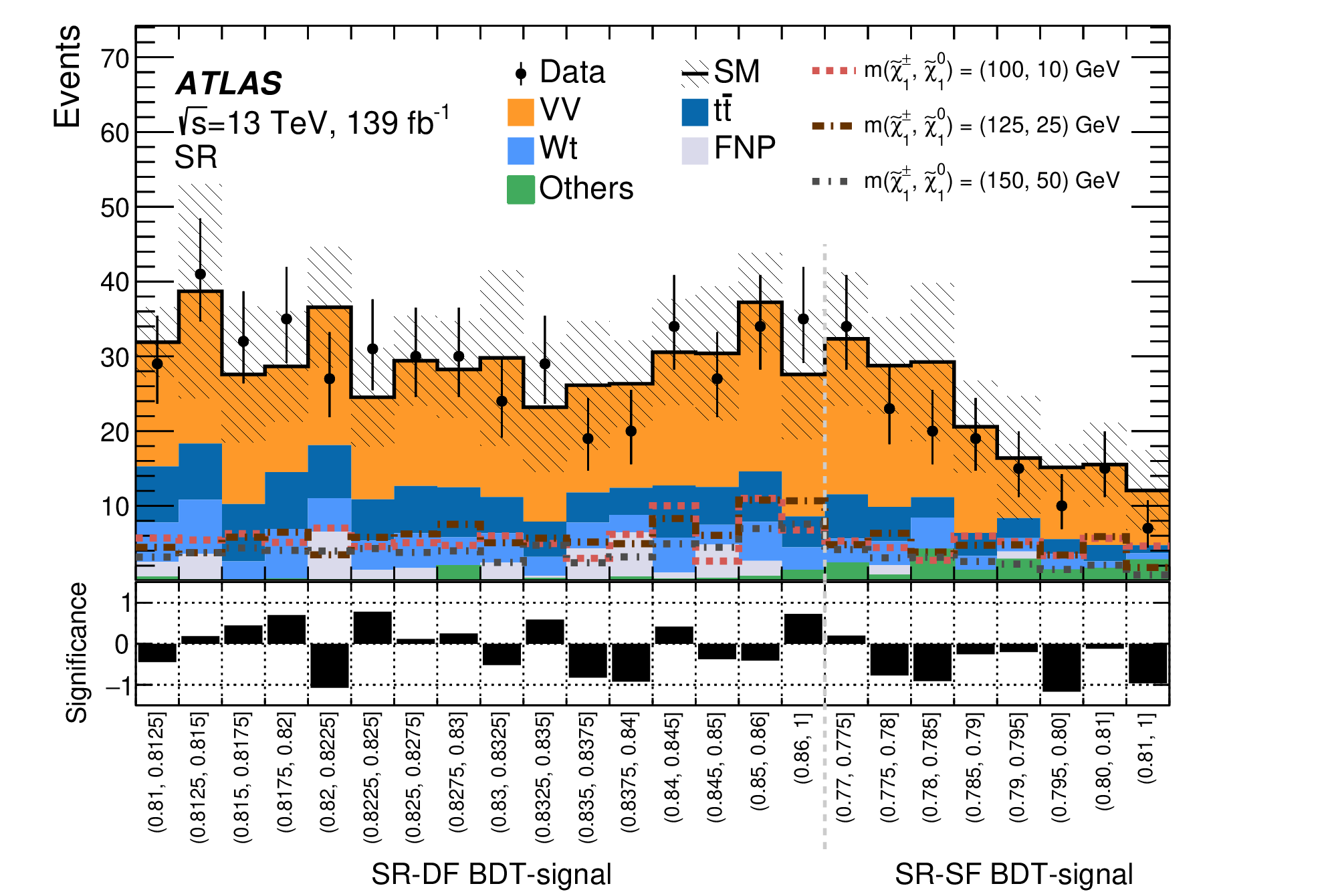}
\caption{The observed and expected number of events along with significance provided by ATLAS~\cite{ATLAS:2022hbt} is presented here. 
}
\label{fig:atlas1}
\end{center}
\end{figure}

In the Ref~\cite{Cornell:2021gut}, the authors have shown that the gradient boosting algorithm can extend the 95\% C.L. exclusion limits derived from traditional cut-based methods using an example of smuon pair production at the 13 TeV LHC.   The authors have also studied the utility of various performance metrics (auc, ams, F-score etc.) using \texttt{XGBoost} toolkit, feature importance using  SHAP package \cite{Shapley+1953+307+318,lundberg2018consistent} in great details~\cite{Cornell:2021gut}. The possibility of constructing a general machine learning model that may be applied to probe a two-dimensional mass plane has also been examined in this work. 
In \cite{Alvestad:2021sje}, the authors pointed out 
that \texttt{XGBoost} algorithm 
can significantly increase the detectability of a SUSY models 
with a gravitino type LSP and a metastable sneutrino NLSP by analyzing the events 
coming from the electroweakinos production along with the slepton productions. 
The possibility of probing the light Higgsinos, which are still allowed by LHC Run-II and LZ experiment data,  at the upcoming HL-LHC run via \texttt{XGBoost} framework has been studied in Ref.~\cite{Barman:2024xlc}

\begin{figure}[!htb]
\begin{center}
\includegraphics[width=0.5\textwidth]{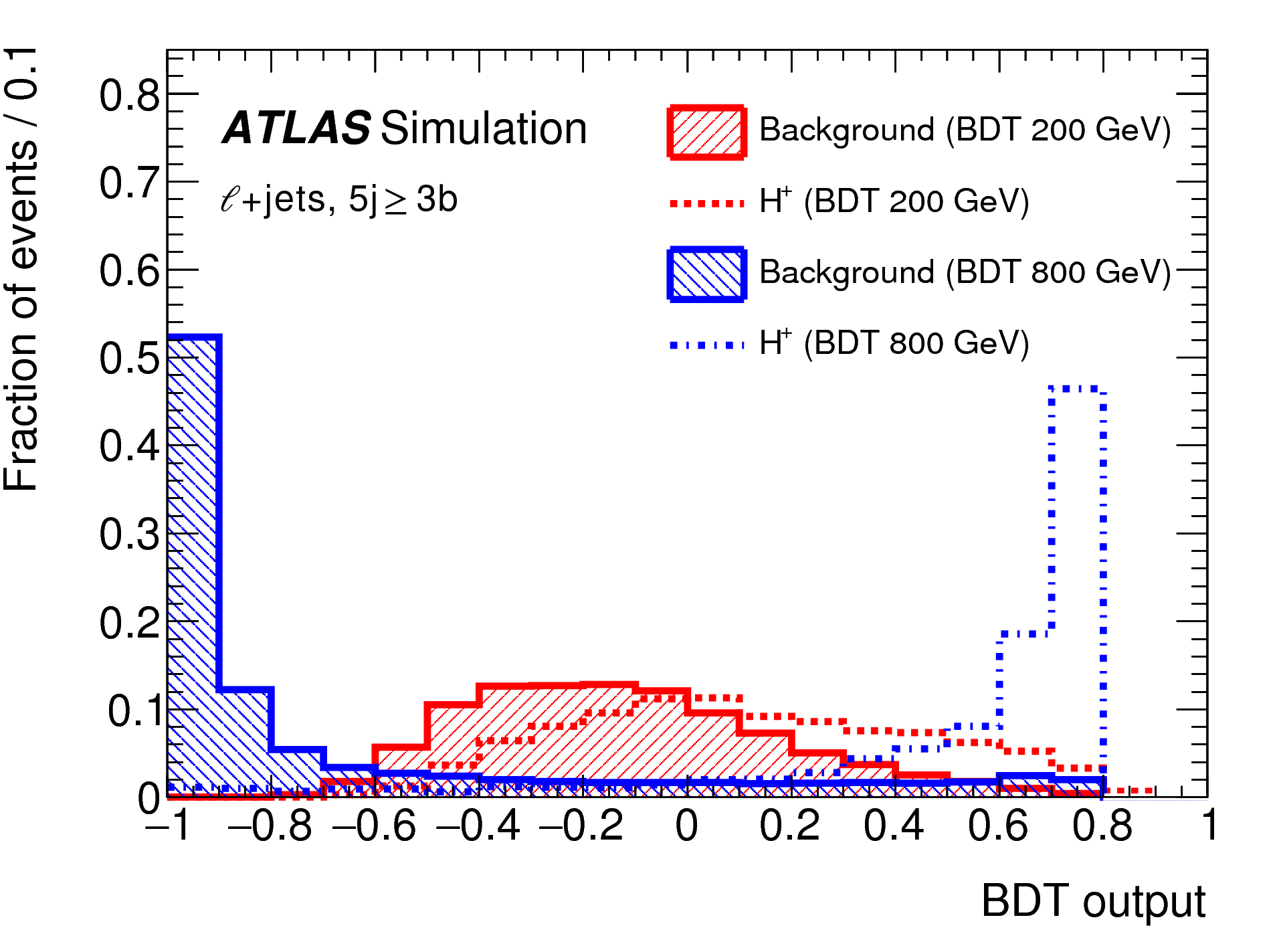}
\includegraphics[width=0.4\textwidth]{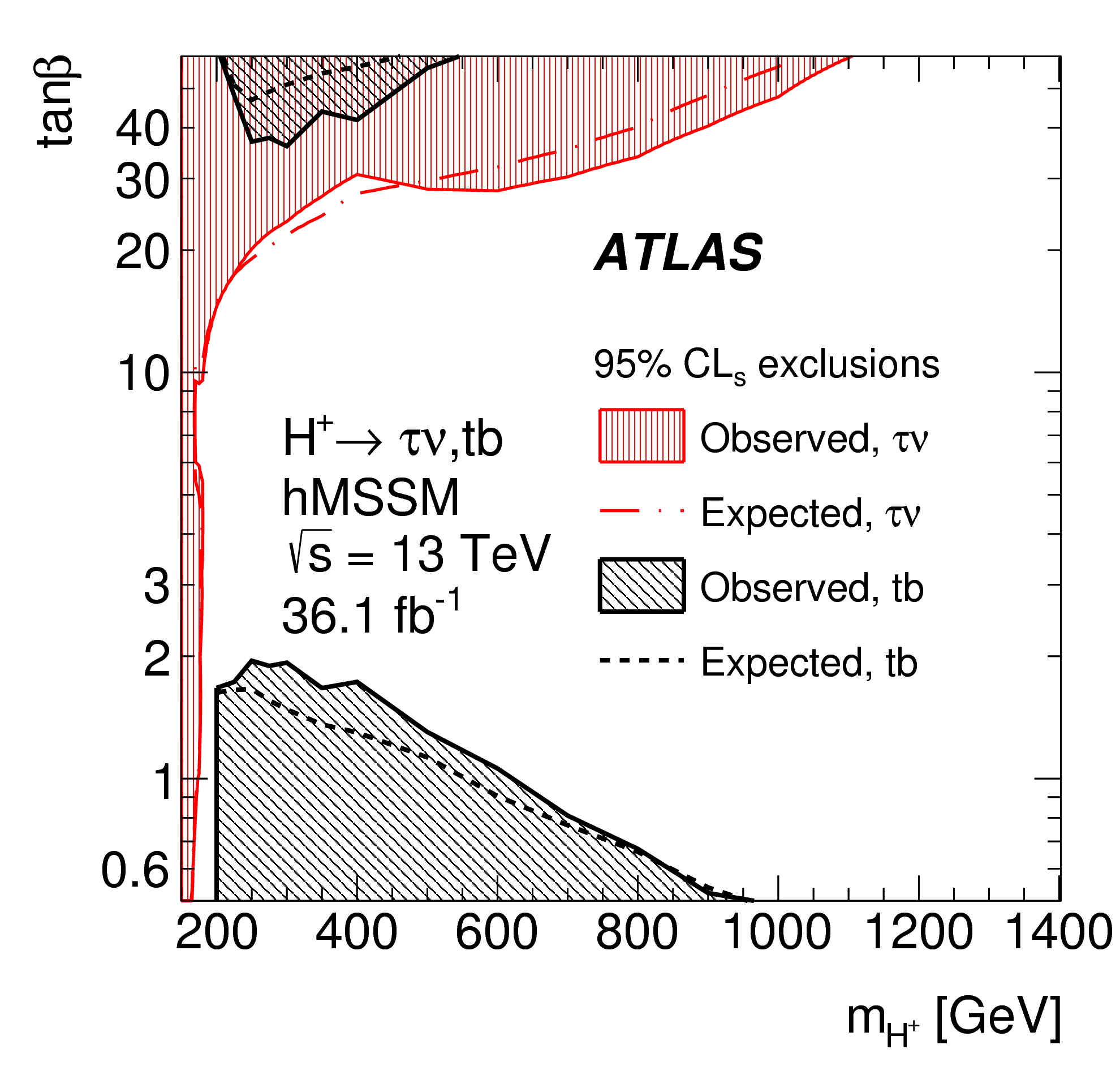}
\caption{
Left: BDT score distribution obtained by ATLAS~\cite{ATLAS:2018ntn} for 
$m_{H^\pm}$ = 200 and 800 GeV along with SM backgrounds. 
Right: Expected (dotted line) and observed (hatched area) limits  on tan$\beta$ as a function $m_{H^\pm}$ 
from the $H^+ \to \tau \nu_\tau $ (red color) \cite{ATLAS:2018gfm} and $H^+ \to t b $  decay (black) \cite{ATLAS:2018ntn} in the hMSSM scenario \cite{Djouadi:2013uqa}
}

\label{fig:atlas3} 
\end{center}
\end{figure}

\noindent\underline{\textbf{Searches of Heavy Higgs:}}
The Minimal Supersymmetric Standard Model (MSSM) consists of two Higgs doublets and after electroweak symmetry breaking, the Higgs sector contains two CP even Higgs (h and H), one CP odd Higgs boson (A) and two charged Higgs bosons $H^\pm$. 
Extensive studies from LHC collaborations and phenomenological groups have already been performed for the Heavy Higgs searches. The implication of LHC results from 
125 GeV Higgs boson searches and other heavy Higgs searches have been analyzed 
in Ref.~\cite{Bagnaschi:2017tru,GAMBIT:2017zdo,Bhattacherjee:2015sga,Barman:2016jov}. The ATLAS collaboration has used the multivariate BDT analysis for the searches of charged Higgs ($H^\pm$) in Ref.~\cite{ATLAS:2018ntn, ATLAS:2018gfm}. For $m_{H^\pm} > m_{top}$ and in the alignment limit, $H^+ \to t b $  is the dominant decay mode and for large values of tan$\beta$, the branching ratio of $H^+ \to \tau \nu_\tau $  becomes $\sim 10\% $. 
ATLAS and CMS both have looked for these final states where the dominant production mode is $p p \to t b H^+$ \cite{ATLAS:2018ntn, ATLAS:2018gfm, CMS:2020imj, CMS:2019bfg}. CMS collaboration has used a a multivariate BDT with gradient boost (BDTG) classifier within TMVA toolkit for $H^+ \to t b $ mode~\cite{CMS:2020imj}. ATLAS has performed the training of the BDTs with the TMVA toolkit for different values of $m_{H^\pm}$ in various signal regions to discriminate the $H^\pm$ signal from the SM backgrounds with $1l (l =e,\mu) + n_j + m_{bjet}$  final states~\cite{ATLAS:2018ntn}. The BDT score distribution obtained from this analysis for signal benchmark points  with $m_{H^\pm}$ = 200 and 800 GeV along with SM backgrounds are shown in Fig.\ref{fig:atlas3} (left panel) from the 
$ 1l+5j\geq$ 3b channel. For the $\tau \nu_\tau$ final states, training of the BDT 
has been performed using the FASTBDT algorithm \cite{Keck:2017gsv}. The results \cite{ATLAS:2018gfm} were interpreted in the $tan\beta$ - $m_{H^\pm}$  mass plane (see the right panel in Fig.~\ref{fig:atlas3}). The limit comparison plot shows that at high tan$\beta$, stringent limits come from the $\tau \nu$ channel and at low tan$\beta$, tb channel is more effective for  $m_{H^\pm} > 200$ GeV.

The authors in Ref.~\cite{Baer:2021qxa} have studied the prospect of Heavy Higgs search in a radiatively-driven natural supersymmetric  model where 
the neutral Heavy Higgs decays to $\lsptwo \lsptwo$ or $\chonepm \champ2$ 
and give rise to $4l+\met$ signature. 
Detecting this signature at the HL-LHC will be very difficult. But it is shown the conventional cut-based method can probe $m_A,H$ upto 1.65 TeV at the future 100 TeV pp collider \cite{Baer:2021qxa} with signal significance greater than 5$\sigma$. Using a multivariate analysis (BDT) within the TMVA toolkit, the authors also pointed out that the BDT algorithm improves the signal significance and heavy Higgs can be probed upto 2 TeV.

\subsubsection{Searches for  RPV SUSY scenarios using BDT}
One of the key motivations for going beyond the RPC MSSM is its inability to provide an explanation for the neutrino oscillation phenomena. The RPV SUSY scenarios can explain the light neutrino masses and mixing~\cite{Grossman:2003gq, Davidson:2000uc, Roy:1996bua, Allanach:2007qc, Diaz:2014jta, Choudhury:2023lbp}. 
The RPV SUSY models can also address the muon (g-2) anomaly~\cite{Chakraborty:2015bsk, Altmannshofer:2020axr, Chakraborti:2022vds, Hundi:2011si} or flavor anomalies~\cite{Trifinopoulos:2019lyo, Domingo:2018qfg, Das:2017kfo}. 
Due to the presence of lepton and/or baryon number violating terms, the LSP becomes unstable and decays to SM particles. As a result, the final state 
contains less missing energy and a higher number of leptons and/or jets, depending upon the coupling. Although the ATLAS and CMS collaboration have mainly presented results in the RPC SUSY scenarios, there exist a few analyses, based on cut based method,  involving strong or electroweak sparticles pair production with subsequent RPV decays \cite{ATLAS:2021yyr,ATLAS:2021fbt,ATLAS:2024kqk,CMS:2017szl,CMS:2021knz,CMS:2018skt}. A recent work \cite{Dreiner:2023bvs} has summarized the possible gaps in
RPV-MSSM searches at the LHC in great detail by meticulously classifying 
the various possible RPV-MSSM signatures at the LHC and analyzing 
both direct and indirect production of various LSPs, the authors have derived limits on sparticle masses.

\begin{figure}[!htb]
\begin{center}
\includegraphics[width=0.49\textwidth]{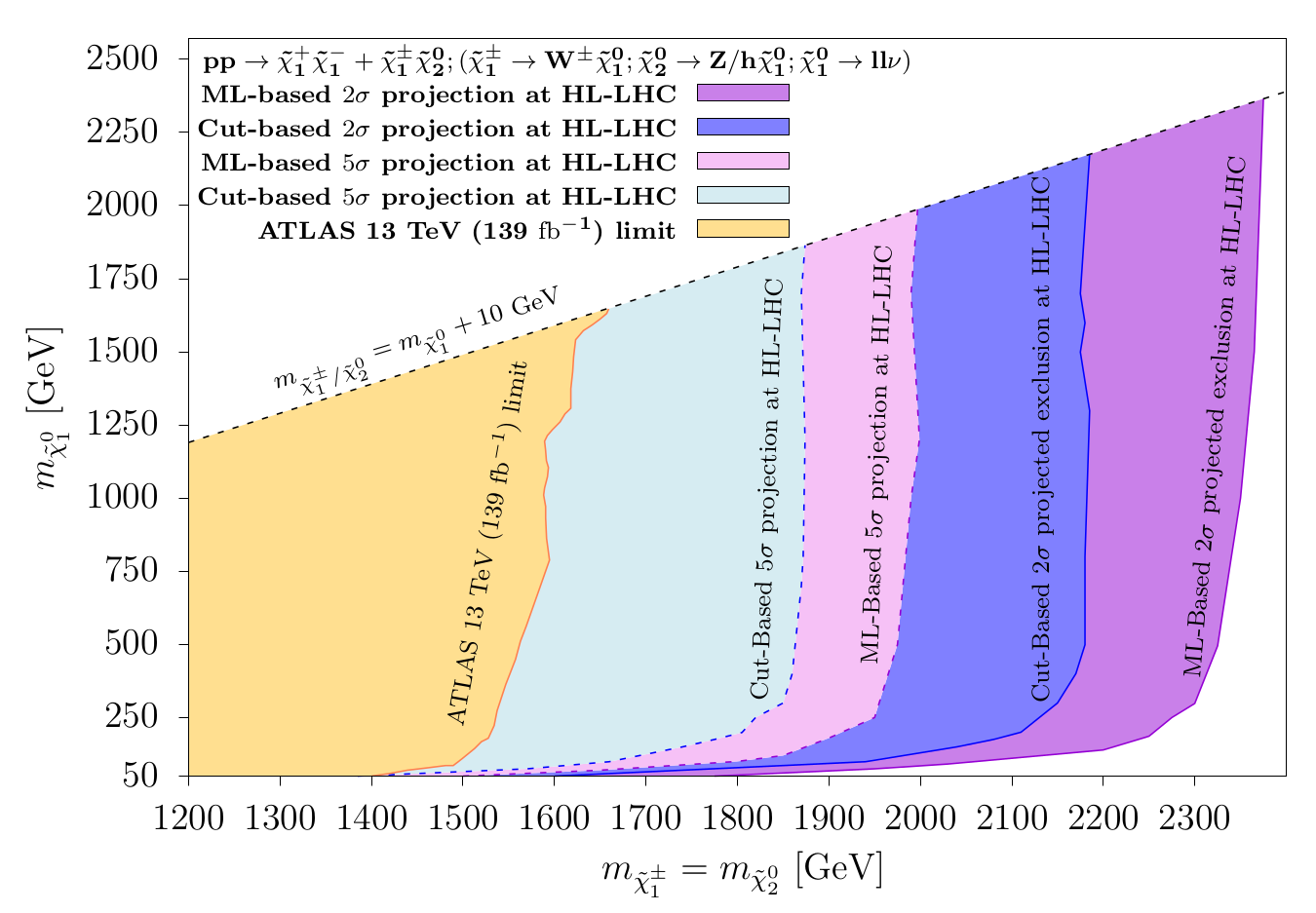}
\includegraphics[width=0.49\textwidth]{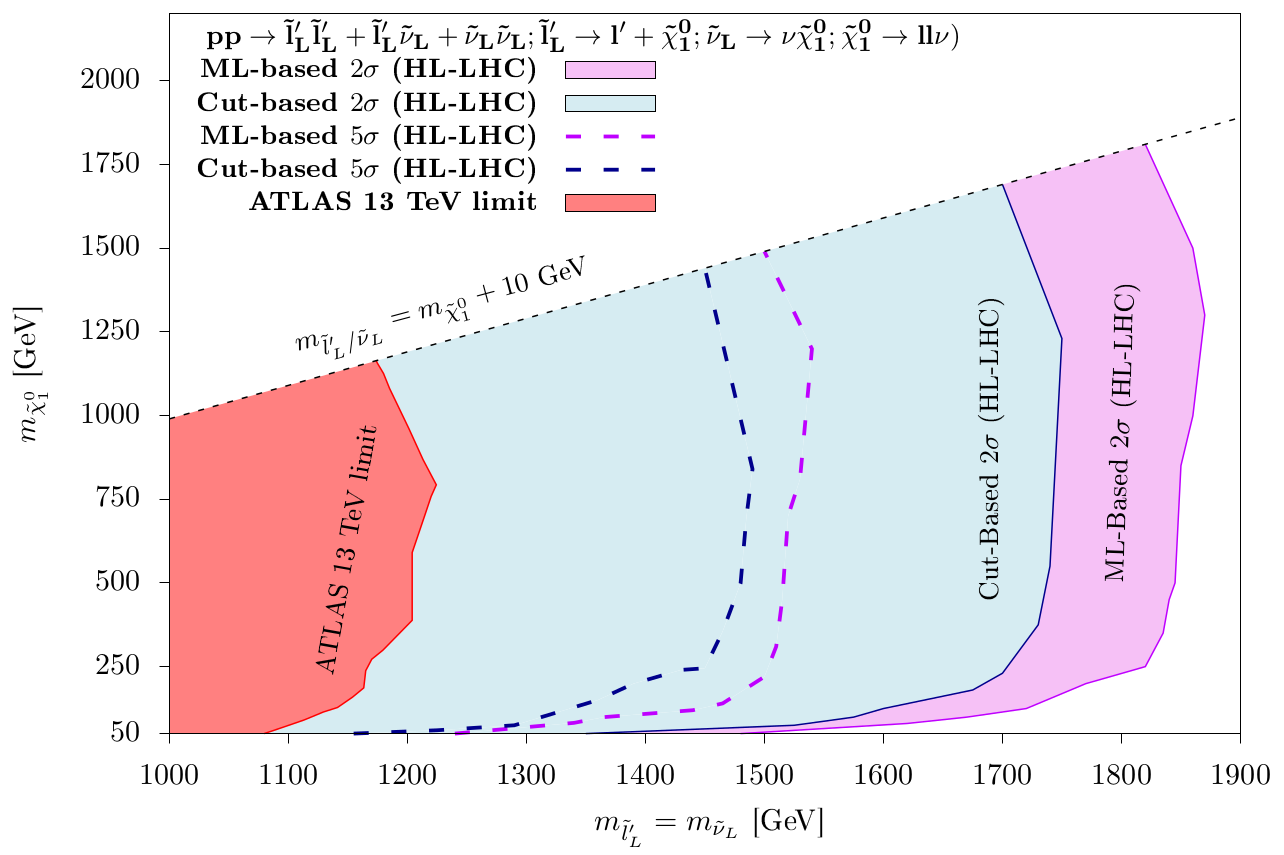}
\caption { Projected discovery ($5\sigma$) and exclusion ($2\sigma$) limits from the wino searches \cite{Choudhury:2023eje} and slepton searches \cite{Choudhury:2023yfg} at the HL-LHC are presented in the left and right figure respectively. In the left panel, the dark blue (violet) color refers to $2\sigma$ reach from the cut-based (ML) analysis and the light blue (violet) color represents $5\sigma$ reach from cut-based (ML) analysis~\cite{Choudhury:2023eje}. In the right panel, the light violet and blue color regions can be probed from the ML and cut-based analysis, respectively, at the $2\sigma$ level~\cite{Choudhury:2023yfg}.}
\label{fig:ew}
\end{center}
\end{figure}

In the presence of non-zero trilinear couplings $\lambda_{ijk}$, the  LSP decays into $l_k^{\prime\pm}l_{i/j}^{\prime\mp}\nu_{j/i}$, where $l^{\prime} = e, \mu, \tau$. Thus, the final states become leptonically enriched. 
For different choices of $\lambda_{ijk}$ couplings, the LHC collaborations have 
already derived limits on chargino and slepton masses~\citep{ATLAS:2021yyr}. 
The yellow and red regions in the left and right panels of Fig.\ref{fig:ew} represent the excluded regions on chargino-LSP and slepton-LSP mass plane. 
The authors in Ref.~\cite{Choudhury:2023eje, Choudhury:2023yfg} have extended the similar $4l + \met$ final states in the context of the electroweak sparticle  
searches at the 14 TeV high-luminosity LHC (HL-LHC) and the proposed
high-energy (27 TeV) upgrade of the LHC (HE-LHC). It has been shown that 
an optimized cut-based analysis will be able to exclude $\mchonepm$ upto 
 2.18 TeV  at the HL-LHC from wino like  $\chonepm\chonemp$ and $\chonepm\lsptwo$ productions \cite{Choudhury:2023eje}. The authors have further explored 
 a multivariate analysis based on an Extreme Gradient BDT algorithm to 
 improve the results further. 
 The projected $2\sigma$ exclusion limit reaches up to
2.37 TeV and 4.0 TeV at the HL-LHC and HE-LHC, respectively, from 
the ML-based analysis~\cite{Choudhury:2023eje} (the dark violet color represents the   $2\sigma$  reach for 
HL-LHC in Fig the left panel of Fig.\ref{fig:ew}). 
In another recent work~\cite{Choudhury:2023yfg}, 
considering the pair and associated production of mass degenerate 
sleptons, the authors have studied the sensitivity of similar  final state 
at the future LHC. The right panel of Fig.\ref{fig:ew} shows the 
discovery and exclusion reach on L-type slepton masses obtained for 
nonzero $\lambda_{121}$ and/or $\lambda_{122}$ coupling values 
\cite{Choudhury:2023yfg}. 
It is observed that 
the projected exclusion limits on slepton and sneutrino masses at the HL-LHC (HE-LHC) are $\sim$ 1.85 (3.0) TeV from ML-based analysis. The ML algorithm 
shows significant improvement over the cut-and-count method (see Fig.\ref{fig:ew}). 
The work in Ref.~\cite{Bhattacherjee:2023kxw} has addressed the searching prospect of long-lived particles in the RPV SUSY scenarios from the electroweakino pair production, 
where the LSP decays via UDD-type couplings. Utilizing the 
Gradient Boosting algorithm within \textbf{XGBoost} toolkit, 
the authors obtained that wino-like (higgsino like) $\lsptwo/\chonepm$ 
with a mass of  1900 (1600) GeV and $\lspone$ with a mass greater than 800 (700) 
GeV can be probed for decay length ranging from 1 cm to 200 cm~\cite{Bhattacherjee:2023kxw}.

\section{Decision Tree algorithms}
\label{sec:decision_tree}
The decision trees~\cite{Breiman:1984jka} are one of the simplest and most widely used 
classification and regression tools, which use an inverted tree like structure with a root on the top, to predict the value of an output by applying binary cuts in the feature space. Breiman et al.~\cite{breiman,Breiman:1984jka} proposed \textit{``Classification and Regression Trees'' (CART)}
algorithm for the implementation of decision trees (DT). The CART algorithm 
begins with the root-node consisting of the entire training dataset. It then proceeds to split the parent node into several branches or child nodes 
until the model can make a decision regarding the class of a given data point 
or it satisfies a predetermined stopping criteria. 
Through this binary splitting, the algorithm effectively partitions the data into subsets. When the splitting criterion is met by the node, it is termed as a leaf.
We demonstrate this process with an example.
Let us consider a gluino pair production channel at the LHC, where each gluino decays into one stable neutralino (LSP) and two quarks ($\tilde{g}\rightarrow q\bar{q}\lspone$), resulting in a final state consisting of four jets associated with a large missing transverse energy ($\met$). It should be noted that for the SUSY signal, the final states will consist of jets with large momentum and large values of effective mass ($m_{eff}$) compared to the SM backgrounds.
\begin{table}[!htb]
\begin{center}
\scriptsize
\begin{tabular}{|c|c|c|c|>{\columncolor[gray]{0.8}}c|}
\hline
\rowcolor{yellow}
$p_T^{j_1}$ GeV &  $N_b$ in GeV & $\met$ in GeV & $m_{eff}$ in GeV & Label \\ \hline
240.2 & 0 & 738.1 & 2112.1 & $s$ \\ \hline
110.4 & 1 & 52.9 & 352.4 & $b$ \\ \hline
98.3 & 0 & 348.0 & 1578.2 & $s$ \\ \hline
87.7 & 1 & 68.4 & 345.7 & $b$ \\ \hline
61.3 & 1 & 27.3 & 130.6 & $b$ \\ \hline
129.8 & 1 & 102.5 & 812.1 & $b$ \\ \hline
\end{tabular}
\caption{Dataset to classify signal and background for four input features. The rightmost column (in gray) represents the label, whether it is a signal ($s$) or background ($b$).}
\label{tab:dt_hep}
\end{center}
\end{table}
\begin{figure}[!htb]
\begin{center}
\includegraphics[scale=0.7]{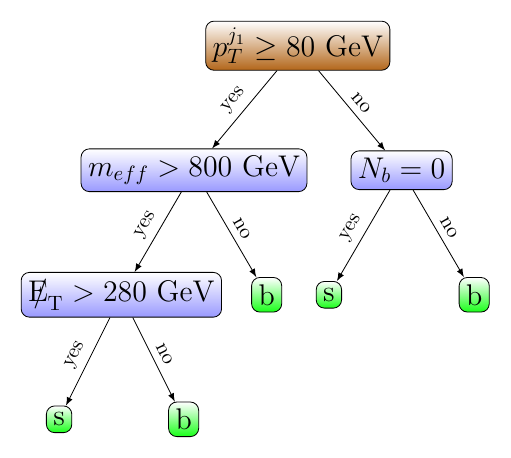}
\caption{A decision tree based on the dataset given in Table. \ref{tab:dt_hep} is shown here. The root node is marked with brown color. Ordinary and leaf nodes are denoted using blue and green colors, respectively.}
\label{fig:dt_hep}
\end{center}
\end{figure}
 Here, $m_{eff}$ denotes the scalar sum of $p_T$ of jets having $p_T\geq50$ GeV added with $\met$. In this case, $t\bar{t}$ + jets act as the most dominant SM background. We label the supersymmetric signal as `$s$' and the SM background as `$b$'. To separate signal and background, four variables (input features) are considered, namely transverse momentum of leading jet ($p_T^{j_1}$), number of b-tagged jets in an event ($N_b$), missing transverse energy ($\met$), and, effective mass ($m_{eff}$). A sample dataset is illustrated in Table~\ref{tab:dt_hep}. 
A decision tree based on Table~\ref{tab:dt_hep} is shown in Fig.~\ref{fig:dt_hep}. In the tree visualization, the root node is depicted in brown, leaf nodes in green, and ordinary nodes in blue.

In the context of HEP, the goal for DT algorithm is mainly to classify an entry as a signal or background.  For the classification problem, the journey from the  \textit{root-node} to the \textit{leaf-node} signifies a series of cuts that determines whether an entry is classified as signal or background depending upon
the characteristics of the \textit{leaf-node}. 
Each leaf can be assigned a purity value, calculated as $p = S/(S+B)$ where $S$ and $B$ represent the sum of weights of the signal and background events contained within that leaf. Depending upon this value of purity, the events in that leaf can be identified as signal or background ( if $p > 0.5$, it is identified as signal; otherwise, it is classified as background.)
At each node, the feature and its corresponding threshold value determine the subsequent splitting of the node. The selection of the feature and its threshold value depends on the decrease in impurity.
The commonly used impurity functions include \textit{the misclassification error}
, \textit{the cross entropy} \cite{breiman,Breiman:1984jka}  and \textit{the Gini index of diversity} \cite{gini,ceriani2012origins}, all of which are determined by the purity of signal and background. Among these impurity functions, the most popular one is the \textit{Gini index}, which performs similarly to entropy. 
In HEP, generally, we try to optimize the signal significance $S/\sqrt{S+B}$ or $S/\sqrt{B}$ by applying 
\textit{Cross section significance} or \textit{excess significance} which is defined as $-S^2/(S+B)$ and $-S^2/B$ respectively. On the other hand, regression tress trees split nodes based on minimizing \textit{ the Sum of Squared Errors (SSE)}. The tree-growing process is stopped using \textit{stopping criteria}.
The \textit{stopping} condition may include several criteria, such as reaching a maximum tree depth, maximum number of leaf nodes, ensuring a minimum number of instances within each leaf node, insufficient improvement through further splitting, or achieving complete splitting, where all events within the node belong to the same class. 
Sometimes, even before reaching the maximum allowed depth or number of nodes, 
early stopping criteria are used to prevent overfitting and improve the generalization.

After constructing the tree, predictions are generated by navigating from the root node to a leaf node that matches the input data. Although decision trees are adept at modeling training data, they are known for their susceptibility to instability. Overfitting to the training sample can make a decision tree overly sensitive to minor variations in inputs, reducing its effectiveness for unfamiliar events or data, as it heavily relies on the training set.
Pruning refers to the process of cutting irrelevant branches or reducing the size of DT by removing nodes and branches that do not significantly improve the predictive performance of unseen data. Pruning a tree is crucial for reducing instability, avoiding overfitting, and enhancing the model's ability to generalize with new data. ``Pre-pruning'' is similar to the early stopping we have already discussed above. ``Post-pruning'' involves constructing an extensive tree initially and subsequently eliminating irrelevant branches. These branches are pruned by converting an internal node and all its offspring into leaves, thereby eliminating the corresponding subtree.
There are several pruning algorithms like expected error pruning, reduced error pruning~\cite{QUINLAN1987221} and cost-complexity pruning which is a part of the CART algorithm \cite{Breiman:1984jka}.

There is another method to handle the instability of the decision trees, 
which is called ``ensemble learning''. Ensemble learning addresses this by combining predictions from multiple trees, potentially boosting discriminatory power. Techniques such as bagging, boosting, and random forests fall under this framework, offering various ways to improve model performance. 
The combination of aggregation and bootstrapping is known as ``\textit{bagging}" or bootstrap aggregating. Initially, a new dataset is created by randomly sampling data from the original dataset, ensuring that the new dataset contains the same number of samples as the original dataset.
 One important point is that individual samples from the original dataset can appear multiple times in the new dataset. This process is known as "bootstrapping." Subsequently, multiple trees are constructed based on the features of the new dataset. This procedure is repeated several times until the algorithm generates and aggregates a considerable number of trees. 
 In the upcoming subsection, we will summarize the following DT algorithms that utilize   ``ensemble learning'' techniques. Specifically, we will focus on Random Forest, which is an extension of bagging and boosting algorithms. Within the boosting category, our attention will be on AdaBoost, as well as two Gradient boosted decision tree (GBDT) algorithms: XGBoost and LightGBM\footnote{It may be noted that there are several other variations of \texttt{AdaBoost} algorithm, e.g., \texttt{BrownBoost} \cite{freund1999adaptive}, \texttt{$\varepsilon$-LogitBoost} \cite{friedman2000additive}, \texttt{$\varepsilon$-HingeBoost} \cite{Yang:2005nz}, etc. Also, a new GBDT algorithm with categorical feature support known as \texttt{CatBoost} \cite{dorogush2018catboost} has been proposed recently.}.

\subsection{Random Forest}
\label{sec:rf}
The Random forest algorithm \cite{Breiman:2001hzm} is a robust ensemble learning method used for regression and classification jobs. The key concept of the Random forest algorithm is to create an ensemble (forest) of decision trees and to select the prediction voted by the majority/averaging of all those trees depending upon the classification/regression task. For a better overall outcome, a subset of input data is passed through each decision tree. Randomizing the training variables helps to obtain better accuracy. 
Initially, a new dataset is generated for a given input dataset with $N$ samples. The algorithm randomly selects a sample and adds it as the first entry of the new dataset. This process iterates for $N$ times. It may be noted that one particular sample can occur more than once in the new dataset. This process is called ``\texttt{bootstrapping}''. For a given input feature set $x_i$ where  $i=1,\dots,m$, decision trees are made taking a random subset of features $n$, such that $n<m$. Based on these features, a tree is made.
This process repeats several times till the algorithm generates and aggregates a large number of trees. The aggregation, together with bootstrapping, is referred to as ``\texttt{bagging}''.  Among all the trees, the outcome with the majority of votes is selected as the model prediction for classification problems. 
For regression tasks, the final outcome is calculated by averaging the predictions of all the decision trees. 
The model outcome depends on the hyperparameters, such as the number of trees the algorithm grows, the maximum depth of a tree, the maximum number of leaf nodes a tree can have, etc. The creation of a new tree is an iterative process, and it does not stop until the algorithm meets its stopping criterion (a pre-determined number of trees the algorithm can grow, which is fixed by the user). A flow chart for the Random Forest algorithm is shown in Fig.~\ref{fig:flow_rf}. 

\begin{figure}[!htb]
\begin{center}
\includegraphics[width=0.98\textwidth]{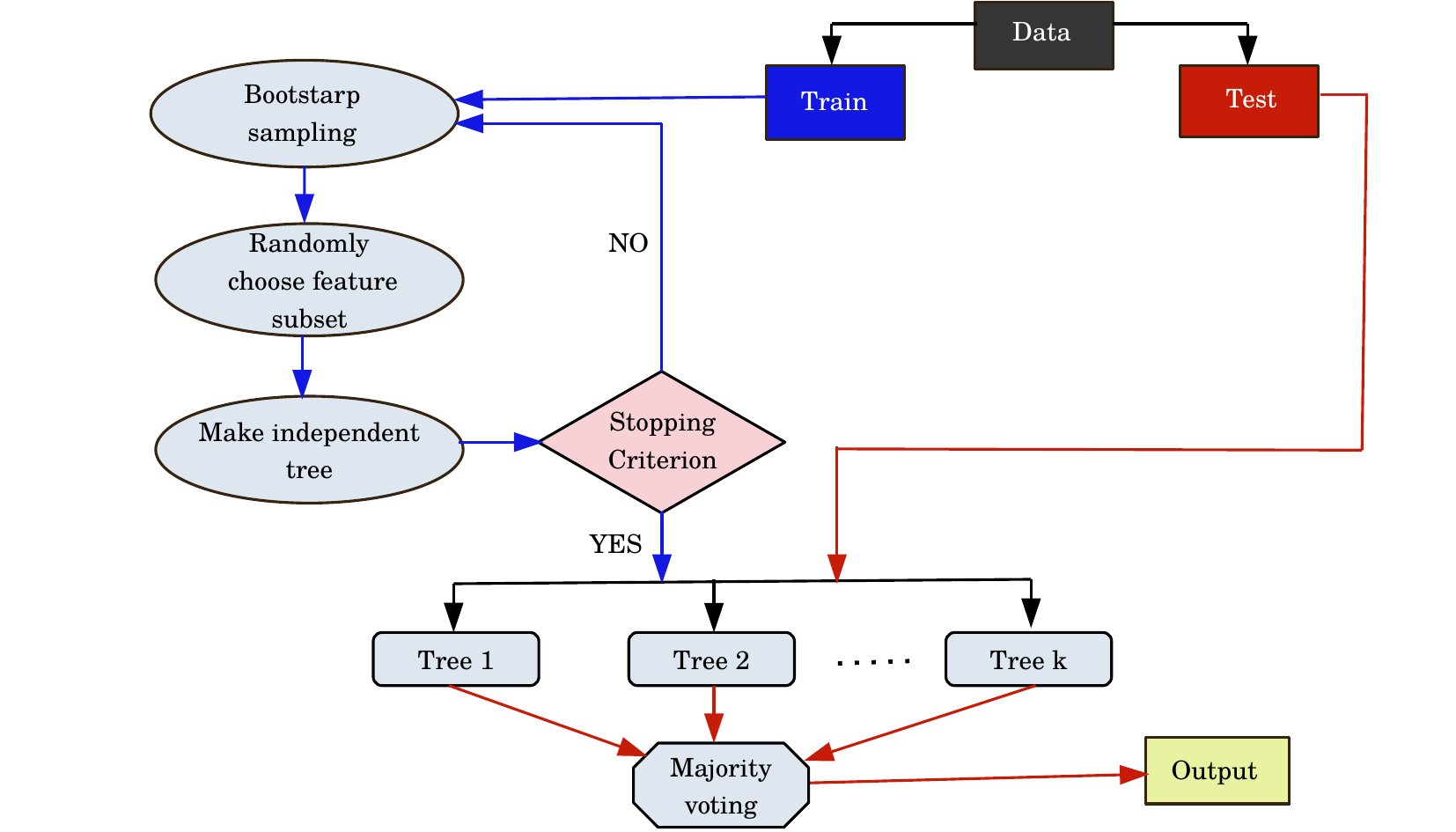}
\caption{Flow-chart for Random Forest algorithm for classification tasks. For regression tasks, the final outcome is calculated by averaging instead of majority voting.}
\label{fig:flow_rf}
\end{center}
\end{figure}

To illustrate this point, let us consider a sample dataset (Table.~\ref{tab:dataset_rf}) for a supersymmetric (SUSY) signal with the Standard Model background, where the signal and background are labeled as 1 and 0, respectively. We have considered the transverse momentum of leading and sub-leading lepton ($p_T^{l_1}$ and $p_T^{l_2}$) and the leading jet ($p_T^{j_1}$), missing transverse energy ($\met$) and $\Delta R$ (where $\Delta R = \sqrt{(\Delta \eta)^2 + (\Delta \phi)^2}$, $\eta$ and $\phi$ are the pseudorapidity and azimuthal angle, respectively.) between leading and sub-leading lepton ($\Delta R(l_1,l_2)$) as input features. 
\begin{table}[h]
    \centering
    \begin{tabular}{|c|c|c|c|c|>{\columncolor[gray]{0.8}}c|}
        \hline
        \rowcolor{yellow}
          $p_T^{l_1}$ & $p_T^{l_2}$ & $p_T^{j_1}$ & $\met$ & $\Delta R(l_1,l_2)$ & Label \\ \hline
         55.1 & 45.3 & 75.2 & 180.4 & 1.5 & 1 \\ \hline
         25.7 & 20.3 & 14.5 & 138.3 & 1.4 & 0 \\ \hline
         65.2 & 52.7 & 88.3 & 201.7 & 1.2 & 1 \\ \hline
         45.9 & 39.6 & 55.7 & 147.2 & 1.4 & 1 \\ \hline
         46.2 & 42.9 & 35.9 & 151.1 & 1.7 & 0 \\ \hline
         40.8 & 31.5 & 41.9 & 130.6 & 1.4 & 0 \\ \hline \hline
         \rowcolor{blue!30}
         40 & 33 & 65 & 155 & 1.6 & 0\\ \hline       
    \end{tabular}
    \caption{Dataset to classify signal and backgrounds for various input features. The output (rightmost column) is 1 for the signal and 0 for the background. The first six labeled entries are training data and the test data is colored in light blue. Here, $p_T^{l_1}$, $p_T^{l_2}$, $p_T^{j_1}$ and $\met$ are in units of GeV.} 
    \label{tab:dataset_rf}  
\end{table}
The label (class) corresponding to each entry is written in the gray-shaded column at the right. The first six entries are training data, and the light-blue colored data is test data. A forest of decision trees is made (Fig.~\ref{fig:rf_tree}) from the training data of Table.~\ref{tab:dataset_rf}.

\begin{figure}[!htb]
\begin{center}
\includegraphics[width=0.32\textwidth]{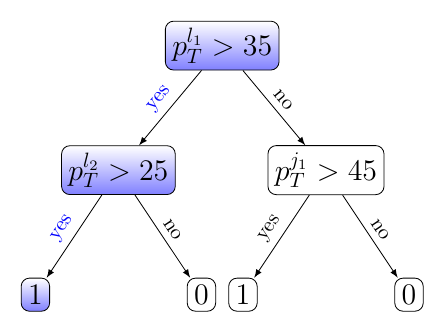}
\includegraphics[width=0.32\textwidth]{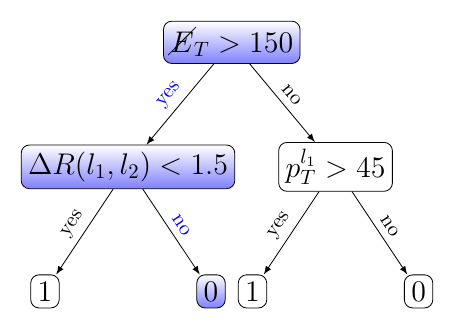}
\includegraphics[width=0.32\textwidth]{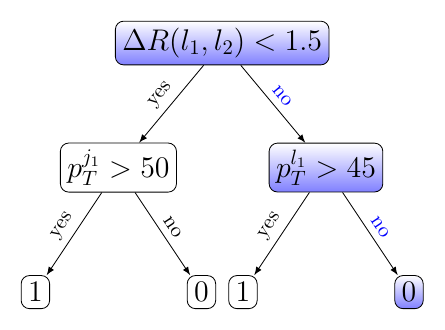}
\caption{A sample forest of three trees made from the training dataset given in Table.~\ref{tab:dataset_rf}. The prediction for the test data (Table.~\ref{tab:dataset_rf}) and its direction of flow is shown in blue color.}
\label{fig:rf_tree}
\end{center}
\end{figure}

Test data generates predictions as it passes through each tree in the forest. For our example, the outcome of each tree is marked with a blue color when the test data from Table.~\ref{tab:dataset_rf} passes through them. In our example, two of the three trees classify that entry as a background ($0$), and one classifies it as a signal ($1$). Since the majority is ``0'', the prediction for this particular entry is ``$0$'' or background. It is worth noting that the predicted output matches the actual data, which signifies the accuracy of the algorithm.

Due to ensemble method and averaging over multiple decision trees, the chances of overfitting get reduced in Random Forest. However, overfitting can still arise due to a very large number of trees, trees with a large number of splits, etc. Fixing the maximum depth of the trees (by setting \texttt{max\_depth} parameter) or reducing the number of features considered at each split (\texttt{max\_features}) can be useful to prevent overfitting by avoiding too complex model. Another way to do so is by controlling the minimum samples required to split a node (\texttt{min\_samples\_leaf}) or by applying penalties on large trees. When the feature space is too large, the model becomes complex. Sometimes, few features merely contribute to the final outcome compared to other features in the input feature set. Omitting those less contributing features and only taking the important ones by employing Principle Component Analysis (PCA) can effectively bring down the overfitting issue. To mitigate the effect of noisy data present, one can diversify the training data, hence reducing the overfitting issue.

Random forest can deal with both classification and regression problems and it can handle high dimensional data very well. The accumulation of numerous decision trees and random feature selection provides better accuracy and reduces overfitting. On the other hand, the random forest algorithm is computationally expensive and time-consuming.

\subsection{AdaBoost}
\label{sec:adb}
Decision trees are the base learner for Random Forest and many other algorithms. The learners can be weak or strong depending on their prediction accuracy. Poor prediction accuracy is associated with weak learners, whereas a strong learner yields better model prediction accuracy. While it is hard to predict an output only through a weak learner, a combination of such weak learners can be made into a strong learner. At each step, the misclassified instances are provided with a larger weightage, resulting in better accuracy at the final step. This process is called ``\texttt{Boosting}". \texttt{AdaBoost}
\cite{freund1999short,wang2012adaboost} works on the basic principle of the boosted decision tree (BDT) algorithm. The term \texttt{AdaBoost} originates from \textit{Adaptive Boosting}, as the misclassified events are given a larger weightage, while the correctly classified events are given a smaller weightage in this algorithm. Freund and Schapire introduced the  \texttt{AdaBoost} algorithm \cite{Freund:1997xna} in 1995, and from then, it has been widely used in many scenarios, including the HEP analysis. 

\begin{figure}[!htb]
\begin{center}
\includegraphics[scale=0.5]{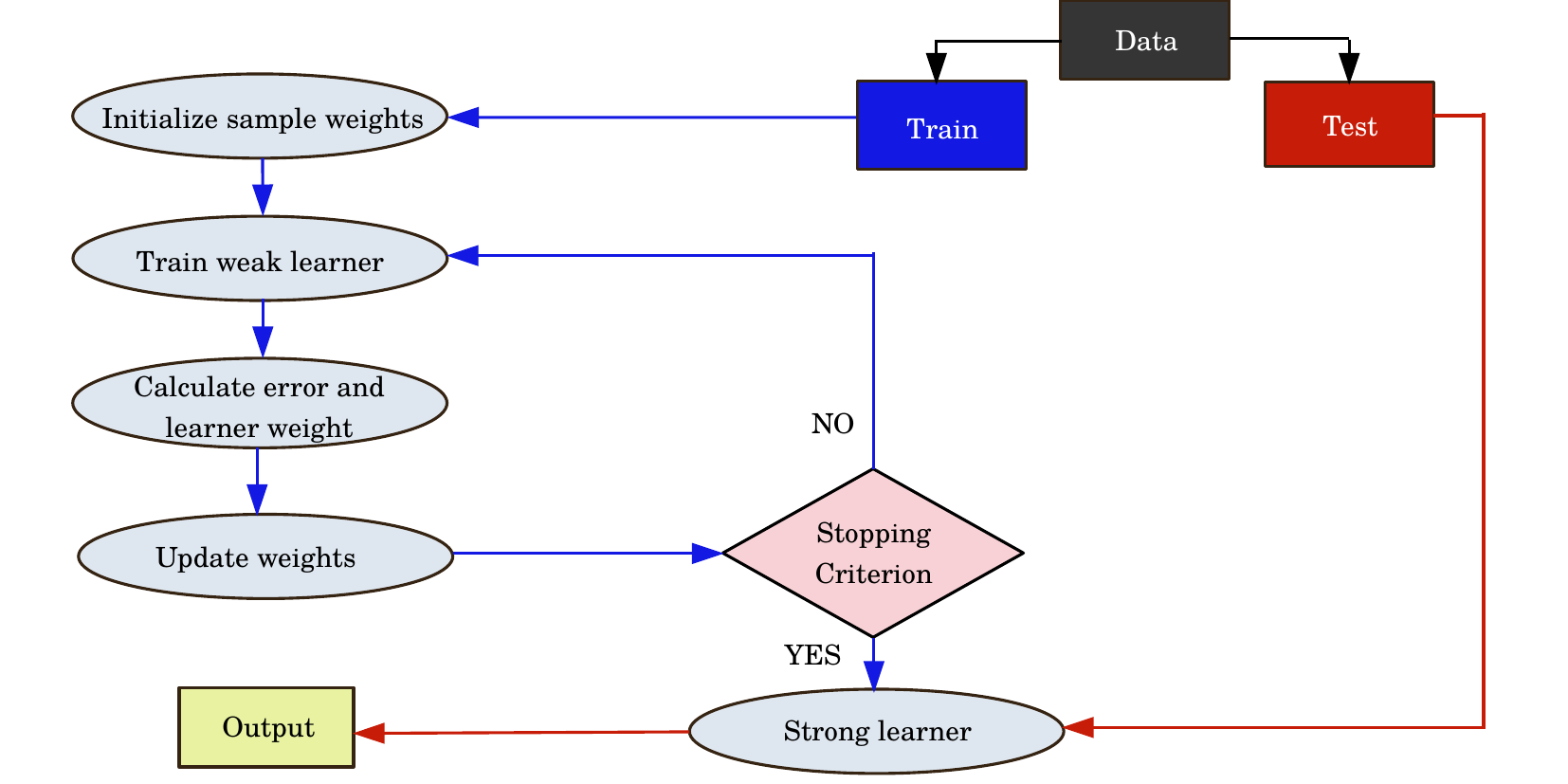}
\caption{Flow chart of  \texttt{AdaBoost} algorithm.}
\label{fig:flow_adb}
\end{center}
\end{figure}

A set of labeled training data $\{x_i,\hat{y}_i \}$ where $x_i$ denotes the input features and $\hat{y}_i$ stands for the label,  is fed into the algorithm. Here $i=1,2,3...m$, and each $x_i$ and $\hat{y}_i$ belong to domains $X$ and $Y$, respectively ($x_i \in X,~\hat{y}_i \in Y=\{-1,+1\}$). For simplicity, here we have chosen binary classification to elaborate on, and the classes are denoted as $+1$ and $-1$. The base learner is repeatedly called by the algorithm for $t=1,\dots,T$ iteration. It may be noted that $T$ denotes the hyperparameter \texttt{number of trees/estimators}. The weight of the $i^{\text{th}}$ sample in $t^{\text{th}}$ iteration is denoted by $w_t(i)$. In the beginning, all samples are assigned the same weight. At each iteration, the learner is compelled to focus on the incorrectly classified event by giving larger weightage to the incorrectly classified events. The learner constructs a weak hypothesis $h_t:X\rightarrow\{-1,+1\}$ for $w_t$ with the lowest error, where the error is given by $\epsilon_t=\sum_{i \colon h_t(x_i)\neq \hat{y}_i} w_t(i)$. The algorithm next assigns a parameter $\alpha_t = \eta\ln(\frac{1-\epsilon_t}{\epsilon_t})$ to the weak hypothesis, $h_t$, which is a measure of performance of $h_t$. Here, $\eta$ acts as the \texttt{learning rate} or \texttt{shrinkage coefficient}, which is a measure of the strength of boosting. It is worth noting that for $\epsilon_t \leq 1/2$, $\alpha_t \geq 0$, and $\alpha_t$ increases with decreasing $\epsilon_t$. The weight, $w_t$, is then modified. A larger weight (multiplying $w_t$ by $e^{\alpha_t}$) is given for the misclassified events ($h_t(x_i) \neq \hat{y}_i$) and a smaller weight (multiplying $w_t$ by $e^{- \alpha_t}$) is given to the correctly classified events ($h_t(x_i) = \hat{y}_i$). After completion of $T^{\text{th}}$ iteration (stopping criterion), the weighted majority of all $T$ weak hypotheses is chosen to be the final hypothesis or strong learner, $H(x)$. The flow chart of the  \texttt{AdaBoost} algorithm is presented in Fig.~\ref{fig:flow_adb}. The ability of  \texttt{AdaBoost} to create a strong learner by combining several weak learners by weighted data sampling makes it more efficient than the Random forest algorithm. Furthermore, the  \texttt{AdaBoost} training process leads to faster convergence in many scenarios. However, 
\texttt{AdaBoost} tends to overfit when the data is too noisy and when the weak learners become too complicated. \texttt{AdaBoost} is prone to assigning too much weight to misclassified instances in the high noise regime, producing skewed data distribution. Accommodating regularization techniques into the algorithm can resolve this issue. Similar to \texttt{Random Forest} model, by tuning the hyperparameters, such as \texttt{max\_depth}, \texttt{max\_features}, can be helpful to reduce overfitting. Also, early stopping, cross-validation, and adjustment of the number of iterations can prevent the algorithm from overfitting.

\subsection{XGBoost}
\label{xgb}
A general gradient boosting decision tree (GBDT) algorithm uses gradient descent to minimize the loss function. In addition to the boosting, it optimizes the algorithm to find the local minima of a differentiable loss function. A loss function signifies the quantitative goodness of the model prediction. The ``eXtreme Gradient Boost'' or \texttt{XGBoost} \cite{Chen:2016btl} is an extension of the GBDT algorithm. Here, the algorithm utilizes the second-order derivatives of the convex loss function, in contrast to a general GBDT algorithm where only the first-order derivative is used. The flow of the \texttt{XGBoost} algorithm can be divided into two parts: the first consists of boosting by obtaining optimal leaf score or weight for tree growth, and the second is optimizing the algorithm for smoother operation. In the following, we discuss them.

\texttt{XGBoost} uses a regularized objective; in simpler terms, one regularization or penalty term is added with the objective function, and it helps to prevent overfitting. For a given dataset $\mathcal{D}=\{x_i,\hat{y}_i\}$ with $m$ features and $n$ entries such that $x_i \in \mathbb{R}^m$, the predicted output for $K$ number of trees is \cite{Chen:2016btl}
\begin{equation}
 y_i = \phi(x_i)=\sum_{k=1}^{K}f_k(x_i), \hspace{1cm}f_k\in \mathcal{F},
\end{equation}
\label{eq:pred_xgb}
where $\mathcal{F}=\{f(x)=w_{q(x)}\}$ is the tree function space. The mapping for a feature $x$ to the corresponding leaf index is denoted by the function $q(x)$ and $w\in\mathbb{R}^T$ where $T$ is the number of leaves at $k^{\text{th}}$ tree. Here, $w_k$ is the score of $k^{\text{th}}$ leaf. A regularization term ($\Omega$) is added to the loss function ($l$) and the new objective function becomes \cite{Chen:2016btl}
\begin{equation}
\mathcal{L}(\phi) = \sum_{i=1}^n l(y_i, \hat{y}_i) + \sum_{k=1}^K \Omega (f_k)
\label{eq:loss_func_xgb}
\end{equation}
where $\Omega(f) = \gamma T + \frac{1}{2}\lambda \Arrowvert w \Arrowvert^2$. Here, $\gamma$ is the minimum loss reduction term used to further divide a leaf node, and $\lambda$ is a ridge regularization term acting on the leaf weights. A higher value of $\lambda$ and $\gamma$ results in a more conservative algorithm. As the number of leaves increases within a tree, the objective function is subsequently increased. Consequently, this makes minimizing the objective more challenging. To overcome this issue, we add $f_t$ to minimize the objective.
\begin{equation}
\mathcal{L}^{(t)} = \sum_{i=1}^m l(\hat{y}_i, y_i^{(t-1)} + f_t(x_i)) + \Omega(f_t)
\label{eq:loss_func2}
\end{equation}
where $\mathcal{L}^{(t)}$ and $y_i^t$ denote the objective and prediction for $i^{\text{th}}$ entry at $t^{\text{th}}$ iteration. The next step is to optimize the objective function, and for that, $\mathcal{L}^{(t)}$ is expanded in the Taylor series up to the second order \cite{Chen:2016btl}  
 \begin{equation}
 \mathcal{L}^{(t)} \simeq \sum_{i=1}^n [l(\hat{y}_i, y_i^{(t-1)})+g_i f_t(x_i)+ \frac{1}{2}h_if_t^2(x_i)] + \Omega (f_t)
 \label{eq:loss_func3}
 \end{equation}
where $g_i = \partial_{y_i^{(t-1)}} l(\hat{y}_i, y_i^{(t-1)})$ is called the gradient and $h_i = \partial^2_{y_i^{(t-1)}} l(\hat{y}_i, y_i^{(t-1)})$ is called the hessian. Let $I_j = \{i|q(x_i)=j \}$ be the set of instances where $i^{\text{th}}$ instance is mapped to $j^{\text{th}}$ leaf with tree structure $q(x_i)$. Eq.(\ref{eq:loss_func3}) can be written as \cite{Chen:2016btl}
 \begin{equation}
  \tilde{\mathcal{L}}^{(t)} (q) = \sum_{j=1}^T[(\sum_{i \in I_j} g_i)w_j + \frac{1}{2}(\sum_{i \in I_j} h_i + \lambda)w_j^2] +\gamma T
  \label{eq:loss_func4}
 \end{equation}
 The optimal weight of the $j^{\text{th}}$ leaf, $w_j^{\ast}$ can be written as
 \begin{equation}
 w_j^{\ast} = \frac{- \sum_{i \in I_j} g_i}{\sum_{i \in I_j} h_i + \lambda}
 \label{eq:w_min}
 \end{equation}
 and the minimum value of $\tilde{\mathcal{L}}^{(t)} (q)$ is 
 \begin{equation}
 \tilde{\mathcal{L}}^{(t)} (q) = -\frac{1}{2} \sum_{j=1}^T \frac{ (\sum_{i \in I_j} g_i)^2}{\sum_{i \in I_j} h_i + \lambda} +\gamma T
 \label{eq:loss_func5}
 \end{equation}
 \begin{figure}[!htb]
\begin{center}
\includegraphics[scale=0.55]{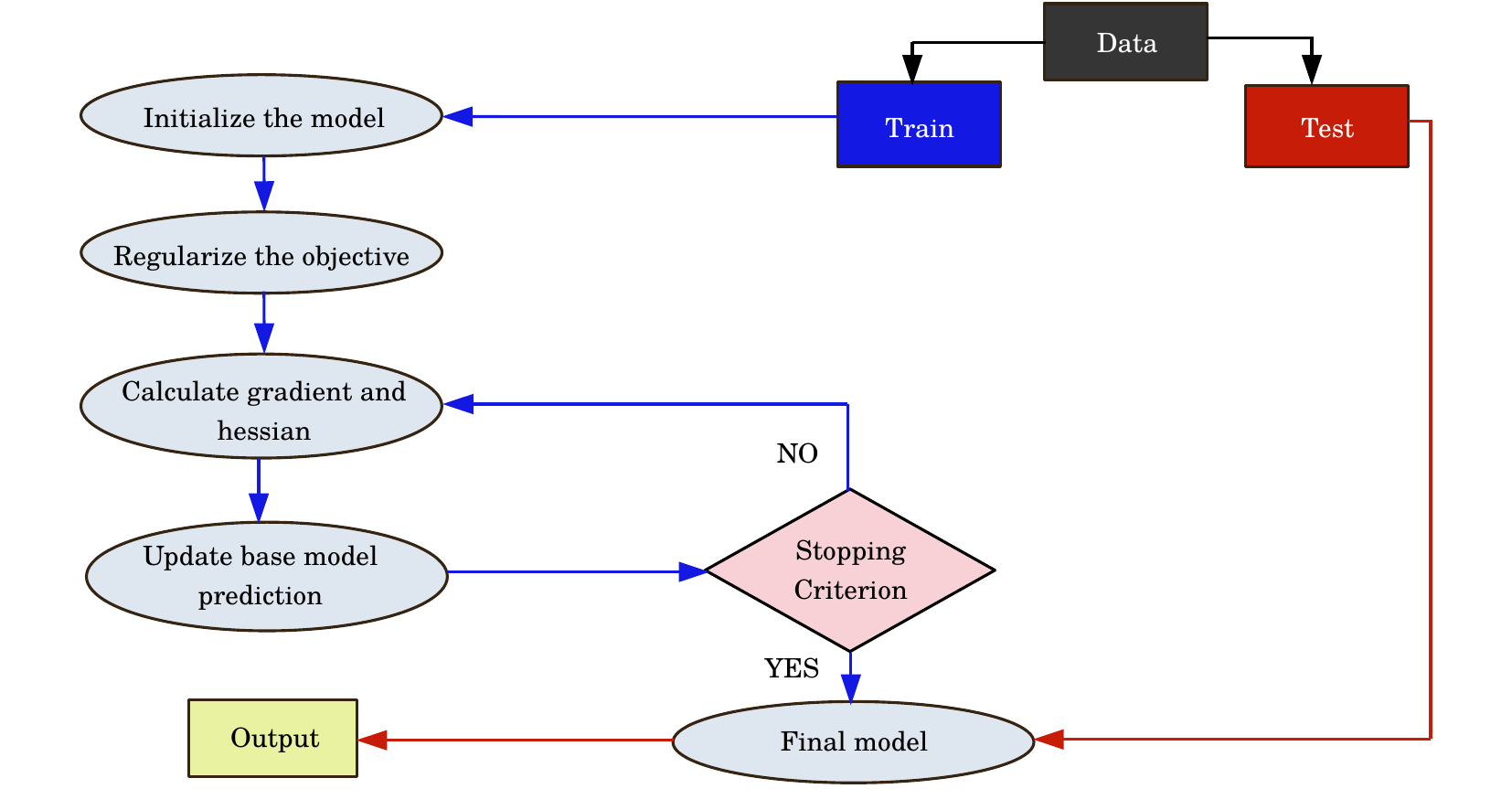}
\caption{Flow-chart for \texttt{XGBoost} algorithm}
\label{fig:flow_xgb}
\end{center}
\end{figure}
and it represents a quantitative measure of the goodness of the tree structure $q$, which is analogous to the impurity score. A smaller value of $\tilde{\mathcal{L}}^{(t)}(q)$ means a better tree structure. While it is non-trivial to count all possible trees to find the tree with the maximum impurity score, the use of a greedy algorithm eases the scenario. From a single leaf with $I$ instances, left and right branches are made with instances $I_L$ and $I_R$, respectively, such that $I=I_L \cup I_R$. It chooses the split with the highest reduction in loss. After the splitting, the loss reduction can be quantified as \cite{Chen:2016btl}
\begin{equation}
\mathcal{L}_{split} = \frac{1}{2}\bigg\lbrack \frac{(\sum_{i\in I_L} g_i)^2}{\sum_{i\in I_L} h_i + \lambda} +  \frac{(\sum_{i\in I_R} g_i)^2}{\sum_{i\in I_R} h_i + \lambda} - \frac{(\sum_{i\in I} g_i)^2}{\sum_{i\in I} h_i + \lambda}\bigg\rbrack - \gamma
\end{equation}

As we can see, \texttt{XGBoost} uses the second order derivative ($h_i$) along with gradient ($g_i$) of the loss function for boosting, which is different from other GBDT algorithms. We present the flow chart of the \texttt{XGBoost} 
algorithm in Fig.\ref{fig:flow_xgb}. Finding the optimal split for a node is a non-trivial task, especially for large and complex datasets. \texttt{XGBoost} uses several split-finding algorithms. For a relatively simple dataset, it uses a ``basic exact greedy algorithm" \cite{Chen:2016btl} where it goes over all possible splits and finds the best one. However, for a large dataset, this technique results in a trade-off with reduced efficiency. To handle this, the algorithm divides the dataset into several percentile or weighted quantile buckets, treats each bucket separately through parallel learning and finds the best split. To deal with sparse data, \texttt{XGBoost} invokes the ``sparsity-aware split algorithm," where it enumerates the gain by going in both directions in a split and chooses the one with maximum gain. These features of \texttt{XGBoost} to tackle such complexities in a dataset make it scalable to almost all scenarios and popular among data scientists.

Similar to other decision tree models, \texttt{XGBoost} is also prone to overfitting for a very large number of trees and/or deep trees, small dataset, high learning rate, etc. One can tweak the hyperparameters like \texttt{max\_depth}, \texttt{learning\_rate} (shrinks the feature weights at each step of boosting), \texttt{colsample\_bytree} (to introduce random subsampling of columns), etc. A quantity of the minimum sum of instance weight is assigned for partition. This is denoted as \texttt{min\_child\_weight}. While splitting a node, if that node comes up with the sum of instances less than the assigned value of \texttt{min\_child\_weight}, the tree-building process will come to a halt. The $\gamma$ parameter, which is the minimum loss reduction for a partition, can also be changed to reduce model complexity. A higher value for \texttt{min\_child\_weight} or $\gamma$ makes the algorithm more conservative. One can add randomness to the data by randomly selecting the fraction of the training sample to be fed into the algorithm during training by adjusting \texttt{subsample} parameter. Like \texttt{AdaBoost}, cross-validation and early stopping techniques are also helpful for reducing overfitting.

\subsection{LightGBM}
\label{lgbm}
Simple gradient-boosting techniques have been quite successful so far. A common attribute of GBDT is to enumerate all instances in the dataset for each feature to find the optimal split. As data complexity and size increase, GBDT faces challenges in efficiency and accuracy. To overcome this issue, novel techniques are implemented upon simple GBDT algorithm, namely, (i) leaf-wise tree growth, (ii) Gradient-based one-sided sampling (GOSS), (iii) histogram splitting, and (iv) Exclusive feature bundling (EFB). The new algorithm that incorporates these new techniques with GBDT, developed by Microsoft, is referred to as Light Gradient Boosting Machine-learning or \texttt{LightGBM} \cite{ke2017lightgbm}. 
 In the following, we discuss the key concepts of \texttt{LightGBM}: 
 
\noindent \textbf{\underline{Leaf-wise tree growth:}} The attribute that fundamentally separates \texttt{LightGBM} from common GBDT is that it is based on ``leaf-wise tree growth'' (demonstrated in the right panel of Fig.\ref{fig:leaf_level}). The leaf node that provides maximum loss is split by the algorithm, and this process is repeated. In this case, one leaf may branch deeper than the other while both of them are coming from the same root node. The general GBDT algorithm is based on ``depth-wise tree growth''  or ``level-wise tree growth''(as shown in the left panel of Fig.\ref{fig:leaf_level}),  where trees are split horizontally at each level, and it expands every node at the same level before going to the next round.  

\begin{figure}[!htb]
\begin{center}
\includegraphics[scale=0.5]{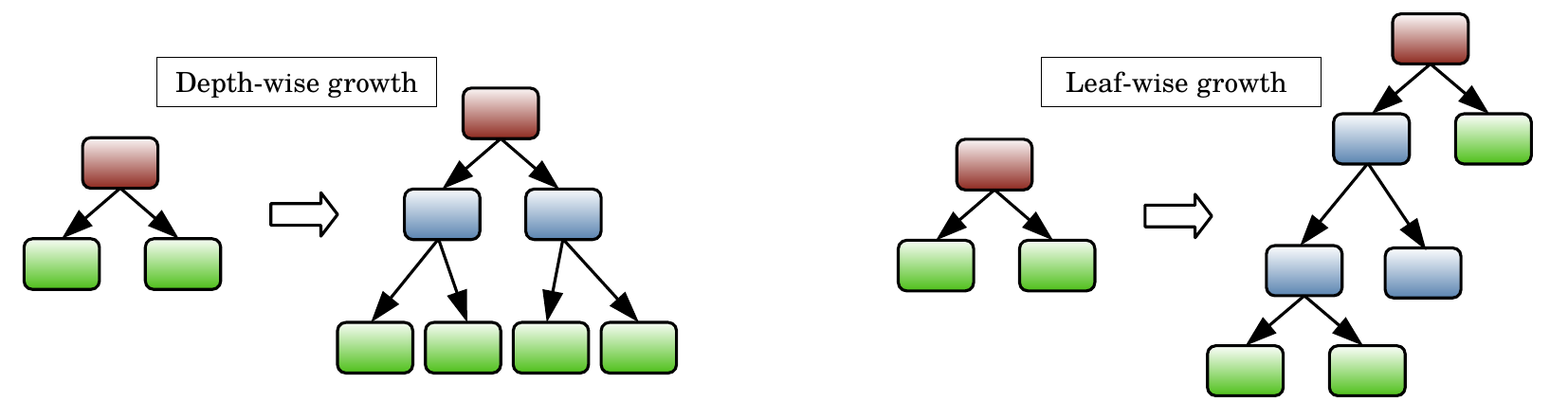}
\caption{Schematics for depth-wise and leaf-wise tree growth. The root node is colored brown, while the leaf nodes are marked green. Ordinary nodes are colored in blue.}
\label{fig:leaf_level}
\end{center}
\end{figure}

\noindent \textbf{\underline{Gradient-based one-sided sampling (GOSS):}} In GBDT, the smaller the gradient is, the better the fit becomes. In this technique, a subset of dataset is created by mostly retaining the instances with higher gradients and discarding (by random sampling) instances with smaller gradients. GOSS retains a percentage ($a\times100\%$) of data with the highest gradient and randomly selects $b\times100\%$ data from the rest and creates a new subsample, given $a, b \in \lbrack 0,1 \rbrack$. While calculating the information gain, the algorithm multiplies a factor $\frac{1-a}{b}$ to the instances with a smaller gradient. The under-trained instances thus get more attention even though no initial weight is associated with the original dataset.

\noindent\textbf{\underline{Histogram splitting:}} 
For a tree branching, \texttt{LightGBM} algorithm bins the feature values into two or more sets rather than evaluating all potential split points for every feature.
Let us suppose for a given dataset, there exists an input feature, ``Age,'' where the entries are 40, 45, 47, 50, 65, 72, 75, 80. The common approach for GBDT while creating a decision tree is to sort the value for which the optimal prediction is achieved. In the case of histogram splitting, the algorithm splits the dataset into two or more bins or buckets, say, age group 40-60 and 61-80 and then proceeds with the analysis. It makes the algorithm faster. 

\noindent \textbf{\underline{Exclusive feature bundling (EFB):}} The complexity of an algorithm increases with increasing input features. To improve efficiency, the algorithm looks for exclusive features in a given dataset (features that do not take non-zero values at the same time). To illustrate this point, let us suppose a binary dataset has a feature column for the gender of the person in a given entry. If we set the male to $1$ then the female is automatically set to $0$ and vice-versa. Both cannot take the same value simultaneously. EFB carefully combines these exclusive features and creates a new bundle, thus effectively reducing the input features without sacrificing accuracy.

\begin{figure}[h]
\begin{center}
\includegraphics[scale=0.5]{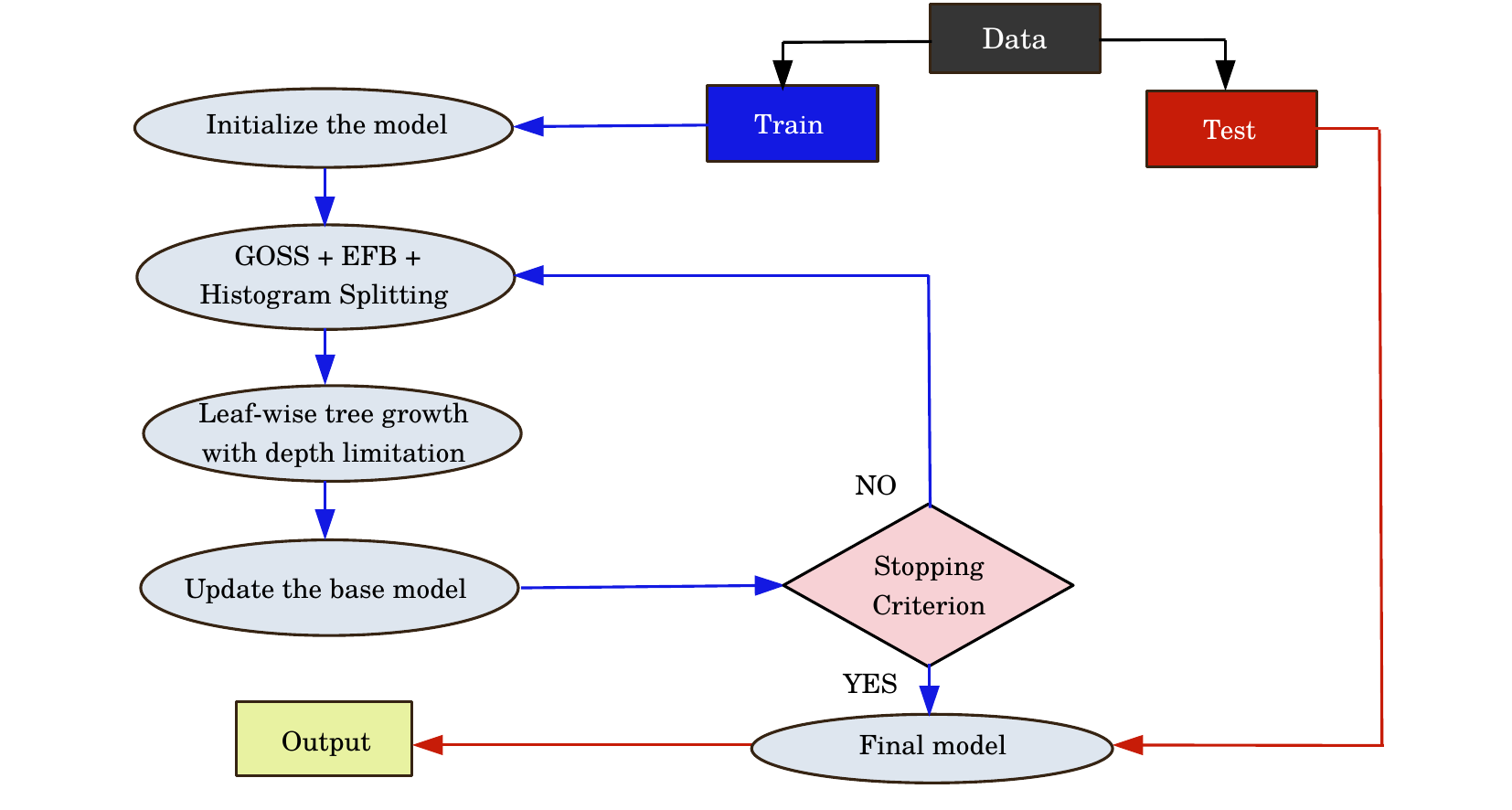}
\caption{Flow-chart for \texttt{LightGBM} algorithm}
\label{fig:flow_lgbm}
\end{center}
\end{figure}

The flow-chart of \texttt{LightGBM} algorithm is shown in  Fig.~\ref{fig:flow_lgbm}.
\texttt{LightGBM} algorithm turns out to be more efficient in terms of time consumption compared with \texttt{XGBoost} while maintaining a competitive accuracy. Due to its histogram-based approach,  \texttt{LightGBM} uses less memory, making it more efficient and scalable while dealing with complex large dataset. Nevertheless, as \texttt{LightGBM} splits the trees leaf-wise, it can produce much more complex tree structures, resulting in overfitting in some cases.
Similar to other BDT models, adjusting the parameters like the number of trees, \texttt{max\_depth}, \texttt{learning\_rate}, \texttt{subsample} is quite useful to prevent overfitting. Controlling the number of bins in which feature values are bucketed can reduce the chances of overfitting. This can be done by setting the \texttt{max\_bin} parameter properly. To avoid complex tree structures, the minimum data in one leaf and the sum of hessian in one leaf are useful, along with the number of leaves in one tree. This procedure can be done using \texttt{min\_data\_in\_leaf}, \texttt{min\_sum\_hessian\_in\_leaf}, and \texttt{num\_leaves} parameters, respectively. While partitioning, the minimum gain to perform a split can be done using \texttt{min\_gain\_to\_split}. In addition, implementing techniques such as regularization and cross-validation also aid in preventing overfitting.
\section{Performance of different Decision Tree based algorithms - a RPC SUSY case study at the HL-LHC }
\label{sec:case_study}
In this section, we study the effectiveness of various decision tree-based techniques for the search for direct pair production of charginos and neutralinos within the RPC MSSM scenario. To accomplish this, we consider a model where $M_1 < M_2 << \mu$, ensuring that both the $\chonepm$ and $\lsptwo$ exhibit wino-like characteristics and become mass degenerate, while the LSP ($\lspone$), remains predominantly bino-like. For such a scenario, the dominant production mode is $pp~\rightarrow~\chonepm\lsptwo$, which gives rise to three lepton final state from subsequent decays as shown in Fig.~\ref{fig:decay}.  
\begin{figure}[!htb]
\begin{center}
\includegraphics[width=0.45\textwidth]{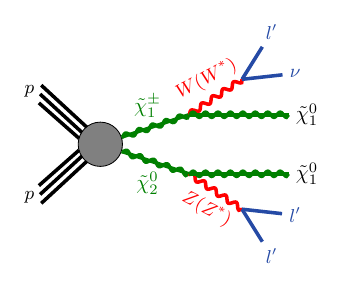}
\caption{The decays of the NLSP via real and virtual $W^{\pm}/Z$ bosons are displayed here.
}
\label{fig:decay}
\end{center}
\end{figure}
We explore both the scenarios where both winos promptly decay to either on-shell or off-shell as follows:
\begin{itemize}
\item On-shell $WZ$ scenarios: the mass difference between the chargino ($\chonepm$) or the second lightest neutralino ($\lsptwo$) and the lightest neutralino ($\lspone$) exceeds the $Z$ boson mass ($\Delta m(\chonepm/\lsptwo,\lspone) > m_Z$), both $\chonepm$ and $\lsptwo$ will undergo decays involving real $W$ and $Z$ bosons, such as $\lsptwo \rightarrow Z\lspone$ and $\chonepm \rightarrow W^{\pm}\lspone$.
\item Off-shell $WZ$ scenarios: In this case the mass difference between $\chonepm/\lsptwo$ and $\lspone$ is less than $W$ boson mass ($\Delta m(\chonepm/\lsptwo,\lspone) < m_W$) , $\chonepm$ and $\lsptwo$ will undergo decay mediated by virtual bosons, such as $\lsptwo~\rightarrow~Z^*\lspone$ and $\chonepm~\rightarrow~W^{\pm*}\lspone$. 
It may be noted that we have not considered the case where the mass gap is greater than $m_W$ and less than $m_Z$. 
\end{itemize} 

These $W/Z$ bosons decay into leptons with an equal branching fraction for each flavor ($l^{\prime} = e, \mu$, and $\tau$), resulting in a final state comprising precisely three light leptons ($l= e, \mu$)\footnote{For the rest of the analysis only electron and muon ($e, \mu$) will be classified as leptons ($l$) unless stated otherwise.} and missing energy coming from two LSPs. The decay processes for both real $WZ$ and virtual $W^*Z^*$ are depicted in Fig.~\ref{fig:decay}\footnote{For earlier works on electroweakino searches in the context of LHC, see Ref.~\cite{Choudhury:2013jpa,Chakraborti:2014gea,Chakraborti:2015mra,Choudhury:2016lku,Barman:2020azo,Datta:2018lup}.}. This $3l+\met$ final state can originate from several Standard Model backgrounds such as $WZ$, $ZZ$, $t\bar{t}V$, $VVV$, $t\bar{t}$, and $Z+jets$. ATLAS has conducted a similar analysis using data from Run-I, Run-II, and high-luminosity LHC (HL-LHC). The current bounds on the electroweakinos from ATLAS Run-II results is 640 GeV for a massless $\lspone$~\cite{ATLAS:2021moa} while the simulation by ATLAS indicated that future HL-LHC will be able to probe $\sim$ 1.1 TeV $\chonepm/\lsptwo$ in such scenario with a cut-based analysis~\cite{TheATLAScollaboration:2014nwe}. 
Our aim is to study the sensitivity of wino searches using optimized cut-based and ML methods. 
Signal events are simulated using \texttt{Pythia6}~\cite{Sjostrand:2006za}, while all background events are simulated using \texttt{MadGraph5-aMC@NLO}~\cite{Alwall:2014hca} at the leading order (LO). The NLO+NLL cross-sections for signals are calculated using \texttt{Resummino-3.1.1}~\cite{Fuks:2013vua}, and the NLO cross-sections for all backgrounds are generated by \texttt{MadGraph5-aMC@NLO} which are listed in two Tables~\ref{tab:sigma_signal} and \ref{tab:sigma_bkg} in the Appendix~\ref{sec:appendix}. For this analysis, we consider a flat $k$ factor = 1.4 for all the SM backgrounds and SUSY signals. For the detector simulation, \texttt{Delphes-3.5.0}~\cite{deFavereau:2013fsa} is utilized. Event reconstruction employs the anti-$k_t$ algorithm with a jet radius of 0.4, imposing criteria of $p_T > 20$ and $|\eta| < 2.8$ for jet selection. Leptons are identified with $p_T > 10$ and $|\eta| < 2.5~(2.7)$ for electrons (muons) respectively. For b-jets, $|\eta| < 2.5$ is considered, with an 85\% b-tagging efficiency and a 25\% miss-tagging rate for light jets. Lepton-lepton isolation, lepton-jet isolation, etc, also have been implemented as described in the Refs.~\cite{ATLAS:2021yyr,Choudhury:2023eje}.

Initially, a cut-based analysis is conducted, followed by machine learning (ML) analysis using various DT algorithms. Then, we look for the improvement in signal significance achieved through ML algorithms, along with a comparison of results from different ML approaches. Additionally, the impact of various hyperparameters such as learning rate ($\eta$) and number of trees is discussed.
Furthermore, the significance of different kinematic features in discriminating signal and background events is illustrated using SHapley Additive ex-Planations (SHAP) values obtained through the SHAP package~\cite{Shapley+1953+307+318,lundberg2018consistent,pedregosa2018scikitlearn}. \\
 To do the further analysis, we choose four different benchmark points, allowed by LHC Run-II data, based on the mass difference between NLSP and LSP ($\Delta m(\chonepm/\lsptwo,\lspone)$) with values $\Delta m(\chonepm/\lsptwo,\lspone) = 50, 70$ and $>> m_Z$. The benchmark points are : \\[-1.2ex]
\hrule 
Benchmark points~~~~~~~~~~~$m_{\chonepm/\lsptwo}$ (GeV)~~~~~~~~~~~~~$m_{\lspone}$ (GeV)~~~~~~~~~~~$\Delta m$ (GeV) \\[-1.6ex] 
\hrule 
\hspace*{8mm} \texttt{BP1}\hspace{35mm} 400 \hspace{30mm} 350 \hspace{30mm} 50 \\
\hspace*{12mm} \texttt{BP2}\hspace{35mm} 500 \hspace{30mm} 430 \hspace{30mm} 70 \\
\hspace*{12mm} \texttt{BP3}\hspace{35mm} 800 \hspace{30mm} 100 \hspace{30mm} 700 \\
\hspace*{12mm} \texttt{BP4}\hspace{33mm} 1200 \hspace{30mm} 100 \hspace{29mm} 1000 \\
\hrule 
\hrule
\hspace*{50mm} Pre-selection Cuts\\[-1.3ex]
\hrule
$N_l = 3$ \hspace{10mm} $\Delta R > 0.3$ \hspace{10mm} $N_{SFOS} \hspace{10mm} \geq 1$ \hspace{10mm} $N_{\tau} = 0$ \hspace{10mm} $N_b = 0$ \\[-1.8ex]
\hrule
\vspace{7mm}
\noindent\underline{\textbf{Pre-selection cuts:}} We select the events with exactly three leptons, which must have separation among them with $\Delta R > 0.3$. Also, we choose the events with at least one same flavor opposite sign (SFOS) pair. We also impose that the events with no tau-jets and no b-tagged jets are selected for the analysis. After passing these selections with the remaining events, we do the cut-based analysis and ML analyses.

\subsection{Cut-and-count analysis} 
\label{sec:cut_based}

\begin{figure}[!htb]
\begin{center}
\includegraphics[width=0.32\textwidth]{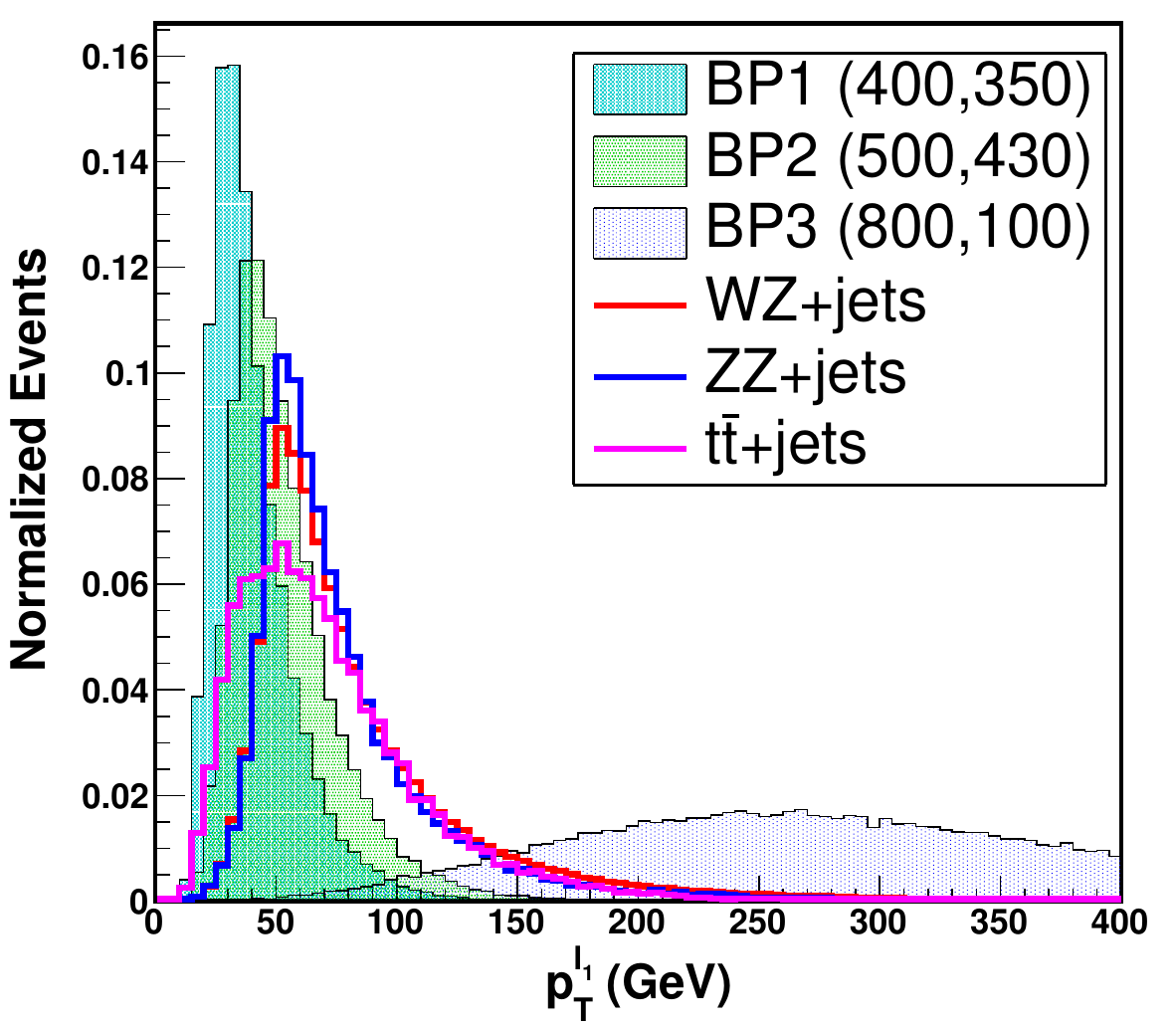}
\includegraphics[width=0.32\textwidth]{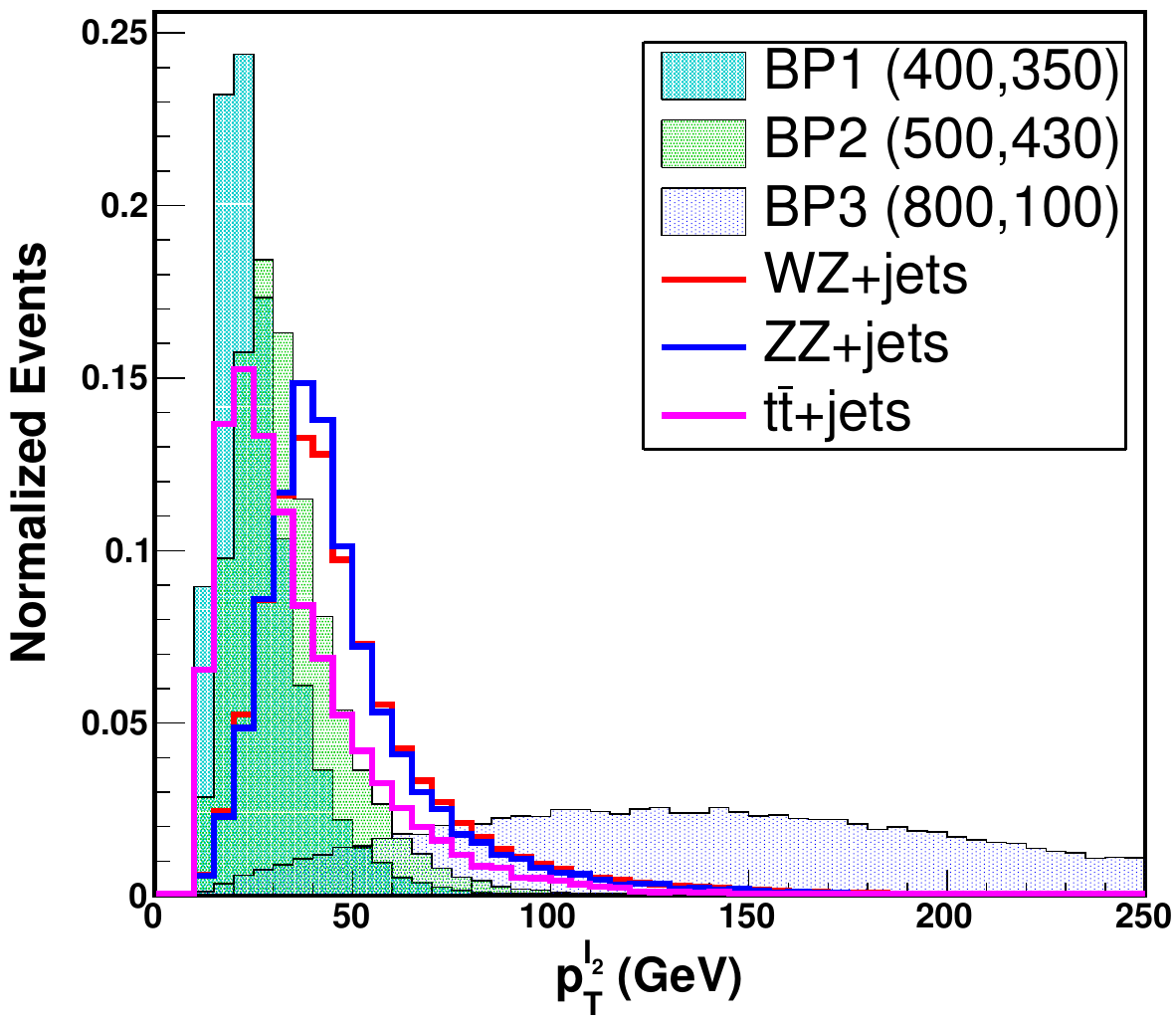}
\includegraphics[width=0.32\textwidth]{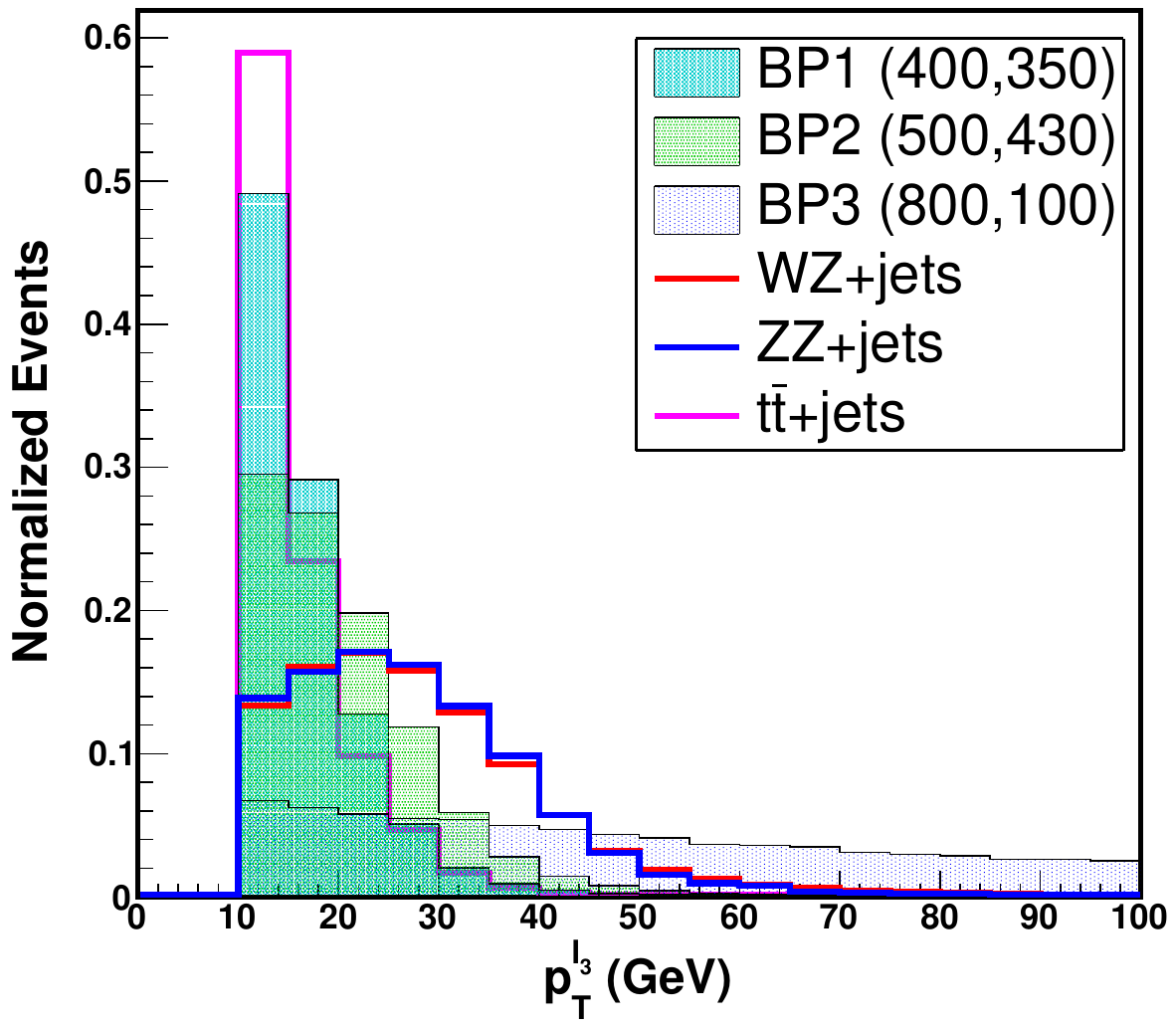}\\
\includegraphics[width=0.32\textwidth]{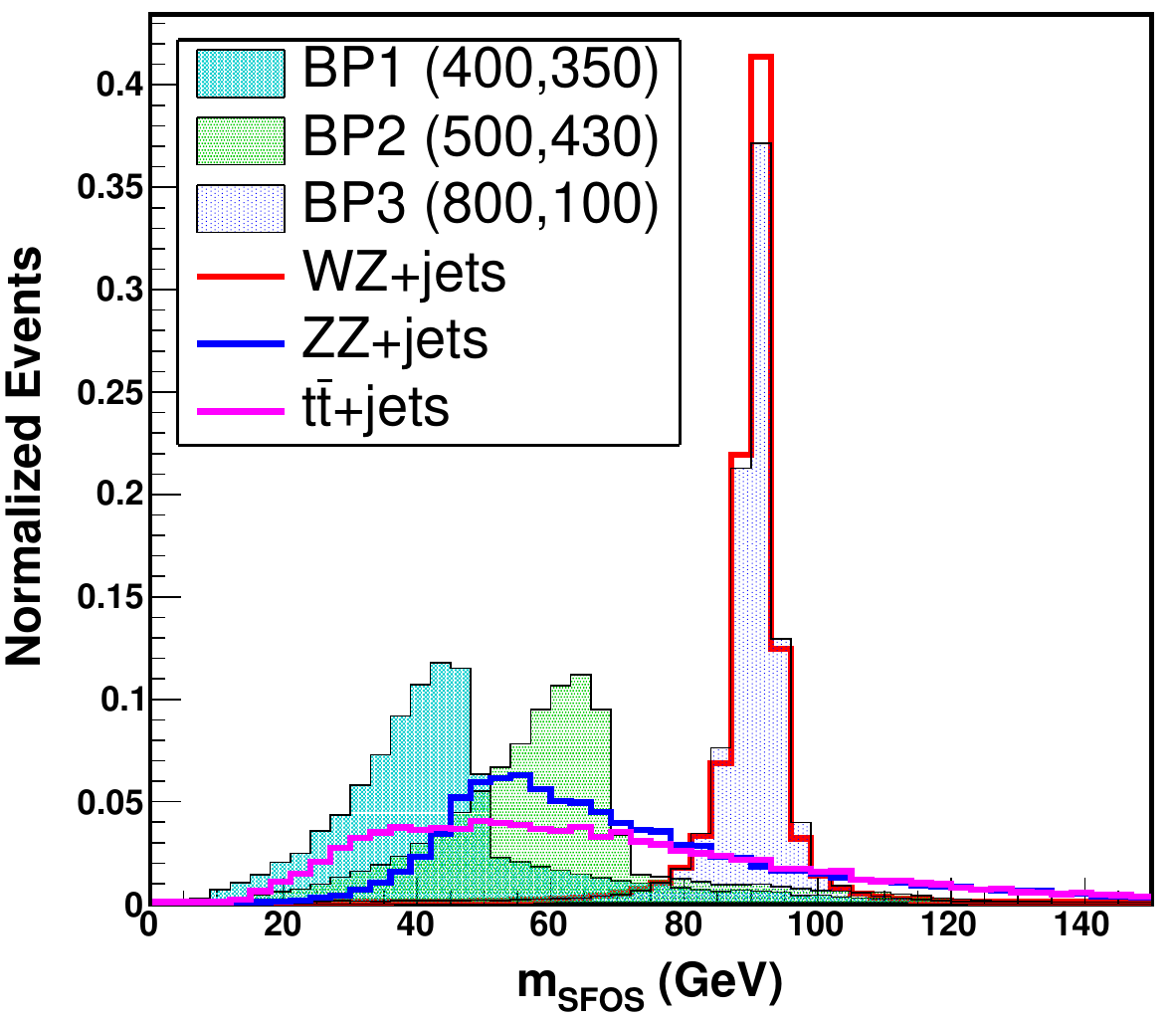}
\includegraphics[width=0.32\textwidth]{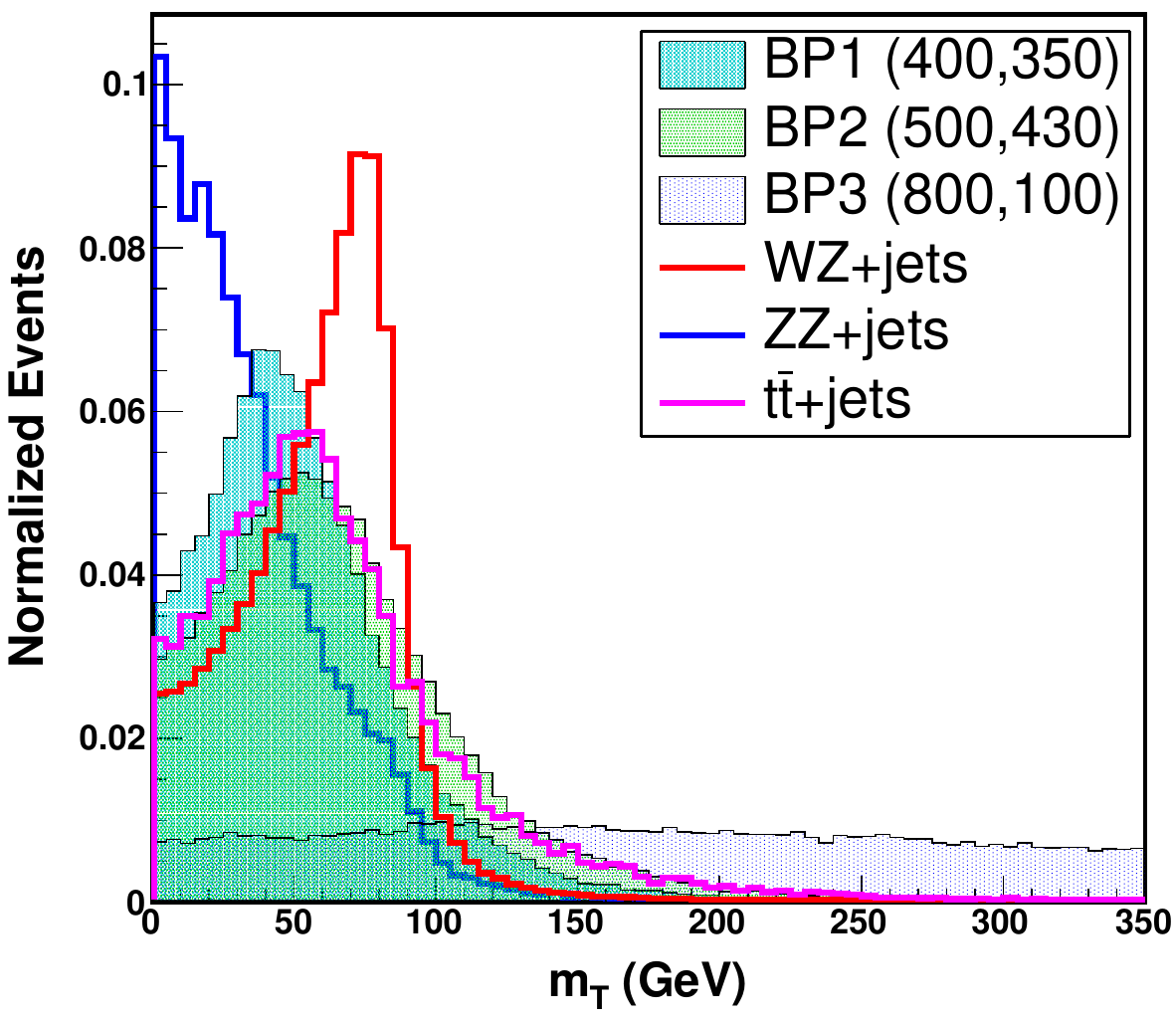}
\includegraphics[width=0.32\textwidth]{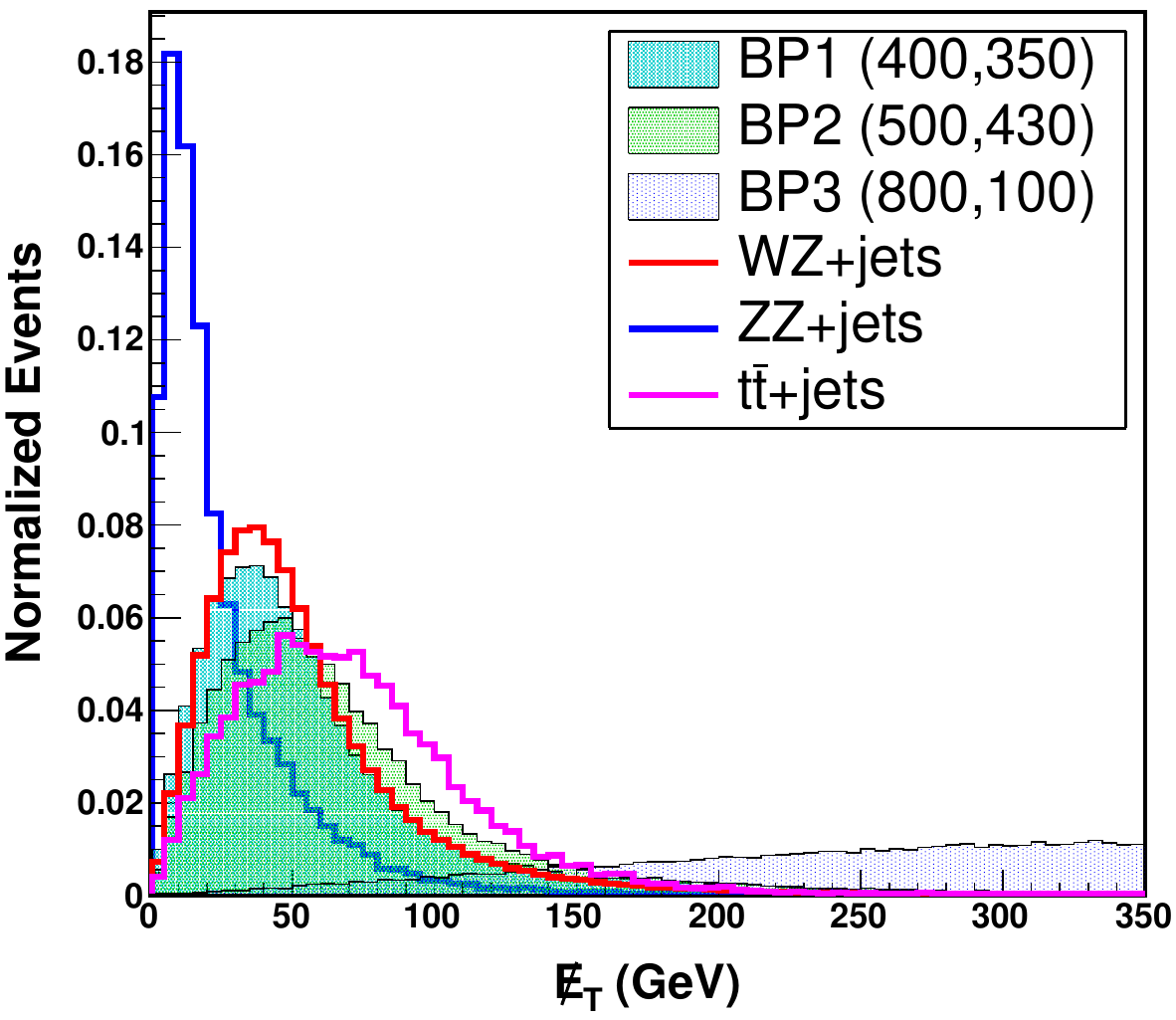}
\caption{The distribution of leading (top left) subleading (top middle) and third leading lepton (top right) transverse momenta and invariant mass of SFOS lepton pair, $m_{SFOS}$ (bottom left) transverse mass, $m_T$ (bottom middle) and missing energy, $\met$ (bottom right) for three different signal benchmark points and three dominant backgrounds are displayed here. The cyan, light green, and light blue colored filled regions correspond to \texttt{BP1}, \texttt{BP2} and \texttt{BP3} respectively in each figure. Also, the red, blue and magenta colored lines refer to $WZ + jets$, $ZZ + jets$ and $t\bar{t} + jets$ respectively for every figure.}
\label{fig:cut_plot}
\end{center}
\end{figure}

In this section, we investigate the search for wino-like $\chonepm\lsptwo$ production at the HL-LHC with $\mathcal{L}$ = 3000 $fb^{-1}$ using the traditional cut-and-count method. We present the transverse momentum ($p_T$) distributions of the three chosen leptons, illustrated in upper panel of Fig.~\ref{fig:cut_plot}, for three distinct benchmark points (\texttt{BP1}, \texttt{BP2}, \texttt{BP3}) corresponding to $\Delta m = 50, 70$, and 700 GeV, alongside three prominent backgrounds: $WZ+jets$, $ZZ+jets$, and $t\bar{t}+jets$. The lepton $p_T$ distribution for signal and backgrounds are quite similar for the compressed benchmark points and distinct in the cases with large $\Delta m$.
We identify the same-flavor opposite-sign (SFOS) lepton pair with mass closest to the $Z$ boson mass and exhibit the distribution of their invariant mass as $m_{SFOS}$, depicted in Fig.~\ref{fig:cut_plot} (bottom left panel). The $m_{SFOS}$ distribution shows that the benchmark points $\Delta m > m_Z$ have a significant overlap with $WZ$ background while the compressed cases have peaked around $\Delta m = 50$ and 70 GeV with a dominant overlap of $t\bar{t}$ background. Subsequently, we proceed to reconstruct the variable transverse mass, such as. 
\begin{equation}
\label{eq:mt}
m_T = \sqrt{2p_T^{l_{w}}\met(1-cos\Delta\phi)}
\end{equation} 
$l_w$ denotes the remaining third lepton subsequent to selecting a lepton pair for the reconstruction of the $Z$ boson, while $\Delta\phi$ signifies the angular disparity between the missing energy ($\met$) and said third lepton. This parameter exploits the disparity between the distribution of the standard model background $WZ$, characterized by a Jacobian peak sharply diminishing around $m_T \sim m_W$.
Additionally, the distributions of $\met$ for both signals and backgrounds demonstrate that signal benchmark points generally possess larger missing energy compared to the standard model backgrounds. We have also defined another parameter as the ratio of the scalar sum of three leptons and the missing energy ($\gamma = \frac{\sum_{i} p_T^{l_i}}{\met}$), which can discriminate the signal and backgrounds.
\begin{table}[!htb]
\centering
\setlength{\tabcolsep}{8.0pt}
\renewcommand{\arraystretch}{1.3}
\begin{tabular}{||c||c|c|c||}
\hline\hline
Selection & \multicolumn{3}{c||}{Signal region} \\
\cline{2-4}
Cuts & \texttt{SR-A} & \texttt{SR-B} & \texttt{SR-C}\\
\hline\hline
$p_T^{l_{1,2,3}}$ (GeV) & $>$ 15, 10, 10 & $>$ 20, 15, 15 & $>$ 40, 30, 30 \\
\hline
$m_{SFOS}$ (GeV) & 10-50 & 30-70 & 81.2-101.2 \\
\hline
$m_T$ (GeV) & $<$ 100 & - & $>$ 140 \\
\hline
$\met$ (GeV) & $>$ 20 & $>$ 15 & $>$ 270 \\
\hline
$N_j$ & = 0 & = 0 & $\le$ 1 \\
\hline
$\gamma$ & $>$ 1 & $>$ 1 & $>$ 0.6 \\
\hline\hline
\end{tabular}
\caption{
Signal regions with different selection cuts, optimized for separate signal points, are presented. \texttt{SR-A} and \texttt{SR-B} are optimized for BPs with $\Delta m = 50$ and 70 GeV respectively. \texttt{SR-C} is optimized for large $\Delta m$ [(\texttt{BP3} (800,100) and \texttt{BP4} (1200,100)].
}
\label{tab:signal_region}
\end{table}

\vspace*{-0.1cm}

From the cut optimization, we have found that distinct sets of cuts become necessary for various benchmark points due to their varying $\Delta m$ values. For benchmark points with substantial $\Delta m$, cuts involving solely $p_T^{l_{1,2,3}}$, $m_{SFOS}$, $m_T$, $\met$, number of jets ($N_j$) and $\gamma$ variables prove adequate. Here, the same cut set is effective for both the benchmark points with $\Delta m = 700, 1100$. Three different signal regions are defined in the Table~\ref{tab:signal_region} as \texttt{SR-A}, \texttt{SR-B} and \texttt{SR-C} corresponding to $\Delta m = 50$ GeV, $\Delta m = 70$ GeV and $\Delta m = 700$ \& 1100 GeV. We provide the signal yield, individual background yields, and total background yield, along with the signal significance without uncertainty ($\sigma_{ss}^{0}$) and with 10\% ($\sigma_{ss}^{10}$) and 20\% uncertainty ($\sigma_{ss}^{20}$) in Table~\ref{tab:cut_analysis}. The signal significance without and with uncertainty are calculated using the formulas mentioned in the Equations.~\ref{eq:ams1} and \ref{eq:ams2} (see section.~\ref{sec:ml_outline}). The $\sigma_{ss}^{0}$ ($\sigma_{ss}^{20}$) values corresponding to BP1, BP2, BP3, and BP4 are 4.72 (0.19), 2.04 (0.07), 10.77 (7.98), and 1.59 (1.30), respectively. The cut-based method shows that \texttt{BP1}, \texttt{BP2} and \texttt{BP3} are within reach of HL-LHC\footnote{Due to the presence of huge background or smaller $S/B$ ratio, the compressed scenario (\texttt{BP1}, \texttt{BP2}) will not be sensitive even we consider a 10\% systematic uncertainty.}. 

\begin{table}[!htb]
\centering
\setlength{\tabcolsep}{6.0pt}
\renewcommand{\arraystretch}{1.3}
\begin{tabular}{||c||c|c|c|c||c|c|c||}
\hline\hline
&  & \texttt{SR-A} & \texttt{SR-B} & \texttt{SR-C} & \multicolumn{3}{c||}{} \\
\cline{2-5}
 & $WZ+jets$ & 2247 & 4669 & 5.38 & \multicolumn{3}{c||}{}\\
\cline{2-5}
& $ZZ+jets$ & 187 & 301 & 2.93 & \multicolumn{3}{c||}{}\\
\cline{2-5}
Backgrounds& $t\bar{t}+jets$ & 13198 & 13176 & 0.00 & \multicolumn{3}{c||}{}\\
\cline{2-5}
& $VVV+jets$ & 116 & 218 & 1.69 & \multicolumn{3}{c||}{}\\
\cline{2-5}
& $t\bar{t}V+jets$ & 37 & 88 & 0.47 & \multicolumn{3}{c||}{}\\
\cline{2-5}
& $Z+jets$ & 0.00 & 0.00 & 0.00 & \multicolumn{3}{c||}{}\\
\cline{2-8}
& Total & 15785 & 18454 & 10.47 & \multicolumn{3}{c|}{Signal Significance} \\
\cline{6-8}
& Background yield & & & & $\sigma_{ss}^{0}$ & $\sigma_{ss}^{10}$ & $\sigma_{ss}^{20}$ \\
\hline
& \texttt{BP1} & 596 & - & - & 4.72 & 0.37 & 0.19 \\
\cline{2-8}
Signal & \texttt{BP2} & - & 277 & - & 2.04 & 0.15 & 0.07 \\
\cline{2-8}
& \texttt{BP3} & - & - & 51.10 & 10.77 & 9.76 & 7.98 \\
\cline{2-8}
& \texttt{BP4} & - & - & 5.55 & 1.59 & 1.50 & 1.30 \\
\hline\hline
\end{tabular}
\caption{The different background yield, total background yield and signal yields calculated at $\sqrt{s} = 14$ TeV and $\mathcal{L} = 3000~fb^{-1}$ corresponding to different signal regions are shown in this table. The signal significance without uncertainty ($\sigma_{ss}^{0}$), with 10\% ($\sigma_{ss}^{10}$) and 20\% systematic uncertainty ($\sigma_{ss}^{20}$) are also presented in the last three column. 
}
\label{tab:cut_analysis}
\end{table}

\subsection{Machine Learning based analysis}
\label{sec:ml_analysis}
We will now conduct a similar analysis employing Machine Learning algorithms. For this purpose, we have compiled a dataset comprising events after the pre-selection cuts. In order to fully harness the benefits of utilizing a boosted decision tree approach over the cut-and-count method for distinguishing signal from background, it's crucial to incorporate not just the primary kinematic variables such as the number of leptons ($n_l$), number of jets ($n_j$), and transverse momenta of leptons ($p_T^l$), but also to generate additional derived variables like the mass of same-flavor opposite-sign (SFOS) pairs, the sum of transverse momenta ($p_T$) of leptons and jets, as well as variables such as effective mass. Therefore, we have developed 21 distinct features tailored for the specific analysis required in machine learning. These features variables are mentioned below :
\begin{itemize}
\item The transverse momenta of the three chosen leptons, denoted as $p_T^{l_1}$, $p_T^{l_2}$, and $p_T^{l_3}$, are represented by three variables. 
\item The angular difference between each pair of the three leptons, denoted as $\Delta R_{l_1l_2}$, $\Delta R_{l_1l_3}$, and $\Delta R_{l_2l_3}$ (three features), is measured. (3 features) 
\item Azimuthal angular difference between leptons and missing energy represented as $\Delta\phi_{l_i\met}$ where i=1,2,3 (3 features)
\item No of jets, $N_j$  and missing energy, $\met$ (2 features)
\item The SFOS pair number ($N_{SFOS}$) and the mass of SFOS pair ($m_{SFOS}$) with mass closest of $Z$ boson mass (2 features)
\item After selecting the SFOS pair, with the remaining lepton, we construct the transverse mass ($m_T$), which is defined in Equation.~\ref{eq:mt}. Also, we calculate the invariant mass of three leptons ($m_{3l}$). Then we calculate the difference between the masses of $m_{3l}$ and $m_{SFOS}$ denoted as $\Delta m(m_{3l},m_{SFOS})$. (3 features) 
\item Next, we compute the scalar sum of the transverse momenta of three leptons ($H_T^{lep}$) and the scalar sum of transverse momenta of jets ($H_T^{jet}$). Additionally, we determine the sum of missing energy and $H_T^{lep}$ as $H_T^{lep}+\met$, and similarly for jets as $H_T^{jet}+\met$. We introduce the effective mass variable, defined as $m_{eff} = H_T^{lep} + H_T^{jet} + \met$, comprising a total of 5 features.
\end{itemize} 
We incorporate the relevant weight into both the signal and background during the preparation of the data file. Subsequently, we proceed with the machine learning analysis using four distinct algorithms.
\subsubsection{Hyperparameter variation for different algorithms}
\label{sec:rpc_xgb}
In this section we present the role of hyperparameters for the performances of different algorithms. In Sec.~\ref{sec:decision_tree}, we have already summarized the mechanics of these algorithms. Each algorithm offers a range of hyperparameters for customization. For instance, with the \texttt{XGBoost} algorithm, one can adjust parameters such as the learning rate ($\eta$), the number of trees, the maximum depth of each tree ($max\_depth$), the minimum weight required for node splitting (\textit{Min. child weight}), a regularization parameter ($\gamma$) to prevent overfitting, an $\alpha$ parameter for further regularization via a penalty term within the loss function, and the subsample size of the dataset used for training, which is randomly selected etc..

\begin{figure}[!htb]
\begin{center}
\includegraphics[scale=0.24]{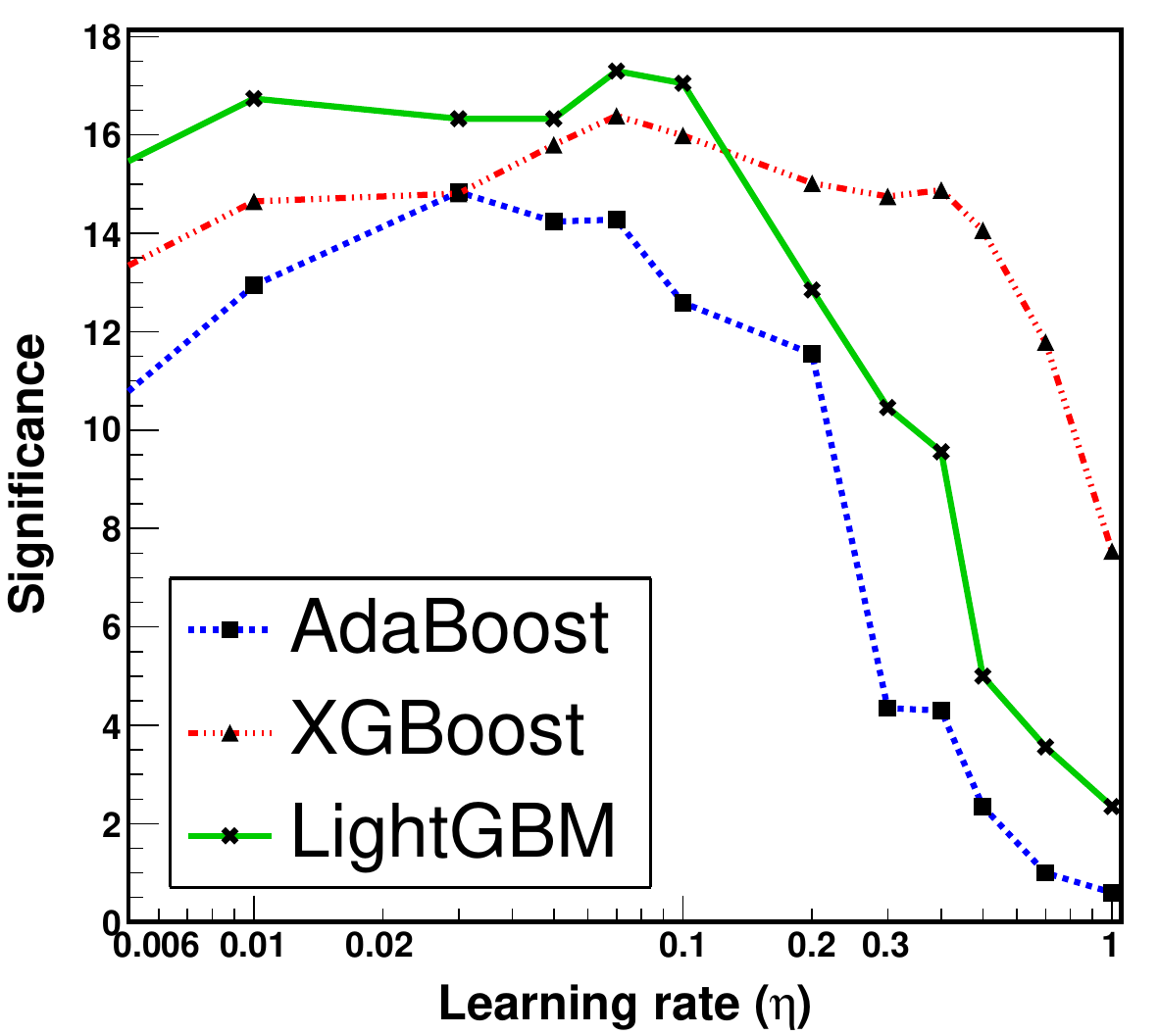}
\includegraphics[scale=0.24]{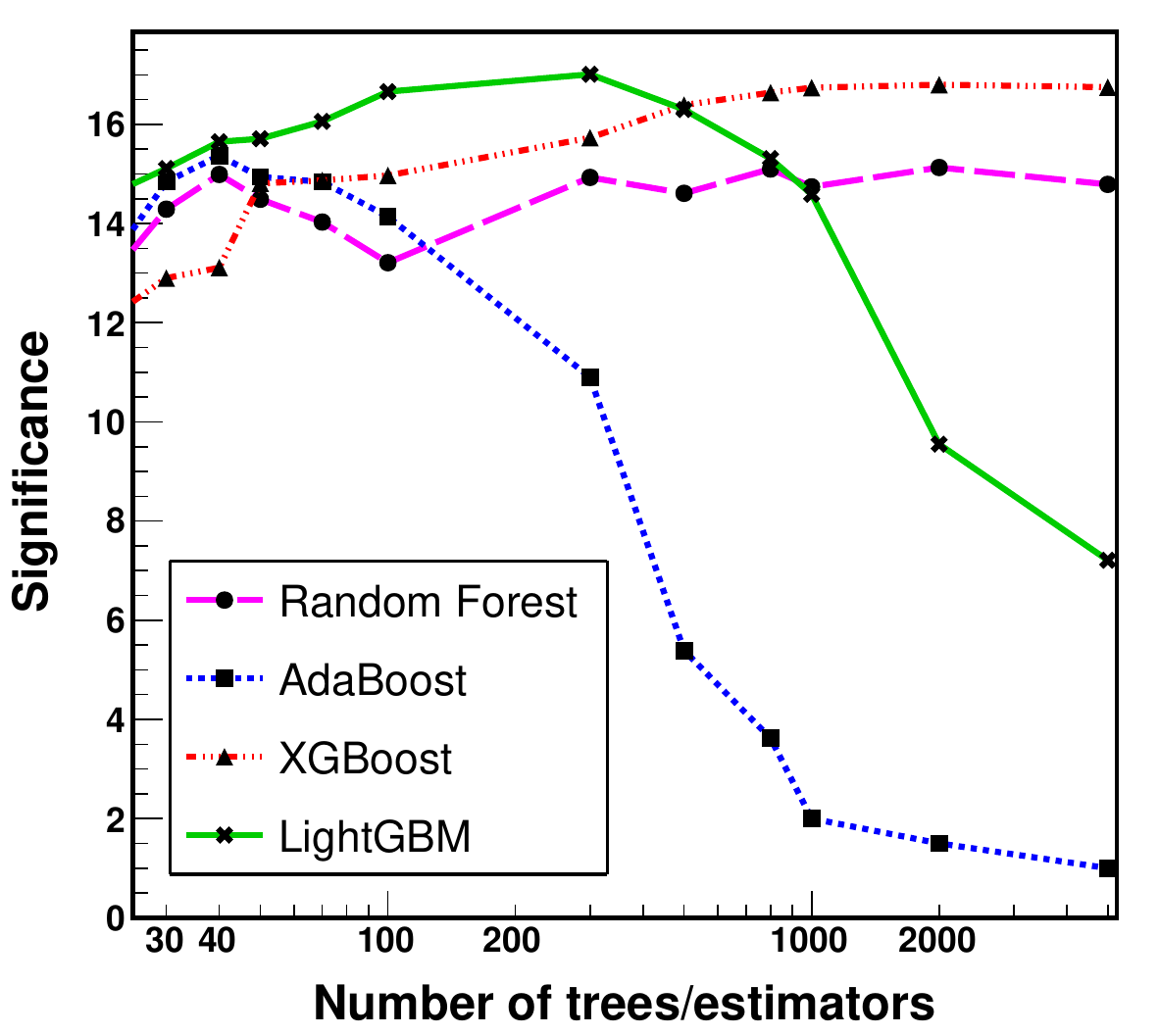}
\includegraphics[scale=0.24]{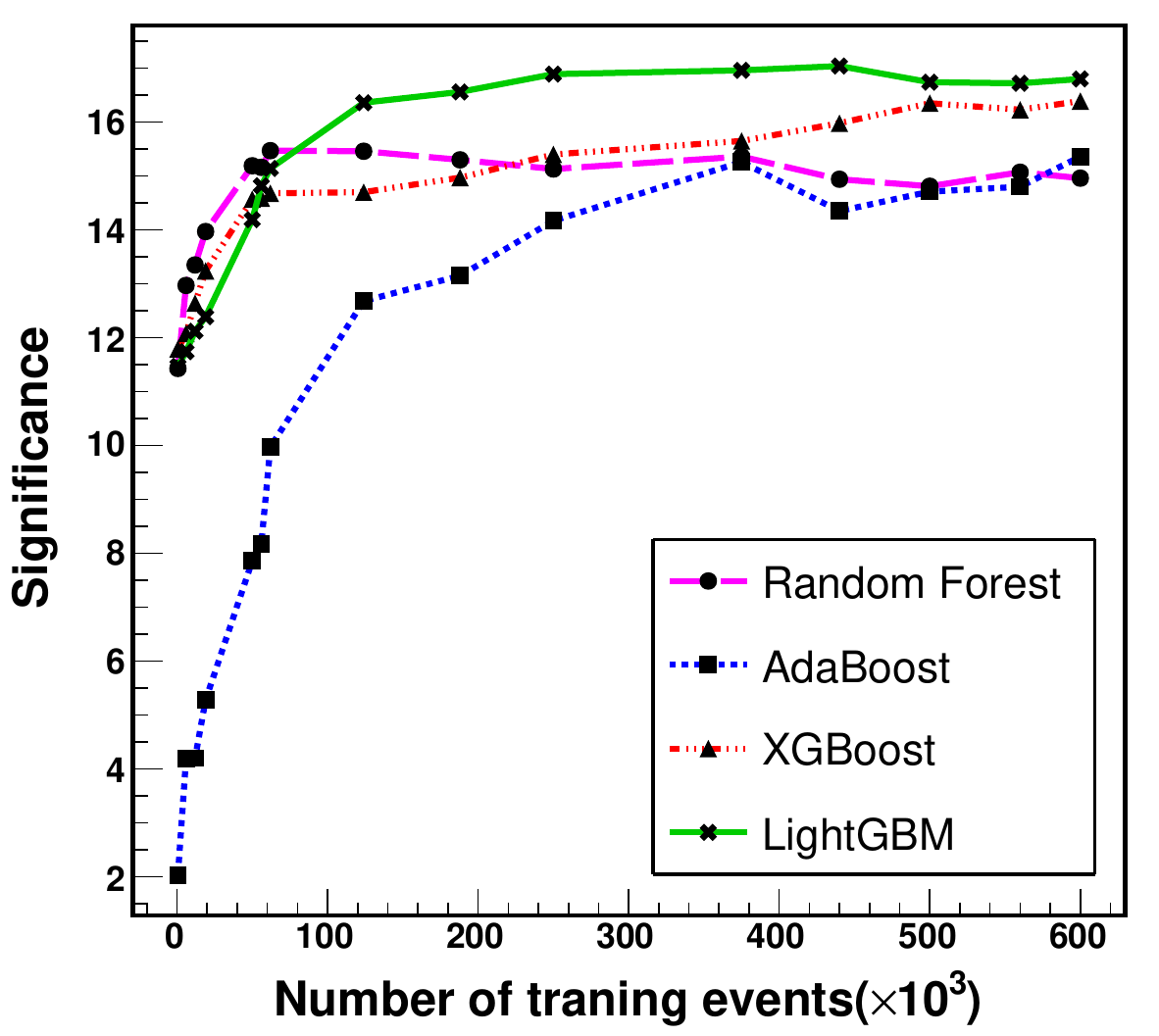}
\caption{The signal significance as a function of learning rate ($\eta$), number of trees or estimators and the number of training events for the \texttt{BP3} (800,100) signal benchmark point are displayed corresponding to the four different ML algorithms in the left, middle and right panel respectively.}
\label{fig:variation_plot}
\end{center}
\end{figure}

Similarly, the \texttt{LightGBM} algorithm shares many of these hyperparameters, alongside additional parameters like the number of leaves, which should be set to a value less than $2^{max\_depth}$ to mitigate overfitting and enhance model accuracy.
\texttt{AdaBoost} follows a similar optimization strategy, with parameters such as the number of estimators, which corresponds to the number of trees. However, unlike other algorithms, \texttt{Random Forest} doesn't include a learning rate parameter in its optimization process.

We have estimated the best set of hyperparameters effective for each algorithm and mentioned in the Table.~\ref{tab:param_set} (see Appendix.~\ref{sec:appendix}). Upon determining the optimal hyperparameters for each algorithm, we investigate how varying the learning rate parameter affects the signal significance across three algorithms: \texttt{AdaBoost}, \texttt{XGBoost}, and \texttt{LightGBM}. We examine the range of the $\eta$ parameter from 0.0001 to 1.0 and depict the resulting variations in Fig.~\ref{fig:variation_plot} (left). The plot clearly indicates that \texttt{AdaBoost} demonstrates peak performance around $\eta$ = 0.02-0.08, \texttt{XGBoost} performs optimally within $\eta$ = 0.05-0.15, and \texttt{LightGBM} shows higher significance around $\eta$ = 0.01-0.1.

Additionally, we explore the impact of the number of trees, ranging from 1 to 5000, on the algorithms, as shown in Fig.~\ref{fig:variation_plot} (middle), while keeping all other parameters constant. It becomes evident from the plot that for \texttt{Random Forest} and \texttt{XGBoost} algorithms, the signal significance reaches its maximum after approximately 300 trees, with minimal variation thereafter. However, for \texttt{AdaBoost} and \texttt{LightGBM} algorithms, the signal significance declines rapidly after 100 and 500 trees, respectively.
Furthermore, we analyze the effect of the number of training events in the dataset on all algorithms, presented in Fig.~\ref{fig:variation_plot} (right), to ensure that we have taken an adequate amount of data for the ML analysis. Upon reaching around 3$\times10^{5}$ training events, all algorithms achieve maximum significance, with marginal variation as the number of training events increases.

\subsubsection{Feature importance with SHapley}
\label{sec:shap}
As discussed in Sec.\ref{sec:ml_analysis}, for this analysis, more than 20 distinct features have been used to train the ML model and the impact of these features in discriminating the signal and background can be determined using SHAP package~\cite{Shapley+1953+307+318,lundberg2018consistent,pedregosa2018scikitlearn}. In the context of tree ensemble methods like gradient boosting methods or random forests, it is common to assign importance values to each input feature in order to comprehend the predictions. These values can be calculated for an individualized prediction or the global prediction of an entire dataset. Popular packages like XGBoost~\cite{Shapley+1953+307+318,lundberg2018consistent,pedregosa2018scikitlearn} offer implementations of tree ensembles that enable users to calculate a metric of feature importance. These metrics aim to condense the complexity of ensemble models and offer an understanding of the key features influencing the model's predictions. Global feature importance is derived through three main methodologies: gain, split count, permutation, and Saabas method. However, most of these methods of calculating feature importance metrics are inconsistent and ineffective.\\
However, SHapley values, which stem from principles of game theory, demonstrate the contribution of an individual player within a group. 
As indicated in Ref.~\cite{lundberg2018consistent}, among various techniques mentioned above, SHapley values emerge as the most dependable measure for selecting feature importance. 
This method of local feature attribution, founded on SHapley values, was pioneered by S. Lundberg and S. Lee, as outlined in \cite{lundberg2017unified}. 
The technique of calculating SHapley values involves training the model on various subsets of features, where each subset is a subset of the full feature set. It assigns an importance score to each feature, indicating its impact on model predictions when included. To calculate this impact, two models are trained: one with the feature in the model and another without it. The difference in predictions between these models, considering the current input, helps quantify the feature's influence. Since the influence of a feature's absence is influenced by other features in the model, these differences are computed across all possible feature subsets, excluding the feature under consideration. Subsequently, SHapley values are computed and utilized to attribute importance to each feature. The formula used for the calculation of SHapley value looks like~\cite{lundberg2018consistent} 
\begin{equation}
\label{eq:shap}
\phi_i = \sum_{S\subseteq N\setminus\{i\}} \frac{|S|!(M-|S| - 1)!}{M!}[f_x(S\cup\{i\}) - f_x(S)]
\end{equation}

\begin{figure}[!htb]
\begin{center}
\includegraphics[scale=0.38]{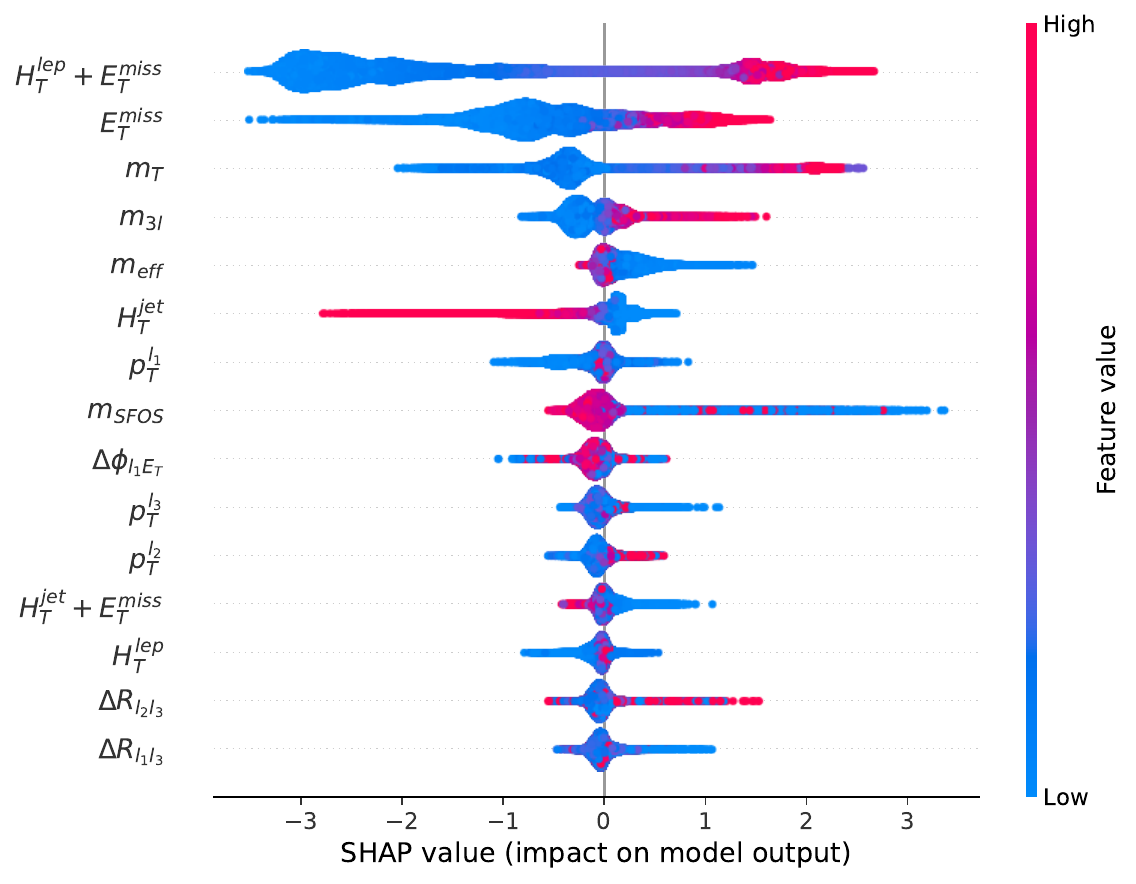}
\includegraphics[scale=0.38]{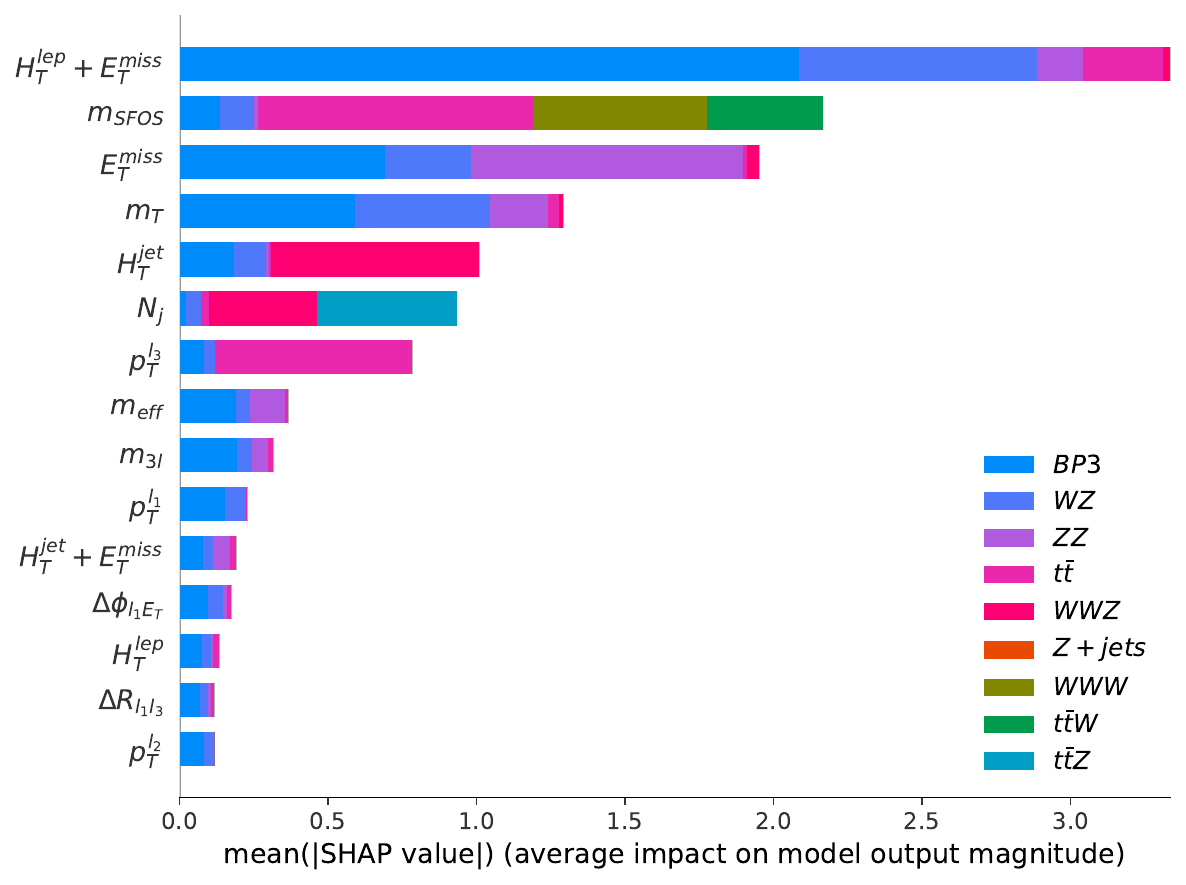}
\caption{Left: SHAP summary plot for then signal point \texttt{BP3} (800,100) \\
Right: SHAP summary bar plot for \texttt{BP3} and all the backgrounds. The most important 15 feature variables are displayed here. }
\label{fig:shap}
\end{center}
\end{figure}

\vspace*{-0.4cm}
In this context, N denotes the complete set of features and ``i'' represents the feature under consideration for SHapley value calculation. M signifies the overall count of features. S denotes a subset of N that excludes i. The function $f_x$ represents the model's prediction. Consequently, the algorithm evaluates a weighted sum of the variances in model predictions when including or excluding the ith feature across all potential combinations of the subset S. \\
Various kinds of visualizations can be generated using SHapley values to illustrate how different features impact the model. SHAP summary plots utilize individualized feature attributions to encapsulate the significance of each feature effectively while maintaining visual simplicity. Initially, features are arranged based on their overall impact, represented by the sum of absolute SHapley values. Subsequently, the corresponding SHapley values are horizontally plotted, with stacking occurring vertically if space becomes insufficient. The SHAP summary plot (dotted) is shown in the Fig.~\ref{fig:shap}. Every point is shaded according to the corresponding feature's value, ranging from a low (blue) hue to a high (red) one. Also, we consider the absolute SHapley values and show a bar-type summary plot where each color represents each class and the bar width shows the effect of each feature on classifying each class. In the left panel of the Fig.~\ref{fig:shap}, we have displayed the SHapley value distribution for the signal benchmark point \texttt{BP3} only. The right panel of Fig.~\ref{fig:shap} represents the combined effect of the signal point and all SM backgrounds. In Fig.~\ref{fig:shap}, we plot only the first 15 feature variables according to importance. The variable ($H_T^{lep}+\met$) is the top most important feature, while the $m_{SFOS}$ variable is the next effective variable when the model tries to discriminate the signal and background (see the right panel of Fig.~\ref{fig:shap}.  
and the combined effect of the signal and all the background (right). It is obvious that when we try to discriminate the signal and background, variable $m_{SFOS}$ will be very effective and that's why it has more importance value in the bar plot (right) than the dotted plot (left). The other variables like $\met$ and $m_T$ show the higher SHapley value, which is quite similar to the cut-based analysis.

\subsubsection{Comparison of results coming from different algorithms}
\label{sec:comparison}

\begin{figure}[!htb]
\begin{center}
\includegraphics[scale=0.45]{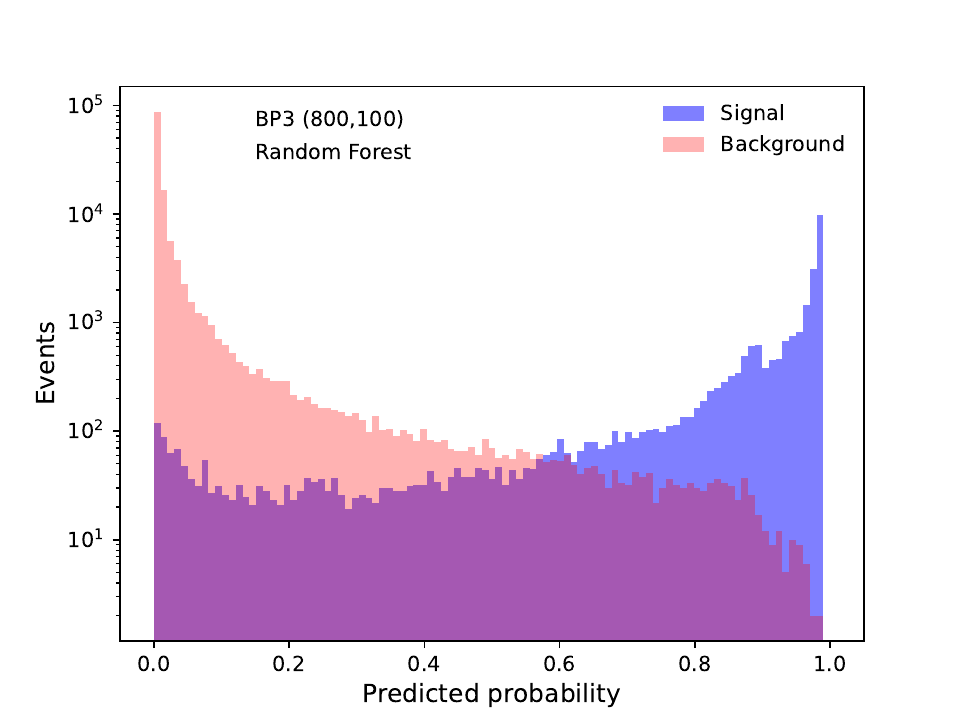}
\includegraphics[scale=0.45]{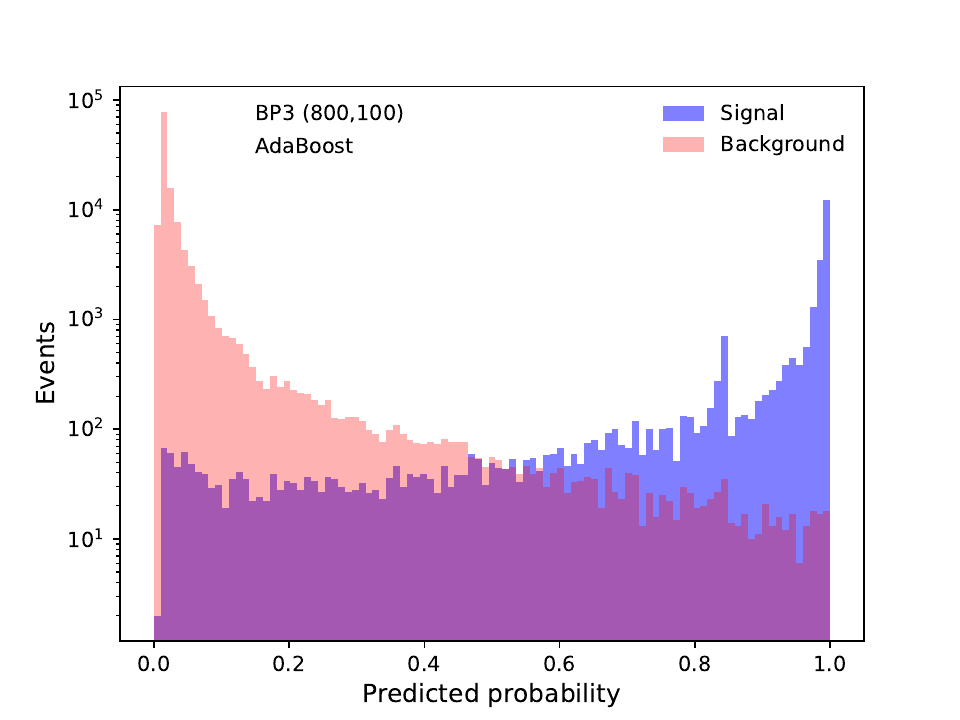}
\includegraphics[scale=0.45]{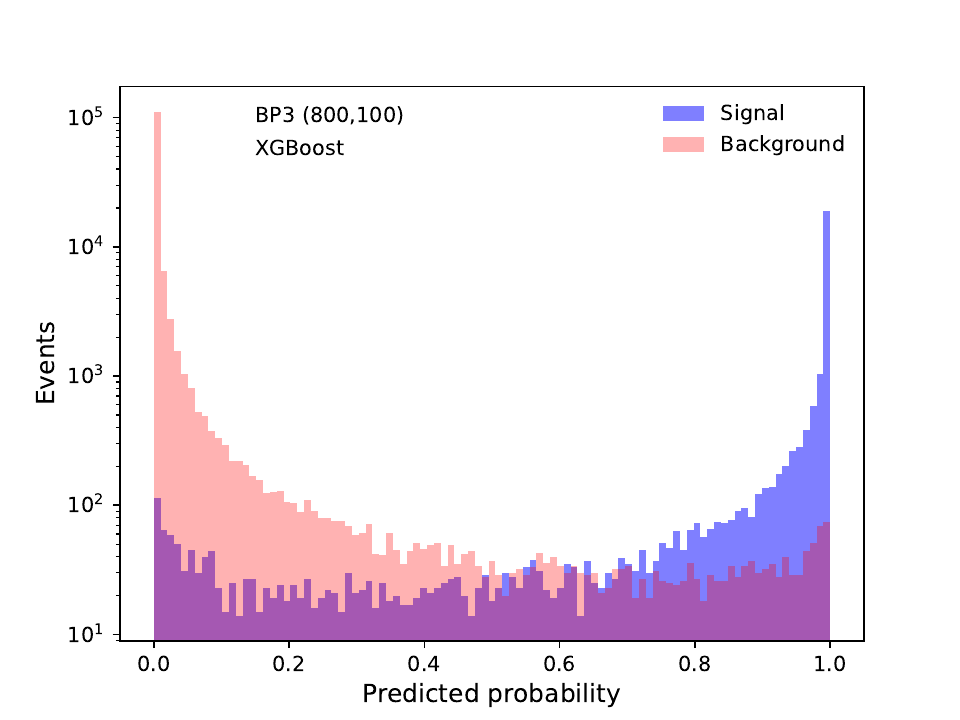}
\includegraphics[scale=0.45]{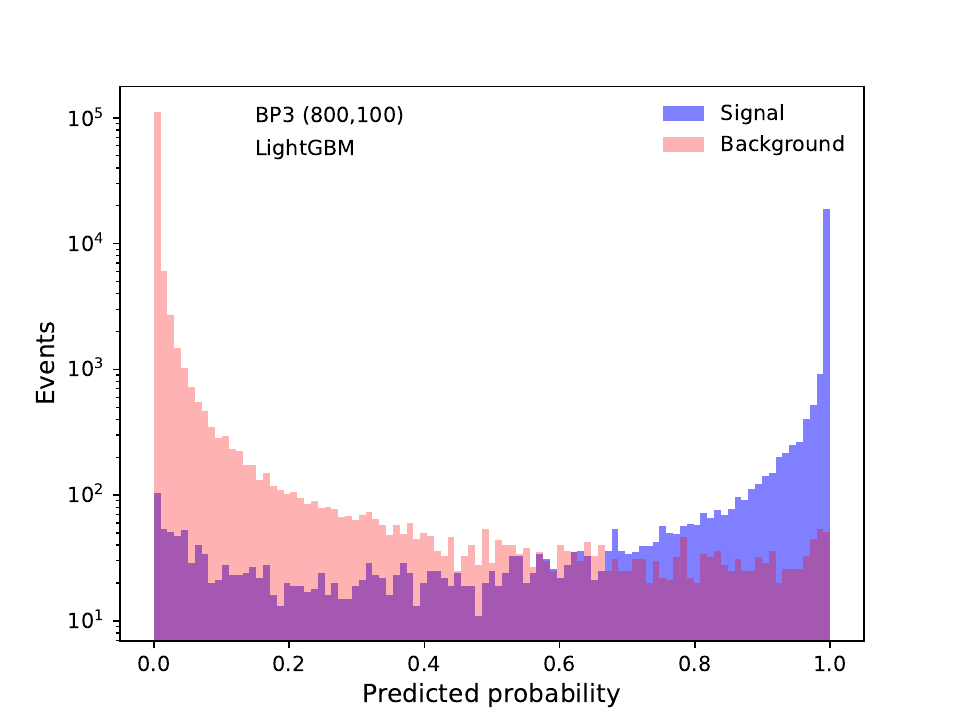}
\caption{The signal and background events distribution as a function of the predicted probability in the test dataset corresponding to the \texttt{BP3} for the \texttt{Random Forest} classifier (upper left), \texttt{AdaBoost} (upper right), \texttt{XGBoost} classifier (lower left) and \texttt{LightGBM} (lower right).}
\label{fig:event_dist}
\end{center}
\end{figure}
 
After optimization of hyperparameters for each algorithm, we proceed to visualize the distribution of signal and background events corresponding to two signal benchmark points \texttt{BP1} and \texttt{BP3}, which represent two different scenarios like small $\Delta m$ (50 GeV) and large $\Delta m$ (700 GeV). The event distributions of \texttt{BP3} (\texttt{BP1}) are displayed in Fig.~\ref{fig:event_dist} (Fig..~\ref{fig:event_dist1} in the Appendix.~\ref{sec:appendix}). Analysis of the distributions corresponding to \texttt{BP3} reveals that the \texttt{XGBoost} and \texttt{LightGBM} algorithms outperform \texttt{Random Forest} and \texttt{AdaBoost} in discriminating signal from background, as evidenced by the higher number of background events at larger probability cuts for the latter two algorithms. For small $\Delta m$ also, these two algorithms work better, but for all the algorithms, there is significant overlapping between signal and background which is evident from the Fig.~\ref{fig:event_dist1}. In the Fig.~\ref{fig:cut_plot}, the distribution of the kinematic variables also displays the same behavior about the significant overlapping.

We also represent the ROC curve for two different benchmark points \texttt{BP1} (left) and  \texttt{BP3} (right) in the Figure.~\ref{fig:roc}. Here we can see that \textit{auc} values are a little bit different for each algorithm corresponding to \texttt{BP1} whereas for \texttt{BP3}, the \textit{auc} values are pretty close. 
\begin{figure}[!htb]
\begin{center}
\includegraphics[scale=0.36]{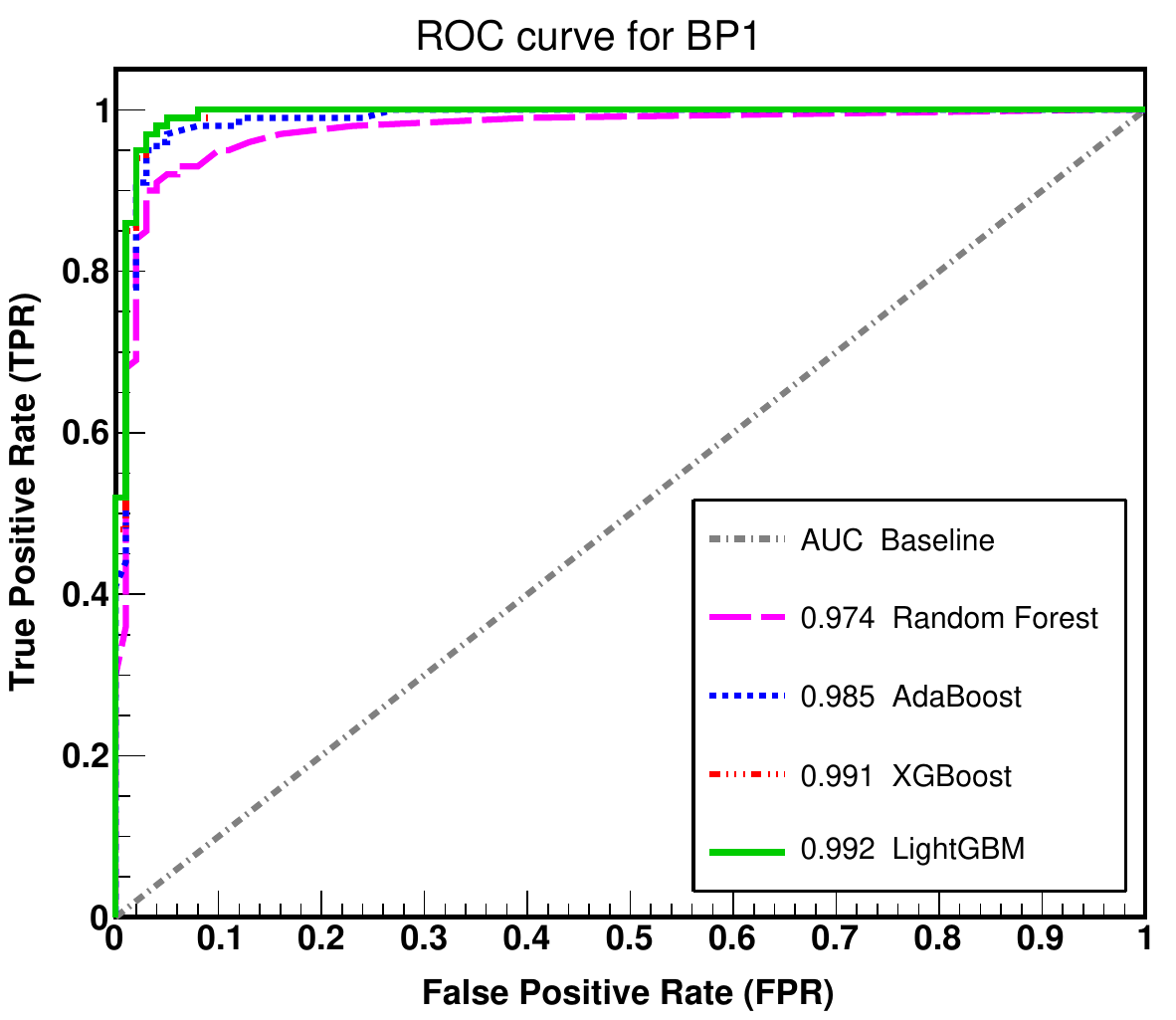}
\includegraphics[scale=0.36]{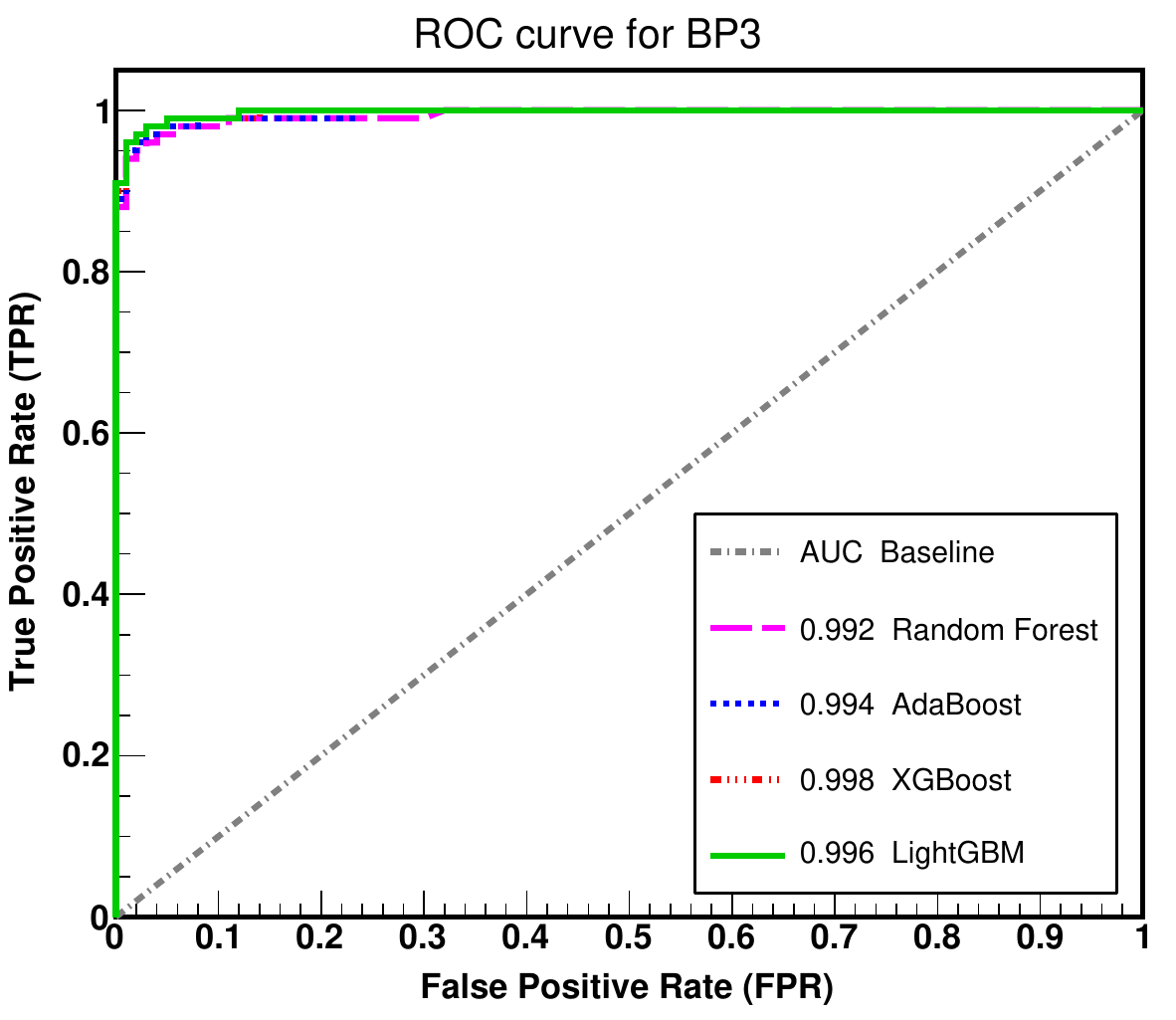}
\caption{ROC curves and \textit{auc} values are shown for each ML algorithm corresponding to two signal benchmark points \texttt{BP1} (left) and \texttt{BP3} (right) are displayed here.}
\label{fig:roc}
\end{center}
\end{figure}
 
\begin{table}[!htb]
\setlength{\tabcolsep}{1.0pt}
\renewcommand{\arraystretch}{1.9}
\centering
\small
\begin{tabular}{||c||c|c|c|c|c|c||c|c|c|c|c|c||}
\hline\hline
  & \multicolumn{6}{c||}{\texttt{BP1 (400,350)}} & \multicolumn{6}{c||}{BP2 (500,430)} \\
\cline{2-13} 
Algorithm& Proba- & Signal & Back- & $\sigma_{ss}^{0}$ & $\sigma_{ss}^{10}$  & $\sigma_{ss}^{20}$ & Proba & Signal & Back- & $\sigma_{ss}^{0}$ & $\sigma_{ss}^{10}$ & $\sigma_{ss}^{20}$ \\
& bility & yield & ground & &  & & bility & yield & ground &  & &  \\
& cut & (S) & yield (B) & &  & & cut & (S) & yield (B) &  &  & \\
\hline\hline 
\texttt{RF} & 0.80 & 1152 & 5772 & 14.70 & 1.86 & 0.94 & 0.70 & 559 & 12158 & 5.03 & 0.45 & 0.23 \\ 
\hline
\texttt{AdaBoost} & 0.86 & 1439 & 6434 & 17.22 & 2.05 & 1.03 & 0.80 & 760 & 13276 & 6.54 & 0.56 & 0.28 \\ 
\hline
\texttt{XGBoost} & 0.92 & 1264 & 3757 & 19.61 & 3.01 & 1.52 & 0.83 & 786 & 10017 & 7.76 & 0.76 & 0.38 \\ 
\hline
\texttt{LightGBM} & 0.92 & 1254 & 3512 & 20.06 & 3.18 & 1.60 & 0.90 & 572 & 4996 & 7.96 & 1.09 & 0.55  \\ 
\hline\hline
& \multicolumn{6}{c||}{\texttt{BP3 (800,100)}} & \multicolumn{6}{c||}{BP4 (1200,100)} \\
\hline
\texttt{RF} & 0.96 & 81.28 & 11.94 & 14.85 & 13.12 & 10.40 & 0.92 & 9.43 & 11.99 & 2.45 & 2.29 & 1.94 \\ 
\hline
\texttt{AdaBoost} & 0.98 & 77.12 & 12.19 & 14.20 & 12.56 & 9.95  & 0.98 & 9.49 & 17.91 & 2.08 & 1.89 & 1.54 \\ 
\hline
\texttt{XGBoost} & 0.99 & 93.78 & 12.38 & 16.39 & 14.35 & 11.27 & 0.99 & 10.24 & 12.05 & 2.63 & 2.45 & 2.08 \\ 
\hline
\texttt{LightGBM} & 0.99 & 99.64 & 11.94 & 17.30 & 15.14 & 11.90 & 0.97 & 10.05 & 11.97 & 2.60 & 2.42 & 2.06 \\ 
\hline\hline
\end{tabular}
\caption{The signal yields and the total background yields calculated at $\sqrt{s} = 14$ TeV and $\mathcal{L} = 3000~fb^{-1}$ corresponding to the four benchmark points are presented here for four different DT algorithms. We also show the signal significance with 0\% ($\sigma_{ss}^{0}$), 10\% ($\sigma_{ss}^{10}$) and 20\% ($\sigma_{ss}^{20}$) systematic uncertainty. Here \texttt{RF} refers to the \texttt{Random Forest} algorithm.}
\label{tab:ml_compare}
\end{table}

In the Table.~\ref{tab:ml_compare} we provide the signal yield, total background yield, and signal significance without systematic uncertainty ($\sigma_{ss}^{0}$), and with uncertainties of 10\% ($\sigma_{ss}^{10}$) and 20\% ($\sigma_{ss}^{20}$). 
For benchmark points with low $\Delta m$ (\texttt{BP1} abd \texttt{BP2}), there are significant overlap between signal and background events and the dominant contribution comes from $WZ+jets$ and $t\bar{t}+jets$ backgrounds. In these compressed scenarios, due to the small $S/B$ ratio ($<1$), we observe a sharp decrease in the significance when we consider the systematic uncertainties. 
For \texttt{BP1} and \texttt{BP2}, the signal significance becomes $\sim$ 3-4 times larger as compared to the cut-and-count method. For \texttt{BP3} and \texttt{BP4}, the signal significance increases by $\sim$ 30-60\% in ML method. Specifically, when considering signal significance with uncertainty, we can exclude the point where $m_{\chonepm/\lsptwo}$ = 400 GeV for $m_{\lspone}$ = 350 GeV. Similarly, we can also exclude $m_{\chonepm/\lsptwo}$ = 1200 GeV for $m_{\lspone}$ = 100 GeV, regardless of uncertainty. It should be noted that the performances of \texttt{XGBoost} and \texttt{LightGBM} are almost similar and \texttt{LightGBM} gives a slightly better signal significance. The Table.~\ref{tab:ml_compare} indicates that even with 20\% systematic uncertainty chargino mass with 800 (1200) GeV will be within $\sim12\sigma$ and $\sim2\sigma$ reach for a LSP mass of 100 GeV.

\section{Summary}
\label{sec:summary}
Over the years, machine learning techniques have significantly boosted efficiency and accuracy in HEP data analysis for both experimental and phenomenological works.
For the analyses of event triggering, jet tagging, particle identification, 
event selection, object reconstruction, event classification etc., the HEP community has been widely using the boosted decision tree (BDT) algorithms for a long time. 
We have briefly summarized the major important analyses performed by the ATLAS and CMS collaboration utilizing LHC data, as well as the works of the phenomenological groups involving the use of ML algorithms with emphasis on DT-based approaches. We have also outlined the basic concepts of machine learning 
along with different kinds of loss functions and their roles, issues of underfitting, overfitting, etc.  
Evaluation of the effectiveness and accuracy of the  ML models in solving a specific task, e.g.,  classifying signal and background events, can be done using 
performance metrics. Different kinds of metrics relevant to particle physics analysis, like ROC curve, F-score, ams score, etc., are discussed in detail. 

In this article, we have focused solely on a prominent machine-learning 
technique, namely Decision Tree (DT), specifically boosted decision tree,  which is primarily used for tasks in supervised classification and regression.
We particularly explore four decision tree-based ML algorithms, namely, \texttt{Random Forest},    \texttt{AdaBoost} and two gradient boosting frameworks such as   \texttt{XGBoost}, and \texttt{LightGBM}, in the context of Supersymmetry which is a well-promising candidate for beyond the Standard model framework. 
We summarize the basic concepts and working principles of these four 
algorithms along with the flowcharts. 
In numerous analyses, both the ATLAS and CMS collaborations, HEP phenomenological groups have employed BDT algorithms to enhance the sensitivity of sparticle searches within the framework of RPC and RPV SUSY models. However, exploring the compressed SUSY parameter space still remains as a significant challenge.

Using an example of wino-type electroweakino productions at the high 
luminosity LHC, we  demonstrate   how these algorithms lead to improvement 
in the search sensitivity   compared to traditional cut-based methods in both 
 compressed and non-compressed   R-Parity conserving SUSY scenarios. 
 The  optimization of hyperparameters and its role in signal significance 
 are studied in detail for these four algorithms. 
 We have also discussed how to find out the individualized and global feature importance in ML methods using \texttt{SHAP} package. We have found that there are $\sim$ 30-60\% gains in the significance for the signal benchmark points with a large mass gap between NLSP-LSP pair, whereas, for the compressed region (mass gap 50 and 70 GeV), the signal significance improves by $\sim$ 3-4 times compared to the cut-based analysis. We have obtained that the \texttt{LightGBM} and \texttt{XGBoost} algorithms perform better than the other two algorithms \texttt{Random Forest} and \texttt{AdaBoost}. 
Although the deep neural networks (DNN) or deep learning (DL) techniques, which are based on multilayer NN, are gaining popularity, BDTs continue to be  
highly relevant and important in High Energy Physics due to their user-friendly nature, interpretability, computational efficiency, and more.

\newpage

\noindent
\textbf{Acknowledgments:} Authors would like to thank Ritik Pal for fruitful suggestions. The work of A.C is supported by the SERB Core Research Grant CRG/2023/008570. 

\vspace{+1.2cm}

\noindent
\textbf{Data availability statement:} No data associated in the manuscript.

\bibliography{reference} 

\providecommand{\href}[2]{#2}\begingroup\raggedright\begin{thebibliography}{100}

\bibitem{Martin:1997ns}
S.~P. Martin, \emph{{A Supersymmetry primer}},
  \href{http://dx.doi.org/10.1142/9789812839657_0001}{\emph{Adv. Ser. Direct.
  High Energy Phys.} {\bf 18} (1998) 1--98},
  [\href{http://arxiv.org/abs/hep-ph/9709356}{{\tt hep-ph/9709356}}].

\bibitem{drees2004theory}
M.~Drees, P.~Roy and R.~Godbole, \emph{Theory and Phenomenology of Sparticles:
  An Account of Four-dimensional N}.
\newblock World Scientific, 2004.

\bibitem{baer2006weak}
H.~Baer and X.~Tata, \emph{Weak Scale Supersymmetry: From Superfields to
  Scattering Events}.
\newblock Cambridge University Press, 2006.

\bibitem{atlas_susy}
``Atlas susy public result.''
  \url{https://twiki.cern.ch/twiki/bin/view/AtlasPublic/SupersymmetryPublicResults}.

\bibitem{cms_susy}
``Cms susy public result.''
  \url{https://cms-results.web.cern.ch/cms-results/public-results/preliminary-results/SUS/index.html}.

\bibitem{Bhattacherjee:2013wna}
B.~Bhattacherjee, A.~Choudhury, K.~Ghosh and S.~Poddar, \emph{{Compressed
  supersymmetry at 14 TeV LHC}},
  \href{http://dx.doi.org/10.1103/PhysRevD.89.037702}{\emph{Phys. Rev. D} {\bf
  89} (2014) 037702}, [\href{http://arxiv.org/abs/1308.1526}{{\tt 1308.1526}}].

\bibitem{Dutta:2015exw}
J.~Dutta, P.~Konar, S.~Mondal, B.~Mukhopadhyaya and S.~K. Rai, \emph{{A Revisit
  to a Compressed Supersymmetric Spectrum with 125 GeV Higgs}},
  \href{http://dx.doi.org/10.1007/JHEP01(2016)051}{\emph{JHEP} {\bf 01} (2016)
  051}, [\href{http://arxiv.org/abs/1511.09284}{{\tt 1511.09284}}].

\bibitem{Chakraborti:2017dpu}
M.~Chakraborti, U.~Chattopadhyay and S.~Poddar, \emph{{How light a higgsino or
  a wino dark matter can become in a compressed scenario of MSSM}},
  \href{http://dx.doi.org/10.1007/JHEP09(2017)064}{\emph{JHEP} {\bf 09} (2017)
  064}, [\href{http://arxiv.org/abs/1702.03954}{{\tt 1702.03954}}].

\bibitem{Chowdhury:2016qnz}
D.~Chowdhury, K.~M. Patel, X.~Tata and S.~K. Vempati, \emph{{Indirect Searches
  of the Degenerate MSSM}},
  \href{http://dx.doi.org/10.1103/PhysRevD.95.075025}{\emph{Phys. Rev. D} {\bf
  95} (2017) 075025}, [\href{http://arxiv.org/abs/1612.06471}{{\tt
  1612.06471}}].

\bibitem{Dutta:2017jpe}
J.~Dutta, P.~Konar, S.~Mondal, B.~Mukhopadhyaya and S.~K. Rai, \emph{{Search
  for a compressed supersymmetric spectrum with a light Gravitino}},
  \href{http://dx.doi.org/10.1007/JHEP09(2017)026}{\emph{JHEP} {\bf 09} (2017)
  026}, [\href{http://arxiv.org/abs/1704.04617}{{\tt 1704.04617}}].

\bibitem{KumarBarman:2020ylm}
R.~Kumar~Barman, G.~Belanger and R.~M. Godbole, \emph{{Status of low mass LSP
  in SUSY}}, \href{http://dx.doi.org/10.1140/epjst/e2020-000198-1}{\emph{Eur.
  Phys. J. ST} {\bf 229} (2020) 3159--3185},
  [\href{http://arxiv.org/abs/2010.11674}{{\tt 2010.11674}}].

\bibitem{Barman:2022jdg}
R.~K. Barman, G.~B\'elanger, B.~Bhattacherjee, R.~M. Godbole and R.~Sengupta,
  \emph{{Is Light Neutralino Thermal Dark Matter in the Phenomenological
  Minimal Supersymmetric Standard Model Ruled Out?}},
  \href{http://dx.doi.org/10.1103/PhysRevLett.131.011802}{\emph{Phys. Rev.
  Lett.} {\bf 131} (2023) 011802}, [\href{http://arxiv.org/abs/2207.06238}{{\tt
  2207.06238}}].

\bibitem{He:2023lgi}
Y.~He, L.~Meng, Y.~Yue and D.~Zhang, \emph{{Impact of the recent measurement of
  (g-2)\ensuremath{\mu}, the LHC search for supersymmetry, and the LZ
  experiment on the minimal supersymmetric standard model}},
  \href{http://dx.doi.org/10.1103/PhysRevD.108.115010}{\emph{Phys. Rev. D} {\bf
  108} (2023) 115010}, [\href{http://arxiv.org/abs/2303.02360}{{\tt
  2303.02360}}].

\bibitem{Chakraborti:2015mra}
M.~Chakraborti, U.~Chattopadhyay, A.~Choudhury, A.~Datta and S.~Poddar,
  \emph{{Reduced LHC constraints for higgsino-like heavier electroweakinos}},
  \href{http://dx.doi.org/10.1007/JHEP11(2015)050}{\emph{JHEP} {\bf 11} (2015)
  050}, [\href{http://arxiv.org/abs/1507.01395}{{\tt 1507.01395}}].

\bibitem{Chakraborti:2014gea}
M.~Chakraborti, U.~Chattopadhyay, A.~Choudhury, A.~Datta and S.~Poddar,
  \emph{{The Electroweak Sector of the pMSSM in the Light of LHC - 8 TeV and
  Other Data}}, \href{http://dx.doi.org/10.1007/JHEP07(2014)019}{\emph{JHEP}
  {\bf 07} (2014) 019}, [\href{http://arxiv.org/abs/1404.4841}{{\tt
  1404.4841}}].

\bibitem{Bhattacharyya:2011se}
N.~Bhattacharyya, A.~Choudhury and A.~Datta, \emph{{Low mass neutralino dark
  matter in mSUGRA and more general models in the light of LHC data}},
  \href{http://dx.doi.org/10.1103/PhysRevD.84.095006}{\emph{Phys. Rev. D} {\bf
  84} (2011) 095006}, [\href{http://arxiv.org/abs/1107.1997}{{\tt 1107.1997}}].

\bibitem{Choudhury:2012tc}
A.~Choudhury and A.~Datta, \emph{{Many faces of low mass neutralino dark matter
  in the unconstrained MSSM, LHC data and new signals}},
  \href{http://dx.doi.org/10.1007/JHEP06(2012)006}{\emph{JHEP} {\bf 06} (2012)
  006}, [\href{http://arxiv.org/abs/1203.4106}{{\tt 1203.4106}}].

\bibitem{Baer:2021aax}
H.~Baer, V.~Barger and H.~Serce, \emph{{Anomalous muon magnetic moment,
  supersymmetry, naturalness, LHC search limits and the landscape}},
  \href{http://dx.doi.org/10.1016/j.physletb.2021.136480}{\emph{Phys. Lett. B}
  {\bf 820} (2021) 136480}, [\href{http://arxiv.org/abs/2104.07597}{{\tt
  2104.07597}}].

\bibitem{Chakraborti:2022vds}
M.~Chakraborti, S.~Iwamoto, J.~S. Kim, R.~Mase\l{}ek and K.~Sakurai,
  \emph{{Supersymmetric explanation of the muon g \textendash{} 2 anomaly with
  and without stable neutralino}},
  \href{http://dx.doi.org/10.1007/JHEP08(2022)124}{\emph{JHEP} {\bf 08} (2022)
  124}, [\href{http://arxiv.org/abs/2202.12928}{{\tt 2202.12928}}].

\bibitem{Athron:2021iuf}
P.~Athron, C.~Balazs, D.~H.~J. Jacob, W.~Kotlarski, D.~Stockinger and
  H.~Stockinger-Kim, \emph{{New physics explanations of a$_{\mu}$ in light of
  the FNAL muon g - 2 measurement}},
  \href{http://dx.doi.org/10.1007/JHEP09(2021)080}{\emph{JHEP} {\bf 09} (2021)
  080}, [\href{http://arxiv.org/abs/2104.03691}{{\tt 2104.03691}}].

\bibitem{Endo:2021zal}
M.~Endo, K.~Hamaguchi, S.~Iwamoto and T.~Kitahara, \emph{{Supersymmetric
  interpretation of the muon g \textendash{} 2 anomaly}},
  \href{http://dx.doi.org/10.1007/JHEP07(2021)075}{\emph{JHEP} {\bf 07} (2021)
  075}, [\href{http://arxiv.org/abs/2104.03217}{{\tt 2104.03217}}].

\bibitem{Chakraborti:2021bmv}
M.~Chakraborti, L.~Roszkowski and S.~Trojanowski, \emph{{GUT-constrained
  supersymmetry and dark matter in light of the new $(g-2)_\mu$
  determination}}, \href{http://dx.doi.org/10.1007/JHEP05(2021)252}{\emph{JHEP}
  {\bf 05} (2021) 252}, [\href{http://arxiv.org/abs/2104.04458}{{\tt
  2104.04458}}].

\bibitem{Choudhury:2017acn}
A.~Choudhury, S.~Rao and L.~Roszkowski, \emph{{Impact of LHC data on muon $g-2$
  solutions in a vectorlike extension of the constrained MSSM}},
  \href{http://dx.doi.org/10.1103/PhysRevD.96.075046}{\emph{Phys. Rev. D} {\bf
  96} (2017) 075046}, [\href{http://arxiv.org/abs/1708.05675}{{\tt
  1708.05675}}].

\bibitem{Choudhury:2017fuu}
A.~Choudhury, L.~Darme, L.~Roszkowski, E.~M. Sessolo and S.~Trojanowski,
  \emph{{Muon g \ensuremath{-} 2 and related phenomenology in constrained
  vector-like extensions of the MSSM}},
  \href{http://dx.doi.org/10.1007/JHEP05(2017)072}{\emph{JHEP} {\bf 05} (2017)
  072}, [\href{http://arxiv.org/abs/1701.08778}{{\tt 1701.08778}}].

\bibitem{Banerjee:2018eaf}
H.~Banerjee, P.~Byakti and S.~Roy, \emph{{Supersymmetric gauged
  U(1)$_{L_{\mu}-L_{\tau}}$ model for neutrinos and the muon $(g-2)$ anomaly}},
  \href{http://dx.doi.org/10.1103/PhysRevD.98.075022}{\emph{Phys. Rev. D} {\bf
  98} (2018) 075022}, [\href{http://arxiv.org/abs/1805.04415}{{\tt
  1805.04415}}].

\bibitem{Banerjee:2020zvi}
H.~Banerjee, B.~Dutta and S.~Roy, \emph{{Supersymmetric gauged $
  \mathrm{U}{(1)}_{L_{\mu }-{L}_{\tau }} $ model for electron and muon $(g −
  2)$ anomaly}}, \href{http://dx.doi.org/10.1007/JHEP03(2021)211}{\emph{JHEP}
  {\bf 03} (2021) 211}, [\href{http://arxiv.org/abs/2011.05083}{{\tt
  2011.05083}}].

\bibitem{Chakraborti:2021dli}
M.~Chakraborti, S.~Heinemeyer and I.~Saha, \emph{{The new
  \textquotedblleft{}MUON G-2\textquotedblright{} result and supersymmetry}},
  \href{http://dx.doi.org/10.1140/epjc/s10052-021-09900-4}{\emph{Eur. Phys. J.
  C} {\bf 81} (2021) 1114}, [\href{http://arxiv.org/abs/2104.03287}{{\tt
  2104.03287}}].

\bibitem{Frank:2021nkq}
M.~Frank, Y.~Hicylmaz, S.~Mondal, O.~Ozdal and C.~S. Un, \emph{{Electron and
  muon magnetic moments and implications for dark matter and model
  characterisation in non-universal U(1) supersymmetric models}},
  \href{http://dx.doi.org/10.1007/JHEP10(2021)063}{\emph{JHEP} {\bf 10} (2021)
  063}, [\href{http://arxiv.org/abs/2107.04116}{{\tt 2107.04116}}].

\bibitem{Ali:2021kxa}
M.~I. Ali, M.~Chakraborti, U.~Chattopadhyay and S.~Mukherjee, \emph{{Muon and
  electron $(g-2)$ anomalies with non-holomorphic interactions in MSSM}},
  \href{http://dx.doi.org/10.1140/epjc/s10052-023-11216-4}{\emph{Eur. Phys. J.
  C} {\bf 83} (2023) 60}, [\href{http://arxiv.org/abs/2112.09867}{{\tt
  2112.09867}}].

\bibitem{Kowalska:2015zja}
K.~Kowalska, L.~Roszkowski, E.~M. Sessolo and A.~J. Williams,
  \emph{{GUT-inspired SUSY and the muon g \ensuremath{-} 2 anomaly: prospects
  for LHC 14 TeV}},
  \href{http://dx.doi.org/10.1007/JHEP06(2015)020}{\emph{JHEP} {\bf 06} (2015)
  020}, [\href{http://arxiv.org/abs/1503.08219}{{\tt 1503.08219}}].

\bibitem{Chakrabortty:2015ika}
J.~Chakrabortty, A.~Choudhury and S.~Mondal, \emph{{Non-universal Gaugino mass
  models under the lamppost of muon (g-2)}},
  \href{http://dx.doi.org/10.1007/JHEP07(2015)038}{\emph{JHEP} {\bf 07} (2015)
  038}, [\href{http://arxiv.org/abs/1503.08703}{{\tt 1503.08703}}].

\bibitem{Choudhury:2016lku}
A.~Choudhury and S.~Mondal, \emph{{Revisiting the Exclusion Limits from Direct
  Chargino-Neutralino Production at the LHC}},
  \href{http://dx.doi.org/10.1103/PhysRevD.94.055024}{\emph{Phys. Rev. D} {\bf
  94} (2016) 055024}, [\href{http://arxiv.org/abs/1603.05502}{{\tt
  1603.05502}}].

\bibitem{Cao:2022htd}
J.~Cao, F.~Li, J.~Lian, Y.~Pan and D.~Zhang, \emph{{Impact of LHC probes of
  SUSY and recent measurement of (g \ensuremath{-} 2)$_{μ}$ on
  \ensuremath{\mathbb{Z}}$_{3}$-NMSSM}},
  \href{http://dx.doi.org/10.1007/s11433-022-1927-9}{\emph{Sci. China Phys.
  Mech. Astron.} {\bf 65} (2022) 291012},
  [\href{http://arxiv.org/abs/2204.04710}{{\tt 2204.04710}}].

\bibitem{Cao:2023juc}
J.~Cao, L.~Meng and Y.~Yue, \emph{{Electron and muon anomalous magnetic moments
  in the Z3-NMSSM}},
  \href{http://dx.doi.org/10.1103/PhysRevD.108.035043}{\emph{Phys. Rev. D} {\bf
  108} (2023) 035043}, [\href{http://arxiv.org/abs/2306.06854}{{\tt
  2306.06854}}].

\bibitem{Samuel1988}
A.~L. Samuel, \emph{Some Studies in Machine Learning Using the Game of
  Checkers. I}, pp.~335--365.
\newblock Springer New York, New York, NY, 1988.
\newblock 10.1007/978-1-4613-8716-9\_14.

\bibitem{CDF:2010eor}
{\scshape CDF} collaboration, T.~Aaltonen et~al., \emph{{Observation of Single
  Top Quark Production and Measurement of |$V_{tb}$| with CDF}},
  \href{http://dx.doi.org/10.1103/PhysRevD.82.112005}{\emph{Phys. Rev. D} {\bf
  82} (2010) 112005}, [\href{http://arxiv.org/abs/1004.1181}{{\tt 1004.1181}}].

\bibitem{D0:2008wma}
{\scshape D0} collaboration, V.~M. Abazov et~al., \emph{{Evidence for
  production of single top quarks}},
  \href{http://dx.doi.org/10.1103/PhysRevD.78.012005}{\emph{Phys. Rev. D} {\bf
  78} (2008) 012005}, [\href{http://arxiv.org/abs/0803.0739}{{\tt 0803.0739}}].

\bibitem{D0:2006ngk}
{\scshape D0} collaboration, V.~M. Abazov et~al., \emph{{Evidence for
  production of single top quarks and first direct measurement of |Vtb|}},
  \href{http://dx.doi.org/10.1103/PhysRevLett.98.181802}{\emph{Phys. Rev.
  Lett.} {\bf 98} (2007) 181802},
  [\href{http://arxiv.org/abs/hep-ex/0612052}{{\tt hep-ex/0612052}}].

\bibitem{CMS:2014afl}
{\scshape CMS} collaboration, V.~Khachatryan et~al., \emph{{Observation of the
  Diphoton Decay of the Higgs Boson and Measurement of Its Properties}},
  \href{http://dx.doi.org/10.1140/epjc/s10052-014-3076-z}{\emph{Eur. Phys. J.
  C} {\bf 74} (2014) 3076}, [\href{http://arxiv.org/abs/1407.0558}{{\tt
  1407.0558}}].

\bibitem{TMVA:2007ngy}
{\scshape TMVA} collaboration, A.~Hocker et~al., \emph{{TMVA - Toolkit for
  Multivariate Data Analysis}},
  \href{http://arxiv.org/abs/physics/0703039}{{\tt physics/0703039}}.

\bibitem{Chen:2016btl}
T.~Chen and C.~Guestrin, \emph{{XGBoost: A Scalable Tree Boosting System}},
  \href{http://arxiv.org/abs/1603.02754}{{\tt 1603.02754}}.

\bibitem{doi:10.1142/S0217751X20020030}
\emph{Preface to special issue on “learning to discover”},
  \href{http://dx.doi.org/10.1142/S0217751X20020030}{\emph{International
  Journal of Modern Physics A} {\bf 35} (2020) 2002003},
  [\href{http://arxiv.org/abs/https://doi.org/10.1142/S0217751X20020030}{{\tt
  https://doi.org/10.1142/S0217751X20020030}}].

\bibitem{dnn}
G.~H. Y.~LeCun, Y.~Bengio, \emph{{Deep Learning}}, .

\bibitem{Bhat:2010zz}
P.~C. Bhat, \emph{{Multivariate Analysis Methods in Particle Physics}},
  \href{http://dx.doi.org/10.1146/annurev.nucl.012809.104427}{\emph{Ann. Rev.
  Nucl. Part. Sci.} {\bf 61} (2011) 281--309}.

\bibitem{Guest:2018yhq}
D.~Guest, K.~Cranmer and D.~Whiteson, \emph{{Deep Learning and its Application
  to LHC Physics}},
  \href{http://dx.doi.org/10.1146/annurev-nucl-101917-021019}{\emph{Ann. Rev.
  Nucl. Part. Sci.} {\bf 68} (2018) 161--181},
  [\href{http://arxiv.org/abs/1806.11484}{{\tt 1806.11484}}].

\bibitem{Bourilkov:2019yoi}
D.~Bourilkov, \emph{{Machine and Deep Learning Applications in Particle
  Physics}}, \href{http://dx.doi.org/10.1142/S0217751X19300199}{\emph{Int. J.
  Mod. Phys. A} {\bf 34} (2020) 1930019},
  [\href{http://arxiv.org/abs/1912.08245}{{\tt 1912.08245}}].

\bibitem{Schwartz:2021ftp}
M.~D. Schwartz, \emph{{Modern Machine Learning and Particle Physics}},
  \href{http://arxiv.org/abs/2103.12226}{{\tt 2103.12226}}.

\bibitem{Carleo:2019ptp}
G.~Carleo, I.~Cirac, K.~Cranmer, L.~Daudet, M.~Schuld, N.~Tishby et~al.,
  \emph{{Machine learning and the physical sciences}},
  \href{http://dx.doi.org/10.1103/RevModPhys.91.045002}{\emph{Rev. Mod. Phys.}
  {\bf 91} (2019) 045002}, [\href{http://arxiv.org/abs/1903.10563}{{\tt
  1903.10563}}].

\bibitem{Shlomi:2020gdn}
J.~Shlomi, P.~Battaglia and J.-R. Vlimant, \emph{{Graph Neural Networks in
  Particle Physics}},  \href{http://arxiv.org/abs/2007.13681}{{\tt
  2007.13681}}.

\bibitem{Abdughani:2019wuv}
M.~Abdughani, J.~Ren, L.~Wu, J.~M. Yang and J.~Zhao, \emph{{Supervised deep
  learning in high energy phenomenology: a mini review}},
  \href{http://dx.doi.org/10.1088/0253-6102/71/8/955}{\emph{Commun. Theor.
  Phys.} {\bf 71} (2019) 955}, [\href{http://arxiv.org/abs/1905.06047}{{\tt
  1905.06047}}].

\bibitem{Hammad:2023sbd}
A.~Hammad, S.~Moretti and M.~Nojiri, \emph{{Multi-scale cross-attention
  transformer encoder for event classification}},
  \href{http://dx.doi.org/10.1007/JHEP03(2024)144}{\emph{JHEP} {\bf 03} (2024)
  144}, [\href{http://arxiv.org/abs/2401.00452}{{\tt 2401.00452}}].

\bibitem{Hammad:2024cae}
A.~Hammad and M.~M. Nojiri, \emph{{Streamlined jet tagging network assisted by
  jet prong structure}},
  \href{http://dx.doi.org/10.1007/JHEP06(2024)176}{\emph{JHEP} {\bf 06} (2024)
  176}, [\href{http://arxiv.org/abs/2404.14677}{{\tt 2404.14677}}].

\bibitem{Arganda:2024eub}
E.~Arganda, M.~Epele, N.~I. Mileo and R.~A. Morales, \emph{{Machine-Learning
  Performance on Higgs-Pair Production associated with Dark Matter at the
  LHC}},  \href{http://arxiv.org/abs/2401.03178}{{\tt 2401.03178}}.

\bibitem{Cornell:2021gut}
A.~S. Cornell, W.~Doorsamy, B.~Fuks, G.~Harmsen and L.~Mason, \emph{{Boosted
  decision trees in the era of new physics: a smuon analysis case study}},
  \href{http://dx.doi.org/10.1007/JHEP04(2022)015}{\emph{JHEP} {\bf 04} (2022)
  015}, [\href{http://arxiv.org/abs/2109.11815}{{\tt 2109.11815}}].

\bibitem{Coadou:2022nsh}
Y.~Coadou, \emph{{Boosted decision trees}},
  \href{http://arxiv.org/abs/2206.09645}{{\tt 2206.09645}}.

\bibitem{inbook}
C.~Bishop, \emph{Pattern Recognition and Machine Learning}, vol.~16,
  pp.~140--155.
\newblock 01, 2006.
\newblock 10.1117/1.2819119.

\bibitem{hastie2009elements}
T.~Hastie, R.~Tibshirani, J.~H. Friedman and J.~H. Friedman, \emph{The elements
  of statistical learning: data mining, inference, and prediction}, vol.~2.
\newblock Springer, 2009.

\bibitem{carbonell1983overview}
J.~G. Carbonell, R.~S. Michalski and T.~M. Mitchell, \emph{An overview of
  machine learning}, {\emph{Machine learning} (1983) 3--23}.

\bibitem{Bardhan:2024zla}
J.~Bardhan, T.~Mandal, S.~Mitra, C.~Neeraj and M.~Patra, \emph{{Unsupervised
  and lightly supervised learning in particle physics}},
  \href{http://arxiv.org/abs/2403.13676}{{\tt 2403.13676}}.

\bibitem{article}
M.~Niazkar, A.~Menapace, B.~Brentan, R.~Piraei, D.~Jimenez, P.~Dhawan et~al.,
  \emph{Applications of xgboost in water resources engineering: A systematic
  literature review (dec 2018 – may 2023)},
  \href{http://dx.doi.org/10.1016/j.envsoft.2024.105971}{\emph{Environmental
  Modelling \& Software} {\bf 174} (02, 2024) 105971}.

\bibitem{bentejac2021comparative}
C.~Bent{\'e}jac, A.~Cs{\"o}rg{\H{o}} and G.~Mart{\'\i}nez-Mu{\~n}oz, \emph{A
  comparative analysis of gradient boosting algorithms}, {\emph{Artificial
  Intelligence Review} {\bf 54} (2021) 1937--1967}.

\bibitem{weiss2003learning}
G.~M. Weiss and F.~Provost, \emph{Learning when training data are costly: The
  effect of class distribution on tree induction}, {\emph{Journal of artificial
  intelligence research} {\bf 19} (2003) 315--354}.

\bibitem{Chawla2005}
N.~V. Chawla, \emph{Data Mining for Imbalanced Datasets: An Overview},
  pp.~853--867.
\newblock Springer US, Boston, MA, 2005.
\newblock 10.1007/0-387-25465-X\_40.

\bibitem{sulaiman}
M.~Hossin and S.~M.N, \emph{A review on evaluation metrics for data
  classification evaluations},
  \href{http://dx.doi.org/10.5121/ijdkp.2015.5201}{\emph{International Journal
  of Data Mining \& Knowledge Management Process} {\bf 5} (03, 2015) 01--11}.

\bibitem{satio}
T.~Saito and M.~Rehmsmeier, \emph{The precision-recall plot is more informative
  than the roc plot when evaluating binary classifiers on imbalanced datasets},
  \href{http://dx.doi.org/10.1371/journal.pone.0118432}{\emph{PLoS One} {\bf
  10} (2015) }.

\bibitem{LUQUE2019216}
A.~Luque, A.~Carrasco, A.~Martín and A.~{de las Heras}, \emph{The impact of
  class imbalance in classification performance metrics based on the binary
  confusion matrix},
  \href{http://dx.doi.org/https://doi.org/10.1016/j.patcog.2019.02.023}{\emph{Pattern
  Recognition} {\bf 91} (2019) 216--231}.

\bibitem{10.1007/978-3-642-04962-0_53}
Q.~Gu, L.~Zhu and Z.~Cai, \emph{Evaluation measures of the classification
  performance of imbalanced data sets},  in \emph{Computational Intelligence
  and Intelligent Systems} (Z.~Cai, Z.~Li, Z.~Kang and Y.~Liu, eds.), (Berlin,
  Heidelberg), pp.~461--471, Springer Berlin Heidelberg, 2009.

\bibitem{Cowan2012DiscoverySF}
G.~Cowan, \emph{Discovery sensitivity for a counting experiment with background
  uncertainty},  2012.

\bibitem{Cowan:2010js}
G.~Cowan, K.~Cranmer, E.~Gross and O.~Vitells, \emph{{Asymptotic formulae for
  likelihood-based tests of new physics}},
  \href{http://dx.doi.org/10.1140/epjc/s10052-011-1554-0}{\emph{Eur. Phys. J.
  C} {\bf 71} (2011) 1554}, [\href{http://arxiv.org/abs/1007.1727}{{\tt
  1007.1727}}].

\bibitem{hepmllivingreview}
{HEP ML Community}, ``{A Living Review of Machine Learning for Particle
  Physics}.''

\bibitem{ROE2005577}
B.~P. Roe, H.-J. Yang, J.~Zhu, Y.~Liu, I.~Stancu and G.~McGregor, \emph{Boosted
  decision trees as an alternative to artificial neural networks for particle
  identification},
  \href{http://dx.doi.org/https://doi.org/10.1016/j.nima.2004.12.018}{\emph{Nuclear
  Instruments and Methods in Physics Research Section A: Accelerators,
  Spectrometers, Detectors and Associated Equipment} {\bf 543} (2005)
  577--584}.

\bibitem{YANG2005370}
H.-J. Yang, B.~P. Roe and J.~Zhu, \emph{Studies of boosted decision trees for
  miniboone particle identification},
  \href{http://dx.doi.org/https://doi.org/10.1016/j.nima.2005.09.022}{\emph{Nuclear
  Instruments and Methods in Physics Research Section A: Accelerators,
  Spectrometers, Detectors and Associated Equipment} {\bf 555} (2005)
  370--385}.

\bibitem{FREUND1997119}
Y.~Freund and R.~E. Schapire, \emph{A decision-theoretic generalization of
  on-line learning and an application to boosting},
  \href{http://dx.doi.org/https://doi.org/10.1006/jcss.1997.1504}{\emph{Journal
  of Computer and System Sciences} {\bf 55} (1997) 119--139}.

\bibitem{CMS:2018wav}
{\scshape CMS} collaboration, D.~Acosta et~al., \emph{{Boosted Decision Trees
  in the Level-1 Muon Endcap Trigger at CMS}},
  \href{http://dx.doi.org/10.1088/1742-6596/1085/4/042042}{\emph{J. Phys. Conf.
  Ser.} {\bf 1085} (2018) 042042}.

\bibitem{Zabi:2020gjd}
{\scshape CMS} collaboration, A.~Zabi, J.~W. Berryhill, E.~Perez and A.~D.
  Tapper, \emph{{The Phase-2 Upgrade of the CMS Level-1 Trigger}}, .

\bibitem{Gligorov:2012qt}
V.~V. Gligorov and M.~Williams, \emph{{Efficient, reliable and fast high-level
  triggering using a bonsai boosted decision tree}},
  \href{http://dx.doi.org/10.1088/1748-0221/8/02/P02013}{\emph{JINST} {\bf 8}
  (2013) P02013}, [\href{http://arxiv.org/abs/1210.6861}{{\tt 1210.6861}}].

\bibitem{Likhomanenko:2015aba}
T.~Likhomanenko, P.~Ilten, E.~Khairullin, A.~Rogozhnikov, A.~Ustyuzhanin and
  M.~Williams, \emph{{LHCb Topological Trigger Reoptimization}},
  \href{http://dx.doi.org/10.1088/1742-6596/664/8/082025}{\emph{J. Phys. Conf.
  Ser.} {\bf 664} (2015) 082025}, [\href{http://arxiv.org/abs/1510.00572}{{\tt
  1510.00572}}].

\bibitem{Bhattacherjee:2023evs}
B.~Bhattacherjee, P.~Konar, V.~S. Ngairangbam and P.~Solanki, \emph{{LLPNet:
  Graph Autoencoder for Triggering Light Long-Lived Particles at HL-LHC}},
  \href{http://arxiv.org/abs/2308.13611}{{\tt 2308.13611}}.

\bibitem{Bendavid:2017zhk}
J.~Bendavid, \emph{{Efficient Monte Carlo Integration Using Boosted Decision
  Trees and Generative Deep Neural Networks}},
  \href{http://arxiv.org/abs/1707.00028}{{\tt 1707.00028}}.

\bibitem{Feroz:2007kg}
F.~Feroz and M.~P. Hobson, \emph{{Multimodal nested sampling: an efficient and
  robust alternative to MCMC methods for astronomical data analysis}},
  \href{http://dx.doi.org/10.1111/j.1365-2966.2007.12353.x}{\emph{Mon. Not.
  Roy. Astron. Soc.} {\bf 384} (2008) 449},
  [\href{http://arxiv.org/abs/0704.3704}{{\tt 0704.3704}}].

\bibitem{Feroz:2008xx}
F.~Feroz, M.~P. Hobson and M.~Bridges, \emph{{MultiNest: an efficient and
  robust Bayesian inference tool for cosmology and particle physics}},
  \href{http://dx.doi.org/10.1111/j.1365-2966.2009.14548.x}{\emph{Mon. Not.
  Roy. Astron. Soc.} {\bf 398} (2009) 1601--1614},
  [\href{http://arxiv.org/abs/0809.3437}{{\tt 0809.3437}}].

\bibitem{Trotta:2008bp}
R.~Trotta, F.~Feroz, M.~P. Hobson, L.~Roszkowski and R.~Ruiz~de Austri,
  \emph{{The Impact of priors and observables on parameter inferences in the
  Constrained MSSM}},
  \href{http://dx.doi.org/10.1088/1126-6708/2008/12/024}{\emph{JHEP} {\bf 12}
  (2008) 024}, [\href{http://arxiv.org/abs/0809.3792}{{\tt 0809.3792}}].

\bibitem{10.1111/j.1365-2966.2009.14548.x}
F.~Feroz, M.~P. Hobson and M.~Bridges, \emph{{MultiNest: an efficient and
  robust Bayesian inference tool for cosmology and particle physics}},
  \href{http://dx.doi.org/10.1111/j.1365-2966.2009.14548.x}{\emph{Monthly
  Notices of the Royal Astronomical Society} {\bf 398} (09, 2009) 1601--1614},
  [\href{http://arxiv.org/abs/https://academic.oup.com/mnras/article-pdf/398/4/1601/3039078/mnras0398-1601.pdf}{{\tt
  https://academic.oup.com/mnras/article-pdf/398/4/1601/3039078/mnras0398-1601.pdf}}].

\bibitem{GAMBIT:2017snp}
{\scshape GAMBIT} collaboration, P.~Athron et~al., \emph{{Global fits of
  GUT-scale SUSY models with GAMBIT}},
  \href{http://dx.doi.org/10.1140/epjc/s10052-017-5167-0}{\emph{Eur. Phys. J.
  C} {\bf 77} (2017) 824}, [\href{http://arxiv.org/abs/1705.07935}{{\tt
  1705.07935}}].

\bibitem{GAMBIT:2017zdo}
{\scshape GAMBIT} collaboration, P.~Athron et~al., \emph{{A global fit of the
  MSSM with GAMBIT}},
  \href{http://dx.doi.org/10.1140/epjc/s10052-017-5196-8}{\emph{Eur. Phys. J.
  C} {\bf 77} (2017) 879}, [\href{http://arxiv.org/abs/1705.07917}{{\tt
  1705.07917}}].

\bibitem{Choudhury:2023lbp}
A.~Choudhury, S.~Mitra, A.~Mondal and S.~Mondal, \emph{{Bilinear R-parity
  violating supersymmetry under the light of neutrino oscillation, Higgs and
  flavor data}}, \href{http://dx.doi.org/10.1007/JHEP02(2024)004}{\emph{JHEP}
  {\bf 02} (2024) 004}, [\href{http://arxiv.org/abs/2305.15211}{{\tt
  2305.15211}}].

\bibitem{Caron:2016hib}
S.~Caron, J.~S. Kim, K.~Rolbiecki, R.~Ruiz~de Austri and B.~Stienen, \emph{{The
  BSM-AI project: SUSY-AI\textendash{}generalizing LHC limits on supersymmetry
  with machine learning}},
  \href{http://dx.doi.org/10.1140/epjc/s10052-017-4814-9}{\emph{Eur. Phys. J.
  C} {\bf 77} (2017) 257}, [\href{http://arxiv.org/abs/1605.02797}{{\tt
  1605.02797}}].

\bibitem{Bridges:2010de}
M.~Bridges, K.~Cranmer, F.~Feroz, M.~Hobson, R.~Ruiz~de Austri and R.~Trotta,
  \emph{{A Coverage Study of the CMSSM Based on ATLAS Sensitivity Using Fast
  Neural Networks Techniques}},
  \href{http://dx.doi.org/10.1007/JHEP03(2011)012}{\emph{JHEP} {\bf 03} (2011)
  012}, [\href{http://arxiv.org/abs/1011.4306}{{\tt 1011.4306}}].

\bibitem{Buckley:2011kc}
A.~Buckley, A.~Shilton and M.~J. White, \emph{{Fast supersymmetry phenomenology
  at the Large Hadron Collider using machine learning techniques}},
  \href{http://dx.doi.org/10.1016/j.cpc.2011.12.026}{\emph{Comput. Phys.
  Commun.} {\bf 183} (2012) 960--970},
  [\href{http://arxiv.org/abs/1106.4613}{{\tt 1106.4613}}].

\bibitem{Kronheim:2020vct}
B.~S. Kronheim, M.~P. Kuchera, H.~B. Prosper and A.~Karbo, \emph{{Bayesian
  Neural Networks for Fast SUSY Predictions}},
  \href{http://dx.doi.org/10.1016/j.physletb.2020.136041}{\emph{Phys. Lett. B}
  {\bf 813} (2021) 136041}, [\href{http://arxiv.org/abs/2007.04506}{{\tt
  2007.04506}}].

\bibitem{Mullin:2019mmh}
A.~Mullin, S.~Nicholls, H.~Pacey, M.~Parker, M.~White and S.~Williams,
  \emph{{Does SUSY have friends? A new approach for LHC event analysis}},
  \href{http://dx.doi.org/10.1007/JHEP02(2021)160}{\emph{JHEP} {\bf 02} (2021)
  160}, [\href{http://arxiv.org/abs/1912.10625}{{\tt 1912.10625}}].

\bibitem{DiSipio:2019imz}
R.~Di~Sipio, M.~Faucci~Giannelli, S.~Ketabchi~Haghighat and S.~Palazzo,
  \emph{{DijetGAN: A Generative-Adversarial Network Approach for the Simulation
  of QCD Dijet Events at the LHC}},
  \href{http://dx.doi.org/10.1007/JHEP08(2019)110}{\emph{JHEP} {\bf 08} (2019)
  110}, [\href{http://arxiv.org/abs/1903.02433}{{\tt 1903.02433}}].

\bibitem{Butter:2019cae}
A.~Butter, T.~Plehn and R.~Winterhalder, \emph{{How to GAN LHC Events}},
  \href{http://dx.doi.org/10.21468/SciPostPhys.7.6.075}{\emph{SciPost Phys.}
  {\bf 7} (2019) 075}, [\href{http://arxiv.org/abs/1907.03764}{{\tt
  1907.03764}}].

\bibitem{Lin:2019htn}
J.~Lin, W.~Bhimji and B.~Nachman, \emph{{Machine Learning Templates for QCD
  Factorization in the Search for Physics Beyond the Standard Model}},
  \href{http://dx.doi.org/10.1007/JHEP05(2019)181}{\emph{JHEP} {\bf 05} (2019)
  181}, [\href{http://arxiv.org/abs/1903.02556}{{\tt 1903.02556}}].

\bibitem{Musella:2018rdi}
P.~Musella and F.~Pandolfi, \emph{{Fast and Accurate Simulation of Particle
  Detectors Using Generative Adversarial Networks}},
  \href{http://dx.doi.org/10.1007/s41781-018-0015-y}{\emph{Comput. Softw. Big
  Sci.} {\bf 2} (2018) 8}, [\href{http://arxiv.org/abs/1805.00850}{{\tt
  1805.00850}}].

\bibitem{Ren:2017ymm}
J.~Ren, L.~Wu, J.~M. Yang and J.~Zhao, \emph{{Exploring supersymmetry with
  machine learning}},
  \href{http://dx.doi.org/10.1016/j.nuclphysb.2019.114613}{\emph{Nucl. Phys. B}
  {\bf 943} (2019) 114613}, [\href{http://arxiv.org/abs/1708.06615}{{\tt
  1708.06615}}].

\bibitem{Caron:2019xkx}
S.~Caron, T.~Heskes, S.~Otten and B.~Stienen, \emph{{Constraining the
  Parameters of High-Dimensional Models with Active Learning}},
  \href{http://dx.doi.org/10.1140/epjc/s10052-019-7437-5}{\emph{Eur. Phys. J.
  C} {\bf 79} (2019) 944}, [\href{http://arxiv.org/abs/1905.08628}{{\tt
  1905.08628}}].

\bibitem{Baruah:2024gwy}
R.~Baruah, S.~Mondal, S.~K. Patra and S.~Roy, \emph{{Probing intractable
  beyond-standard-model parameter spaces armed with Machine Learning}},
  \href{http://arxiv.org/abs/2404.02698}{{\tt 2404.02698}}.

\bibitem{ATLAS:2018wis}
{\scshape ATLAS} collaboration, M.~Aaboud et~al., \emph{{Performance of
  top-quark and $W$-boson tagging with ATLAS in Run 2 of the LHC}},
  \href{http://dx.doi.org/10.1140/epjc/s10052-019-6847-8}{\emph{Eur. Phys. J.
  C} {\bf 79} (2019) 375}, [\href{http://arxiv.org/abs/1808.07858}{{\tt
  1808.07858}}].

\bibitem{Cogan:2014oua}
J.~Cogan, M.~Kagan, E.~Strauss and A.~Schwarztman, \emph{{Jet-Images: Computer
  Vision Inspired Techniques for Jet Tagging}},
  \href{http://dx.doi.org/10.1007/JHEP02(2015)118}{\emph{JHEP} {\bf 02} (2015)
  118}, [\href{http://arxiv.org/abs/1407.5675}{{\tt 1407.5675}}].

\bibitem{deOliveira:2015xxd}
L.~de~Oliveira, M.~Kagan, L.~Mackey, B.~Nachman and A.~Schwartzman,
  \emph{{Jet-images \textemdash{} deep learning edition}},
  \href{http://dx.doi.org/10.1007/JHEP07(2016)069}{\emph{JHEP} {\bf 07} (2016)
  069}, [\href{http://arxiv.org/abs/1511.05190}{{\tt 1511.05190}}].

\bibitem{Baldi:2016fql}
P.~Baldi, K.~Bauer, C.~Eng, P.~Sadowski and D.~Whiteson, \emph{{Jet
  Substructure Classification in High-Energy Physics with Deep Neural
  Networks}}, \href{http://dx.doi.org/10.1103/PhysRevD.93.094034}{\emph{Phys.
  Rev. D} {\bf 93} (2016) 094034}, [\href{http://arxiv.org/abs/1603.09349}{{\tt
  1603.09349}}].

\bibitem{Komiske:2016rsd}
P.~T. Komiske, E.~M. Metodiev and M.~D. Schwartz, \emph{{Deep learning in
  color: towards automated quark/gluon jet discrimination}},
  \href{http://dx.doi.org/10.1007/JHEP01(2017)110}{\emph{JHEP} {\bf 01} (2017)
  110}, [\href{http://arxiv.org/abs/1612.01551}{{\tt 1612.01551}}].

\bibitem{Chakraborty:2019imr}
A.~Chakraborty, S.~H. Lim and M.~M. Nojiri, \emph{{Interpretable deep learning
  for two-prong jet classification with jet spectra}},
  \href{http://dx.doi.org/10.1007/JHEP07(2019)135}{\emph{JHEP} {\bf 07} (2019)
  135}, [\href{http://arxiv.org/abs/1904.02092}{{\tt 1904.02092}}].

\bibitem{ATLAS:2017dfg}
{\scshape ATLAS} collaboration, \emph{{Quark versus Gluon Jet Tagging Using Jet
  Images with the ATLAS Detector}}, .

\bibitem{CMS:2017wtu}
{\scshape CMS} collaboration, A.~M. Sirunyan et~al., \emph{{Identification of
  heavy-flavour jets with the CMS detector in pp collisions at 13 TeV}},
  \href{http://dx.doi.org/10.1088/1748-0221/13/05/P05011}{\emph{JINST} {\bf 13}
  (2018) P05011}, [\href{http://arxiv.org/abs/1712.07158}{{\tt 1712.07158}}].

\bibitem{Bols:2020bkb}
E.~Bols, J.~Kieseler, M.~Verzetti, M.~Stoye and A.~Stakia, \emph{{Jet Flavour
  Classification Using DeepJet}},
  \href{http://dx.doi.org/10.1088/1748-0221/15/12/P12012}{\emph{JINST} {\bf 15}
  (2020) P12012}, [\href{http://arxiv.org/abs/2008.10519}{{\tt 2008.10519}}].

\bibitem{2020HeavyFI}
\emph{Heavy flavor identification at cms with deep neural networks},  2020.

\bibitem{deepflv}
``Cms phase 1 heavy flavour identification performance anddevelopments, cms
  detector performance note, cms-dp-2017-013, 2017.''

\bibitem{ATLAS:2019bwq}
{\scshape ATLAS} collaboration, G.~Aad et~al., \emph{{ATLAS b-jet
  identification performance and efficiency measurement with $t{\bar{t}}$
  events in pp collisions at $\sqrt{s}=13$ TeV}},
  \href{http://dx.doi.org/10.1140/epjc/s10052-019-7450-8}{\emph{Eur. Phys. J.
  C} {\bf 79} (2019) 970}, [\href{http://arxiv.org/abs/1907.05120}{{\tt
  1907.05120}}].

\bibitem{Krizhevsky2012ImageNetCW}
A.~Krizhevsky, I.~Sutskever and G.~E. Hinton, \emph{Imagenet classification
  with deep convolutional neural networks}, {\emph{Communications of the ACM}
  {\bf 60} (2012) 84 -- 90}.

\bibitem{ATLAS:2020hpj}
{\scshape ATLAS} collaboration, G.~Aad et~al., \emph{{Evidence for
  $t\bar{t}t\bar{t}$ production in the multilepton final state in
  proton\textendash{}proton collisions at $\sqrt{s}=13$ $\text {TeV}$ with the
  ATLAS detector}},
  \href{http://dx.doi.org/10.1140/epjc/s10052-020-08509-3}{\emph{Eur. Phys. J.
  C} {\bf 80} (2020) 1085}, [\href{http://arxiv.org/abs/2007.14858}{{\tt
  2007.14858}}].

\bibitem{ATLAS:2012yve}
{\scshape ATLAS} collaboration, G.~Aad et~al., \emph{{Observation of a new
  particle in the search for the Standard Model Higgs boson with the ATLAS
  detector at the LHC}},
  \href{http://dx.doi.org/10.1016/j.physletb.2012.08.020}{\emph{Phys. Lett. B}
  {\bf 716} (2012) 1--29}, [\href{http://arxiv.org/abs/1207.7214}{{\tt
  1207.7214}}].

\bibitem{CMS:2012qbp}
{\scshape CMS} collaboration, S.~Chatrchyan et~al., \emph{{Observation of a New
  Boson at a Mass of 125 GeV with the CMS Experiment at the LHC}},
  \href{http://dx.doi.org/10.1016/j.physletb.2012.08.021}{\emph{Phys. Lett. B}
  {\bf 716} (2012) 30--61}, [\href{http://arxiv.org/abs/1207.7235}{{\tt
  1207.7235}}].

\bibitem{ATLAS:2017fak}
{\scshape ATLAS} collaboration, M.~Aaboud et~al., \emph{{Search for the
  standard model Higgs boson produced in association with top quarks and
  decaying into a $b\bar{b}$ pair in $pp$ collisions at $\sqrt{s}$ = 13 TeV
  with the ATLAS detector}},
  \href{http://dx.doi.org/10.1103/PhysRevD.97.072016}{\emph{Phys. Rev. D} {\bf
  97} (2018) 072016}, [\href{http://arxiv.org/abs/1712.08895}{{\tt
  1712.08895}}].

\bibitem{Barbier:2004ez}
R.~Barbier et~al., \emph{{R-parity violating supersymmetry}},
  \href{http://dx.doi.org/10.1016/j.physrep.2005.08.006}{\emph{Phys. Rept.}
  {\bf 420} (2005) 1--202}, [\href{http://arxiv.org/abs/hep-ph/0406039}{{\tt
  hep-ph/0406039}}].

\bibitem{Choudhury:2024ggy}
A.~Choudhury, A.~Mondal and S.~Mondal, \emph{{Status of R-parity violating
  SUSY}},  \href{http://arxiv.org/abs/2402.04040}{{\tt 2402.04040}}.

\bibitem{ATLAS:2020syg}
{\scshape ATLAS} collaboration, G.~Aad et~al., \emph{{Search for squarks and
  gluinos in final states with jets and missing transverse momentum using 139
  fb$^{-1}$ of $\sqrt{s}$ =13 TeV $pp$ collision data with the ATLAS
  detector}}, \href{http://dx.doi.org/10.1007/JHEP02(2021)143}{\emph{JHEP} {\bf
  02} (2021) 143}, [\href{http://arxiv.org/abs/2010.14293}{{\tt 2010.14293}}].

\bibitem{ATLAS:2024fub}
{\scshape ATLAS} collaboration, G.~Aad et~al., \emph{{Search for electroweak
  production of supersymmetric particles in final states with two
  $\tau$-leptons in $\sqrt{s}$ = 13 TeV $pp$ collisions with the ATLAS
  detector}},  \href{http://arxiv.org/abs/2402.00603}{{\tt 2402.00603}}.

\bibitem{CMS:2023ktc}
{\scshape CMS} collaboration, A.~Tumasyan et~al., \emph{{Search for top squarks
  in the four-body decay mode with single lepton final states in proton-proton
  collisions at $ \sqrt{s} $ = 13 TeV}},
  \href{http://dx.doi.org/10.1007/JHEP06(2023)060}{\emph{JHEP} {\bf 06} (2023)
  060}, [\href{http://arxiv.org/abs/2301.08096}{{\tt 2301.08096}}].

\bibitem{Balazs:2004bu}
C.~Balazs, M.~Carena and C.~E.~M. Wagner, \emph{{Dark matter, light stops and
  electroweak baryogenesis}},
  \href{http://dx.doi.org/10.1103/PhysRevD.70.015007}{\emph{Phys. Rev. D} {\bf
  70} (2004) 015007}, [\href{http://arxiv.org/abs/hep-ph/0403224}{{\tt
  hep-ph/0403224}}].

\bibitem{CMS:2013phu}
{\scshape CMS} collaboration, S.~Chatrchyan et~al., \emph{{Search for
  Top-Squark Pair Production in the Single-Lepton Final State in pp Collisions
  at $\sqrt{s}$ = 8 TeV}},
  \href{http://dx.doi.org/10.1140/epjc/s10052-013-2677-2}{\emph{Eur. Phys. J.
  C} {\bf 73} (2013) 2677}, [\href{http://arxiv.org/abs/1308.1586}{{\tt
  1308.1586}}].

\bibitem{Jorge:2021vpo}
F.~Jorge, R.~Ronald, S.~Jesus, M.~Juan and A.~Carlos, \emph{{Top squark signal
  significance enhancement by different machine learning algorithms}},
  \href{http://dx.doi.org/10.1142/S0217751X22501974}{\emph{Int. J. Mod. Phys.
  A} {\bf 37} (2022) 2250197}, [\href{http://arxiv.org/abs/2106.06813}{{\tt
  2106.06813}}].

\bibitem{ATLAS:2020xzu}
{\scshape ATLAS} collaboration, G.~Aad et~al., \emph{{Search for new phenomena
  with top quark pairs in final states with one lepton, jets, and missing
  transverse momentum in $pp$ collisions at $ \sqrt{s} $ = 13 TeV with the
  ATLAS detector}},
  \href{http://dx.doi.org/10.1007/JHEP04(2021)174}{\emph{JHEP} {\bf 04} (2021)
  174}, [\href{http://arxiv.org/abs/2012.03799}{{\tt 2012.03799}}].

\bibitem{ATLAS:2022hbt}
{\scshape ATLAS} collaboration, G.~Aad et~al., \emph{{Search for direct pair
  production of sleptons and charginos decaying to two leptons and neutralinos
  with mass splittings near the W-boson mass in $ \sqrt{s} $ = 13 TeV pp
  collisions with the ATLAS detector}},
  \href{http://dx.doi.org/10.1007/JHEP06(2023)031}{\emph{JHEP} {\bf 06} (2023)
  031}, [\href{http://arxiv.org/abs/2209.13935}{{\tt 2209.13935}}].

\bibitem{NIPS2017_6449f44a}
G.~Ke, Q.~Meng, T.~Finley, T.~Wang, W.~Chen, W.~Ma et~al., \emph{Lightgbm: A
  highly efficient gradient boosting decision tree},  in \emph{Advances in
  Neural Information Processing Systems} (I.~Guyon, U.~V. Luxburg, S.~Bengio,
  H.~Wallach, R.~Fergus, S.~Vishwanathan et~al., eds.), vol.~30, Curran
  Associates, Inc., 2017.

\bibitem{Shapley+1953+307+318}
L.~S. Shapley, \emph{17. A Value for n-Person Games}, pp.~307--318.
\newblock Princeton University Press, Princeton, 1953.
\newblock doi:10.1515/9781400881970-018.

\bibitem{lundberg2018consistent}
S.~M. Lundberg, G.~G. Erion and S.-I. Lee, \emph{Consistent individualized
  feature attribution for tree ensembles}, {\emph{arXiv preprint
  arXiv:1802.03888} (2018) }.

\bibitem{Alvestad:2021sje}
D.~Alvestad, N.~Fomin, J.~Kersten, S.~Maeland and I.~Str\"umke, \emph{{Beyond
  cuts in small signal scenarios: Enhanced sneutrino detectability using
  machine learning}},
  \href{http://dx.doi.org/10.1140/epjc/s10052-023-11532-9}{\emph{Eur. Phys. J.
  C} {\bf 83} (2023) 379}, [\href{http://arxiv.org/abs/2108.03125}{{\tt
  2108.03125}}].

\bibitem{Barman:2024xlc}
R.~K. Barman, G.~B\'elanger, B.~Bhattacherjee, R.~Godbole and R.~Sengupta,
  \emph{{Current status of the light neutralino thermal dark matter in the
  phenomenological MSSM}},  \href{http://arxiv.org/abs/2402.07991}{{\tt
  2402.07991}}.

\bibitem{ATLAS:2018ntn}
{\scshape ATLAS} collaboration, M.~Aaboud et~al., \emph{{Search for charged
  Higgs bosons decaying into top and bottom quarks at $\sqrt{s}$ = 13 TeV with
  the ATLAS detector}},
  \href{http://dx.doi.org/10.1007/JHEP11(2018)085}{\emph{JHEP} {\bf 11} (2018)
  085}, [\href{http://arxiv.org/abs/1808.03599}{{\tt 1808.03599}}].

\bibitem{ATLAS:2018gfm}
{\scshape ATLAS} collaboration, M.~Aaboud et~al., \emph{{Search for charged
  Higgs bosons decaying via $H^{\pm} \to \tau^{\pm}\nu_{\tau}$ in the
  $\tau$+jets and $\tau$+lepton final states with 36 fb$^{-1}$ of $pp$
  collision data recorded at $\sqrt{s} = 13$ TeV with the ATLAS experiment}},
  \href{http://dx.doi.org/10.1007/JHEP09(2018)139}{\emph{JHEP} {\bf 09} (2018)
  139}, [\href{http://arxiv.org/abs/1807.07915}{{\tt 1807.07915}}].

\bibitem{Djouadi:2013uqa}
A.~Djouadi, L.~Maiani, G.~Moreau, A.~Polosa, J.~Quevillon and V.~Riquer,
  \emph{{The post-Higgs MSSM scenario: Habemus MSSM?}},
  \href{http://dx.doi.org/10.1140/epjc/s10052-013-2650-0}{\emph{Eur. Phys. J.
  C} {\bf 73} (2013) 2650}, [\href{http://arxiv.org/abs/1307.5205}{{\tt
  1307.5205}}].

\bibitem{Bagnaschi:2017tru}
E.~Bagnaschi et~al., \emph{{Likelihood Analysis of the pMSSM11 in Light of LHC
  13-TeV Data}},
  \href{http://dx.doi.org/10.1140/epjc/s10052-018-5697-0}{\emph{Eur. Phys. J.
  C} {\bf 78} (2018) 256}, [\href{http://arxiv.org/abs/1710.11091}{{\tt
  1710.11091}}].

\bibitem{Bhattacherjee:2015sga}
B.~Bhattacherjee, A.~Chakraborty and A.~Choudhury, \emph{{Status of the MSSM
  Higgs sector using global analysis and direct search bounds, and future
  prospects at the High Luminosity LHC}},
  \href{http://dx.doi.org/10.1103/PhysRevD.92.093007}{\emph{Phys. Rev. D} {\bf
  92} (2015) 093007}, [\href{http://arxiv.org/abs/1504.04308}{{\tt
  1504.04308}}].

\bibitem{Barman:2016jov}
R.~K. Barman, B.~Bhattacherjee, A.~Choudhury, D.~Chowdhury, J.~Lahiri and
  S.~Ray, \emph{{Current status of MSSM Higgs sector with LHC 13 TeV data}},
  \href{http://dx.doi.org/10.1140/epjp/i2019-12566-5}{\emph{Eur. Phys. J. Plus}
  {\bf 134} (2019) 150}, [\href{http://arxiv.org/abs/1608.02573}{{\tt
  1608.02573}}].

\bibitem{CMS:2020imj}
{\scshape CMS} collaboration, A.~M. Sirunyan et~al., \emph{{Search for charged
  Higgs bosons decaying into a top and a bottom quark in the all-jet final
  state of pp collisions at $ \sqrt{s} $ = 13 TeV}},
  \href{http://dx.doi.org/10.1007/JHEP07(2020)126}{\emph{JHEP} {\bf 07} (2020)
  126}, [\href{http://arxiv.org/abs/2001.07763}{{\tt 2001.07763}}].

\bibitem{CMS:2019bfg}
{\scshape CMS} collaboration, A.~M. Sirunyan et~al., \emph{{Search for charged
  Higgs bosons in the H$^{\pm}$ $\to$ $\tau^{\pm}\nu_\tau$ decay channel in
  proton-proton collisions at $\sqrt{s} =$ 13 TeV}},
  \href{http://dx.doi.org/10.1007/JHEP07(2019)142}{\emph{JHEP} {\bf 07} (2019)
  142}, [\href{http://arxiv.org/abs/1903.04560}{{\tt 1903.04560}}].

\bibitem{Keck:2017gsv}
T.~Keck, \emph{{FastBDT: A Speed-Optimized Multivariate Classification
  Algorithm for the Belle II Experiment}},
  \href{http://dx.doi.org/10.1007/s41781-017-0002-8}{\emph{Comput. Softw. Big
  Sci.} {\bf 1} (2017) 2}.

\bibitem{Baer:2021qxa}
H.~Baer, C.~Kao, V.~Barger, R.~Jain, D.~Sengupta and X.~Tata, \emph{{Detecting
  heavy Higgs bosons from natural SUSY at a 100~TeV hadron collider}},
  \href{http://dx.doi.org/10.1103/PhysRevD.105.095039}{\emph{Phys. Rev. D} {\bf
  105} (2022) 095039}, [\href{http://arxiv.org/abs/2112.02232}{{\tt
  2112.02232}}].

\bibitem{Grossman:2003gq}
Y.~Grossman and S.~Rakshit, \emph{{Neutrino masses in R-parity violating
  supersymmetric models}},
  \href{http://dx.doi.org/10.1103/PhysRevD.69.093002}{\emph{Phys. Rev. D} {\bf
  69} (2004) 093002}, [\href{http://arxiv.org/abs/hep-ph/0311310}{{\tt
  hep-ph/0311310}}].

\bibitem{Davidson:2000uc}
S.~Davidson and M.~Losada, \emph{{Neutrino masses in the R(p) violating MSSM}},
  \href{http://dx.doi.org/10.1088/1126-6708/2000/05/021}{\emph{JHEP} {\bf 05}
  (2000) 021}, [\href{http://arxiv.org/abs/hep-ph/0005080}{{\tt
  hep-ph/0005080}}].

\bibitem{Roy:1996bua}
S.~Roy and B.~Mukhopadhyaya, \emph{{Some implications of a supersymmetric model
  with R-parity breaking bilinear interactions}},
  \href{http://dx.doi.org/10.1103/PhysRevD.55.7020}{\emph{Phys. Rev. D} {\bf
  55} (1997) 7020--7029}, [\href{http://arxiv.org/abs/hep-ph/9612447}{{\tt
  hep-ph/9612447}}].

\bibitem{Allanach:2007qc}
B.~C. Allanach and C.~H. Kom, \emph{{Lepton number violating mSUGRA and
  neutrino masses}},
  \href{http://dx.doi.org/10.1088/1126-6708/2008/04/081}{\emph{JHEP} {\bf 04}
  (2008) 081}, [\href{http://arxiv.org/abs/0712.0852}{{\tt 0712.0852}}].

\bibitem{Diaz:2014jta}
M.~A. D\'\i{}az, M.~Rivera and N.~Rojas, \emph{{On Neutrino Masses in the MSSM
  with BRpV}},
  \href{http://dx.doi.org/10.1016/j.nuclphysb.2014.08.012}{\emph{Nucl. Phys. B}
  {\bf 887} (2014) 338--357}, [\href{http://arxiv.org/abs/1401.7357}{{\tt
  1401.7357}}].

\bibitem{Chakraborty:2015bsk}
A.~Chakraborty and S.~Chakraborty, \emph{{Probing $(g-2)_{\mu}$ at the LHC in
  the paradigm of $R$-parity violating MSSM}},
  \href{http://dx.doi.org/10.1103/PhysRevD.93.075035}{\emph{Phys. Rev. D} {\bf
  93} (2016) 075035}, [\href{http://arxiv.org/abs/1511.08874}{{\tt
  1511.08874}}].

\bibitem{Altmannshofer:2020axr}
W.~Altmannshofer, P.~S.~B. Dev, A.~Soni and Y.~Sui, \emph{{Addressing
  R$_{D^{(*)}}$, R$_{K^{(*)}}$, muon $g-2$ and ANITA anomalies in a minimal
  $R$-parity violating supersymmetric framework}},
  \href{http://dx.doi.org/10.1103/PhysRevD.102.015031}{\emph{Phys. Rev. D} {\bf
  102} (2020) 015031}, [\href{http://arxiv.org/abs/2002.12910}{{\tt
  2002.12910}}].

\bibitem{Hundi:2011si}
R.~S. Hundi, \emph{{Constraints from neutrino masses and muon (g-2) in the
  bilinear R-parity violating supersymmetric model}},
  \href{http://dx.doi.org/10.1103/PhysRevD.83.115019}{\emph{Phys. Rev. D} {\bf
  83} (2011) 115019}, [\href{http://arxiv.org/abs/1101.2810}{{\tt 1101.2810}}].

\bibitem{Trifinopoulos:2019lyo}
S.~Trifinopoulos, \emph{{B -physics anomalies: The bridge between R -parity
  violating supersymmetry and flavored dark matter}},
  \href{http://dx.doi.org/10.1103/PhysRevD.100.115022}{\emph{Phys. Rev. D} {\bf
  100} (2019) 115022}, [\href{http://arxiv.org/abs/1904.12940}{{\tt
  1904.12940}}].

\bibitem{Domingo:2018qfg}
F.~Domingo, H.~K. Dreiner, J.~S. Kim, M.~E. Krauss, M.~Lozano and Z.~S. Wang,
  \emph{{Updating Bounds on $R$-Parity Violating Supersymmetry from Meson
  Oscillation Data}},
  \href{http://dx.doi.org/10.1007/JHEP02(2019)066}{\emph{JHEP} {\bf 02} (2019)
  066}, [\href{http://arxiv.org/abs/1810.08228}{{\tt 1810.08228}}].

\bibitem{Das:2017kfo}
D.~Das, C.~Hati, G.~Kumar and N.~Mahajan, \emph{{Scrutinizing $R$-parity
  violating interactions in light of $R_{K^{(\ast)}}$ data}},
  \href{http://dx.doi.org/10.1103/PhysRevD.96.095033}{\emph{Phys. Rev. D} {\bf
  96} (2017) 095033}, [\href{http://arxiv.org/abs/1705.09188}{{\tt
  1705.09188}}].

\bibitem{ATLAS:2021yyr}
{\scshape ATLAS} collaboration, G.~Aad et~al., \emph{{Search for supersymmetry
  in events with four or more charged leptons in 139 fb$^{−1}$ of $ \sqrt{s}
  $ = 13 TeV pp collisions with the ATLAS detector}},
  \href{http://dx.doi.org/10.1007/JHEP07(2021)167}{\emph{JHEP} {\bf 07} (2021)
  167}, [\href{http://arxiv.org/abs/2103.11684}{{\tt 2103.11684}}].

\bibitem{ATLAS:2021fbt}
{\scshape ATLAS} collaboration, G.~Aad et~al., \emph{{Search for
  R-parity-violating supersymmetry in a final state containing leptons and many
  jets with the ATLAS experiment using $\sqrt{s} = 13 { TeV}$
  proton\textendash{}proton collision data}},
  \href{http://dx.doi.org/10.1140/epjc/s10052-021-09761-x}{\emph{Eur. Phys. J.
  C} {\bf 81} (2021) 1023}, [\href{http://arxiv.org/abs/2106.09609}{{\tt
  2106.09609}}].

\bibitem{ATLAS:2024kqk}
{\scshape ATLAS} collaboration, G.~Aad et~al., \emph{{A search for
  R-parity-violating supersymmetry in final states containing many jets in $pp$
  collisions at $\sqrt{s} = 13\,\text{TeV}$ with the ATLAS detector}},
  \href{http://arxiv.org/abs/2401.16333}{{\tt 2401.16333}}.

\bibitem{CMS:2017szl}
{\scshape CMS} collaboration, A.~M. Sirunyan et~al., \emph{{Search for
  $R$-parity violating supersymmetry in pp collisions at $\sqrt{s} = $ 13 TeV
  using b jets in a final state with a single lepton, many jets, and high sum
  of large-radius jet masses}},
  \href{http://dx.doi.org/10.1016/j.physletb.2018.06.028}{\emph{Phys. Lett. B}
  {\bf 783} (2018) 114--139}, [\href{http://arxiv.org/abs/1712.08920}{{\tt
  1712.08920}}].

\bibitem{CMS:2021knz}
{\scshape CMS} collaboration, A.~M. Sirunyan et~al., \emph{{Search for top
  squarks in final states with two top quarks and several light-flavor jets in
  proton-proton collisions at $\sqrt {s}$ = 13\,\,TeV}},
  \href{http://dx.doi.org/10.1103/PhysRevD.104.032006}{\emph{Phys. Rev. D} {\bf
  104} (2021) 032006}, [\href{http://arxiv.org/abs/2102.06976}{{\tt
  2102.06976}}].

\bibitem{CMS:2018skt}
{\scshape CMS} collaboration, A.~M. Sirunyan et~al., \emph{{Search for resonant
  production of second-generation sleptons with same-sign dimuon events in
  proton-proton collisions at $\sqrt{s} =$ 13 TeV}},
  \href{http://dx.doi.org/10.1140/epjc/s10052-019-6800-x}{\emph{Eur. Phys. J.
  C} {\bf 79} (2019) 305}, [\href{http://arxiv.org/abs/1811.09760}{{\tt
  1811.09760}}].

\bibitem{Dreiner:2023bvs}
H.~K. Dreiner, Y.~S. Koay, D.~K\"ohler, V.~M. Lozano, J.~Montejo~Berlingen,
  S.~Nangia et~al., \emph{{The ABC of RPV: classification of R-parity violating
  signatures at the LHC for small couplings}},
  \href{http://dx.doi.org/10.1007/JHEP07(2023)215}{\emph{JHEP} {\bf 07} (2023)
  215}, [\href{http://arxiv.org/abs/2306.07317}{{\tt 2306.07317}}].

\bibitem{Choudhury:2023eje}
A.~Choudhury, A.~Mondal, S.~Mondal and S.~Sarkar, \emph{{Improving sensitivity
  of trilinear R-parity violating SUSY searches using machine learning at the
  LHC}}, \href{http://dx.doi.org/10.1103/PhysRevD.109.035001}{\emph{Phys. Rev.
  D} {\bf 109} (2024) 035001}, [\href{http://arxiv.org/abs/2308.02697}{{\tt
  2308.02697}}].

\bibitem{Choudhury:2023yfg}
A.~Choudhury, A.~Mondal, S.~Mondal and S.~Sarkar, \emph{{Slepton searches in
  the trilinear RPV SUSY scenarios at the HL-LHC and HE-LHC}},
  \href{http://arxiv.org/abs/2310.07532}{{\tt 2310.07532}}.

\bibitem{Bhattacherjee:2023kxw}
B.~Bhattacherjee and P.~Solanki, \emph{{Search for electroweakinos in R-parity
  violating SUSY with long-lived particles at HL-LHC}},
  \href{http://dx.doi.org/10.1007/JHEP12(2023)148}{\emph{JHEP} {\bf 12} (2023)
  148}, [\href{http://arxiv.org/abs/2308.05804}{{\tt 2308.05804}}].

\bibitem{Breiman:1984jka}
L.~Breiman, J.~Friedman, R.~A. Olshen and C.~J. Stone, \emph{{Classification
  and regression trees}}.
\newblock Chapman and Hall/CRC, 1984.

\bibitem{breiman}
L.~Breiman, J.~Friedman, R.~Olshen and C.~Stone, \emph{Classification And
  Regression Trees}.
\newblock 10, 2017,
  \href{http://dx.doi.org/10.1201/9781315139470}{10.1201/9781315139470}.

\bibitem{gini}
``C. gini, variabilità e mutabilità, (reprinted in memorie di metodologica
  statistica, eds. e. pizetti and t. salvemini, libreria eredi virgilio veschi,
  rome, 1955), 1912.''

\bibitem{ceriani2012origins}
L.~Ceriani and P.~Verme, \emph{The origins of the gini index: extracts from
  variabilit e mutabilit (1912) by corrado gini}, {\emph{The Journal of
  Economic Inequality} {\bf 10} (2012) 421--443}.

\bibitem{QUINLAN1987221}
J.~Quinlan, \emph{Simplifying decision trees},
  \href{http://dx.doi.org/https://doi.org/10.1016/S0020-7373(87)80053-6}{\emph{International
  Journal of Man-Machine Studies} {\bf 27} (1987) 221--234}.

\bibitem{freund1999adaptive}
Y.~Freund, \emph{An adaptive version of the boost by majority algorithm},  in
  \emph{Proceedings of the twelfth annual conference on Computational learning
  theory}, pp.~102--113, 1999.

\bibitem{friedman2000additive}
J.~Friedman, T.~Hastie and R.~Tibshirani, \emph{Additive logistic regression: a
  statistical view of boosting (with discussion and a rejoinder by the
  authors)}, {\emph{The annals of statistics} {\bf 28} (2000) 337--407}.

\bibitem{Yang:2005nz}
H.-J. Yang, B.~P. Roe and J.~Zhu, \emph{{Studies of boosted decision trees for
  MiniBooNE particle identification}},
  \href{http://dx.doi.org/10.1016/j.nima.2005.09.022}{\emph{Nucl. Instrum.
  Meth. A} {\bf 555} (2005) 370--385},
  [\href{http://arxiv.org/abs/physics/0508045}{{\tt physics/0508045}}].

\bibitem{dorogush2018catboost}
A.~V. Dorogush, V.~Ershov and A.~Gulin, \emph{Catboost: gradient boosting with
  categorical features support}, {\emph{arXiv preprint arXiv:1810.11363} (2018)
  }.

\bibitem{Breiman:2001hzm}
L.~Breiman, \emph{{Random Forests}},
  \href{http://dx.doi.org/10.1023/A:1010933404324}{\emph{Machine Learning} {\bf
  45} (2001) 5--32}.

\bibitem{freund1999short}
Y.~Freund, R.~Schapire and N.~Abe, \emph{A short introduction to boosting},
  {\emph{Journal-Japanese Society For Artificial Intelligence} {\bf 14} (1999)
  1612}.

\bibitem{wang2012adaboost}
R.~Wang, \emph{Adaboost for feature selection, classification and its relation
  with svm, a review}, {\emph{Physics Procedia} {\bf 25} (2012) 800--807}.

\bibitem{Freund:1997xna}
Y.~Freund and R.~E. Schapire, \emph{{A Decision-Theoretic Generalization of
  On-Line Learning and an Application to Boosting}},
  \href{http://dx.doi.org/10.1006/jcss.1997.1504}{\emph{J. Comput. Syst. Sci.}
  {\bf 55} (1997) 119--139}.

\bibitem{ke2017lightgbm}
G.~Ke, Q.~Meng, T.~Finley, T.~Wang, W.~Chen, W.~Ma et~al., \emph{Lightgbm: A
  highly efficient gradient boosting decision tree}, {\emph{Advances in neural
  information processing systems} {\bf 30} (2017) 3146--3154}.

\bibitem{Choudhury:2013jpa}
A.~Choudhury and A.~Datta, \emph{{Neutralino dark matter confronted by the LHC
  constraints on Electroweak SUSY signals}},
  \href{http://dx.doi.org/10.1007/JHEP09(2013)119}{\emph{JHEP} {\bf 09} (2013)
  119}, [\href{http://arxiv.org/abs/1305.0928}{{\tt 1305.0928}}].

\bibitem{Barman:2020azo}
R.~K. Barman, B.~Bhattacherjee, I.~Chakraborty, A.~Choudhury and N.~Khan,
  \emph{{Electroweakino searches at the HL-LHC in the baryon number violating
  MSSM}}, \href{http://dx.doi.org/10.1103/PhysRevD.103.015003}{\emph{Phys. Rev.
  D} {\bf 103} (2021) 015003}, [\href{http://arxiv.org/abs/2003.10920}{{\tt
  2003.10920}}].

\bibitem{Datta:2018lup}
A.~Datta and N.~Ganguly, \emph{{The past, present and future of the heavier
  electroweakinos in the light of LHC and other data}},
  \href{http://dx.doi.org/10.1007/JHEP01(2019)103}{\emph{JHEP} {\bf 01} (2019)
  103}, [\href{http://arxiv.org/abs/1809.05129}{{\tt 1809.05129}}].

\bibitem{ATLAS:2021moa}
{\scshape ATLAS} collaboration, G.~Aad et~al., \emph{{Search for
  chargino\textendash{}neutralino pair production in final states with three
  leptons and missing transverse momentum in $\sqrt{s} = 13$~TeV pp collisions
  with the ATLAS detector}},
  \href{http://dx.doi.org/10.1140/epjc/s10052-021-09749-7}{\emph{Eur. Phys. J.
  C} {\bf 81} (2021) 1118}, [\href{http://arxiv.org/abs/2106.01676}{{\tt
  2106.01676}}].

\bibitem{TheATLAScollaboration:2014nwe}
\emph{{Search for Supersymmetry at the high luminosity LHC with the ATLAS
  experiment, ATL-PHYS-PUB-2014-010, 2014}}, .

\bibitem{Sjostrand:2006za}
T.~Sjostrand, S.~Mrenna and P.~Z. Skands, \emph{{PYTHIA 6.4 Physics and
  Manual}}, \href{http://dx.doi.org/10.1088/1126-6708/2006/05/026}{\emph{JHEP}
  {\bf 05} (2006) 026}, [\href{http://arxiv.org/abs/hep-ph/0603175}{{\tt
  hep-ph/0603175}}].

\bibitem{Alwall:2014hca}
J.~Alwall, R.~Frederix, S.~Frixione, V.~Hirschi, F.~Maltoni, O.~Mattelaer
  et~al., \emph{{The automated computation of tree-level and next-to-leading
  order differential cross sections, and their matching to parton shower
  simulations}}, \href{http://dx.doi.org/10.1007/JHEP07(2014)079}{\emph{JHEP}
  {\bf 07} (2014) 079}, [\href{http://arxiv.org/abs/1405.0301}{{\tt
  1405.0301}}].

\bibitem{Fuks:2013vua}
B.~Fuks, M.~Klasen, D.~R. Lamprea and M.~Rothering, \emph{{Precision
  predictions for electroweak superpartner production at hadron colliders with
  Resummino}},
  \href{http://dx.doi.org/10.1140/epjc/s10052-013-2480-0}{\emph{Eur. Phys. J.
  C} {\bf 73} (2013) 2480}, [\href{http://arxiv.org/abs/1304.0790}{{\tt
  1304.0790}}].

\bibitem{deFavereau:2013fsa}
{\scshape DELPHES 3} collaboration, J.~de~Favereau, C.~Delaere, P.~Demin,
  A.~Giammanco, V.~Lema\^\i{}tre, A.~Mertens et~al., \emph{{DELPHES 3, A
  modular framework for fast simulation of a generic collider experiment}},
  \href{http://dx.doi.org/10.1007/JHEP02(2014)057}{\emph{JHEP} {\bf 02} (2014)
  057}, [\href{http://arxiv.org/abs/1307.6346}{{\tt 1307.6346}}].

\bibitem{pedregosa2018scikitlearn}
F.~Pedregosa, G.~Varoquaux, A.~Gramfort, V.~Michel, B.~Thirion, O.~Grisel
  et~al., \emph{Scikit-learn: Machine learning in python},  2018.

\bibitem{lundberg2017unified}
S.~Lundberg and S.-I. Lee, \emph{A unified approach to interpreting model
  predictions},  2017.

\end{thebibliography}\endgroup

\newpage 

\section{Appendix}
\label{sec:appendix}
\begin{table}[!htb]
\renewcommand{\arraystretch}{1.1}
\centering
\begin{tabular}{||c||c|c||}
\hline\hline
Benchmark point & Cross-section, $\sigma$ (fb)  & $\sigma^{\prime} = \sigma\times 3l^{\prime}$ (fb) \\
\hline\hline
\texttt{BP1}(400,350) & 144.26 & 4.76 \\
\hline
\texttt{BP2}(500,430) & 56.42 & 1.86 \\
\hline
\texttt{BP3}(800,100) & 6.19 & 0.20 \\
\hline
\texttt{BP4}(1200,100) & 0.6 & 0.02 \\
\hline\hline
\end{tabular}
\caption{The NLL+NLO cross-section of the signal benchmark points are listed in this table.}
\label{tab:sigma_signal}
\end{table}

\begin{table}[!htb]
\centering
\setlength{\tabcolsep}{1.6pt}
\renewcommand{\arraystretch}{1.1}
\begin{tabular}{||c||c|c|c||}
\hline\hline
Background & Cross-section, $\sigma$ (fb)  & $\sigma^{\prime} = \sigma\times nl^{\prime}$ (fb)  & Total event generated\\
\hline\hline
$WZ+jets$ & $3.619\times10^4$ & 1196 ($n=3$) & 3.7$\times10^6$\\
\hline
$ZZ+jets$ & $1.067\times10^4$ & 111 ($n=4$)& 1.6$\times10^5$ \\
\hline
$t\bar{t}+jets$ & $5.930\times10^5$ & 62250 ($n=2$) & 1.2$\times10^7$\\
\hline
$WWZ +jets$ & $1.488\times10^2$ & 3.3251 ($n=3$) & 9.8$\times10^4$\\
\hline
$WWW +jets$ & $1.925\times10^2$& 6.546 ($n=3$) & 5.3$\times10^4$  \\
\hline
$t\bar{t}W+jets$ &$5.510\times10^2$ & 18.74 ($n=3$) & 1.6$\times10^5$\\
\hline
$t\bar{t}Z+jets$ & $6.893\times10^2$& 15.4 ($n=3$) & 1.3$\times10^5$ \\
\hline
$Z+jets$ & $4.516\times10^7$& 4606000 ($n=2$) & 1.3$\times10^6$\\
\hline\hline
\end{tabular}

\caption{The matched cross-sections and the total event generated corresponding to each SM background are displayed in this table.}
\label{tab:sigma_bkg}
\end{table}

\begin{figure}[!htb]
\begin{center}
\includegraphics[scale=0.45]{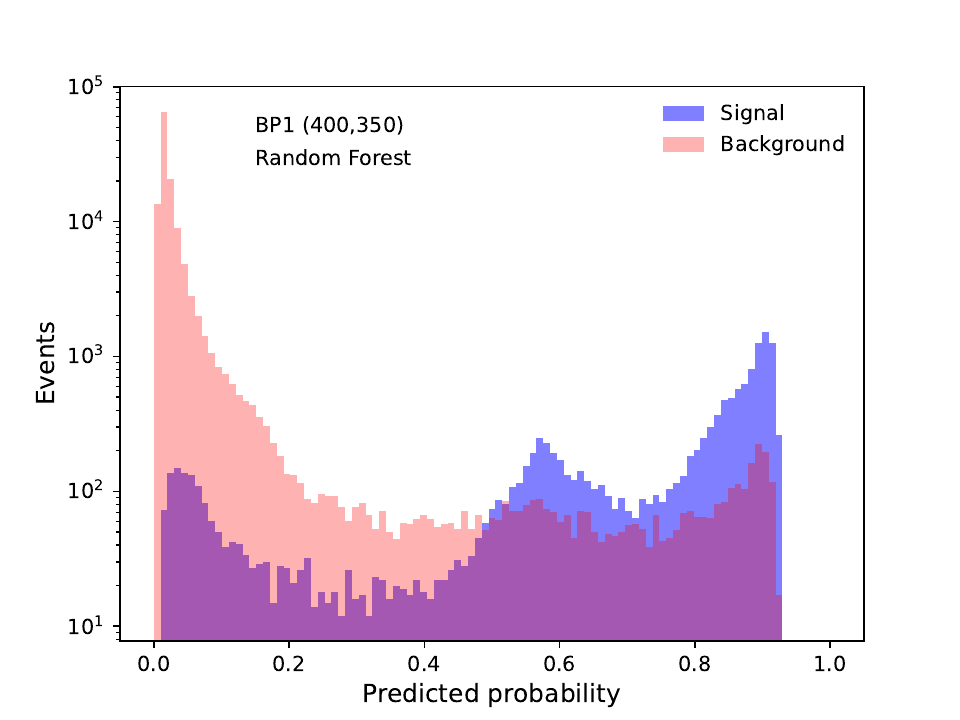}
\includegraphics[scale=0.45]{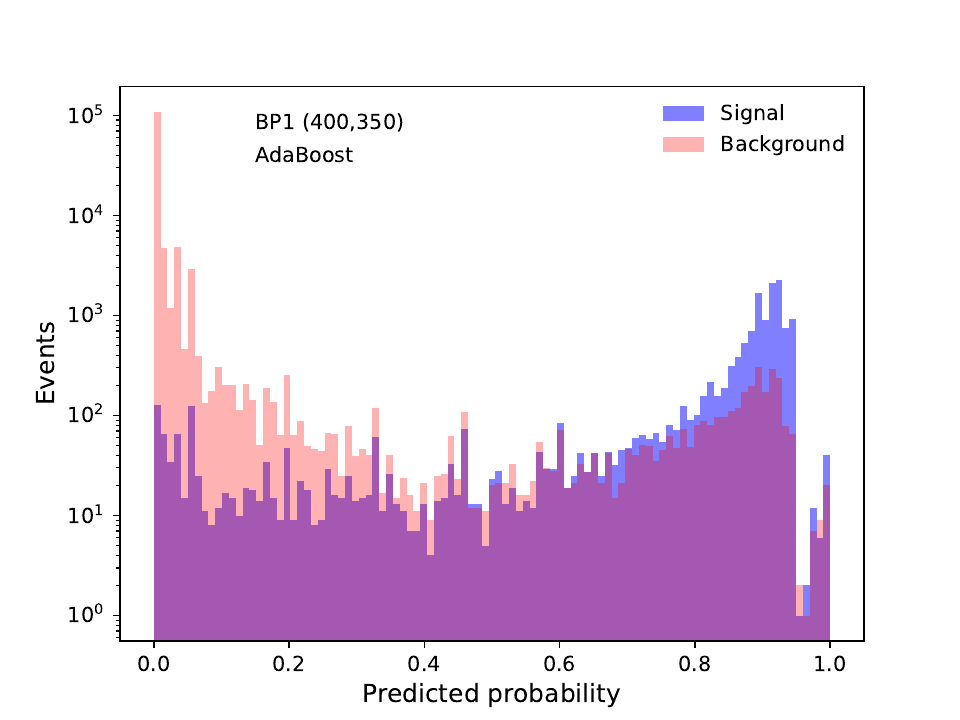}
\includegraphics[scale=0.45]{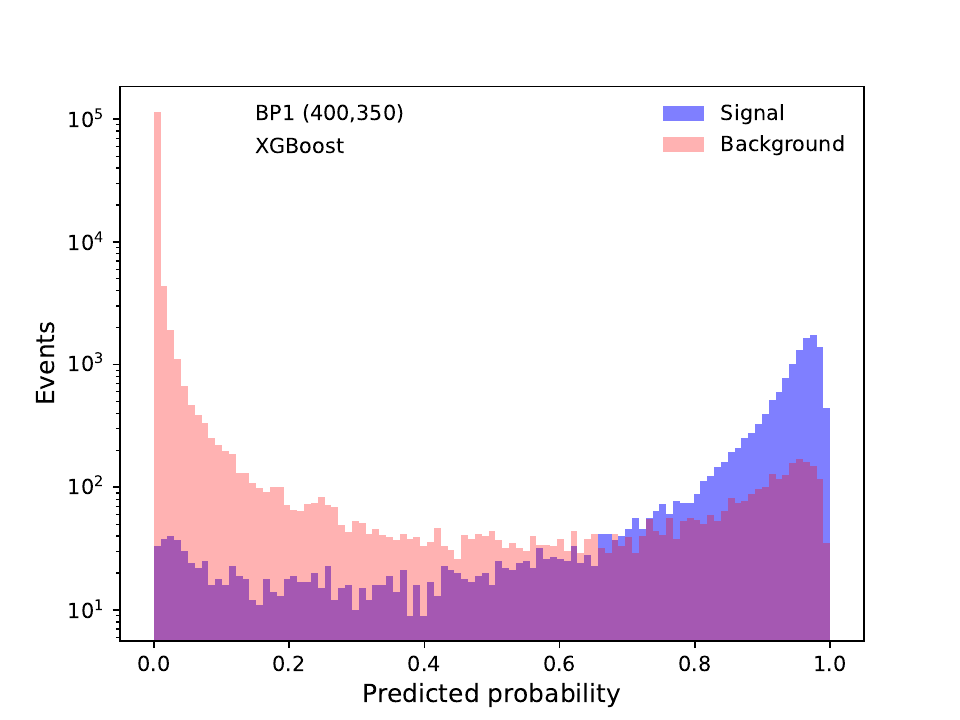}
\includegraphics[scale=0.45]{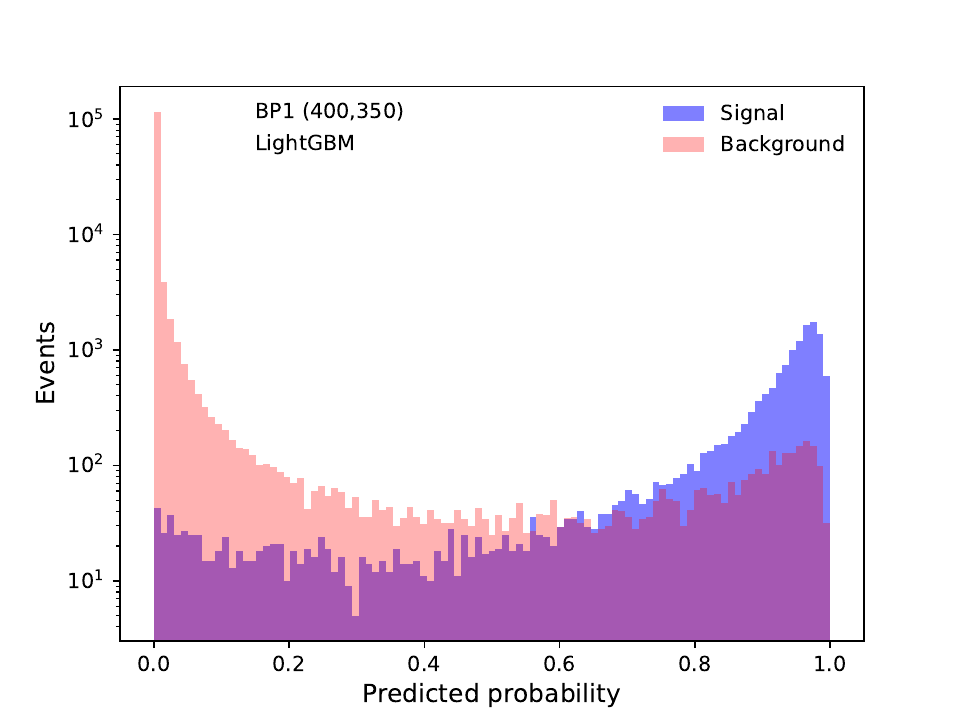}
\caption{The signal and background events distribution with the probability cut in the test dataset corresponding to the \texttt{BP1} for the \texttt{Random Forest} classifier (upper left), \texttt{AdaBoost} (upper right), \texttt{XGBoost} classifier (lower left) and \texttt{LightGBM} (lower right).}
\label{fig:event_dist1}
\end{center}
\end{figure}

\begin{figure}[!htb]
\begin{center}
\includegraphics[scale=0.35]{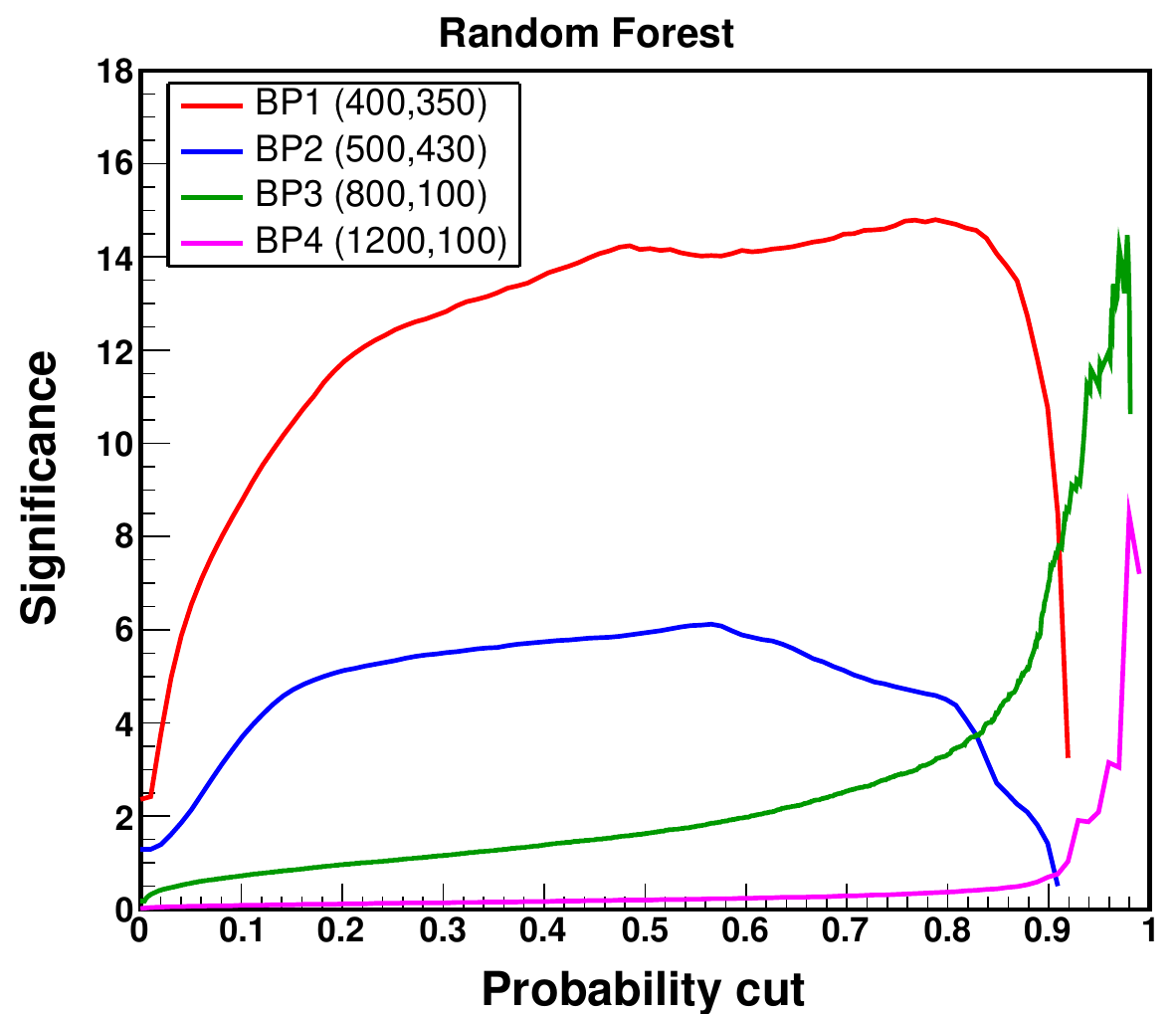}
\includegraphics[scale=0.35]{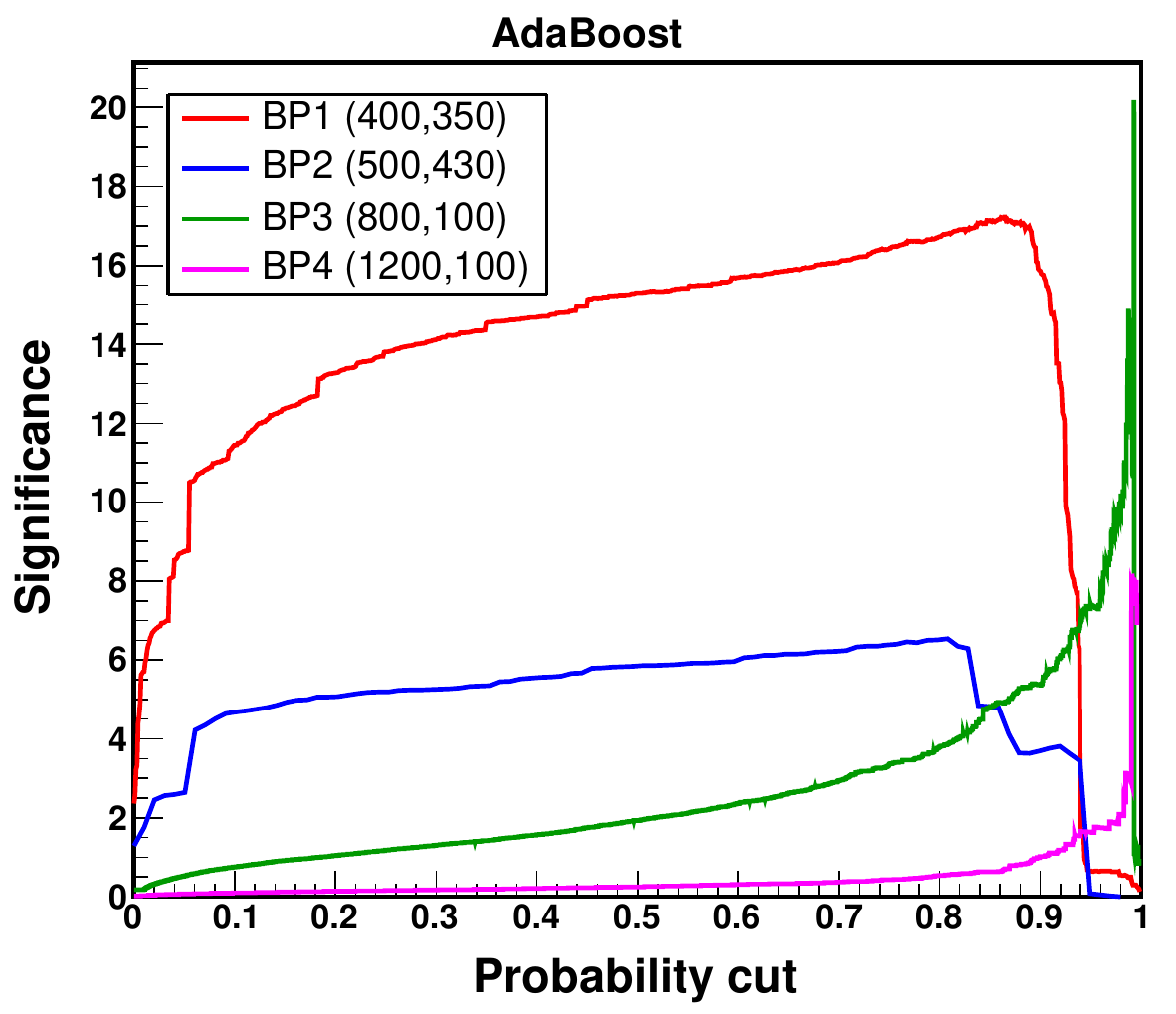}
\includegraphics[scale=0.35]{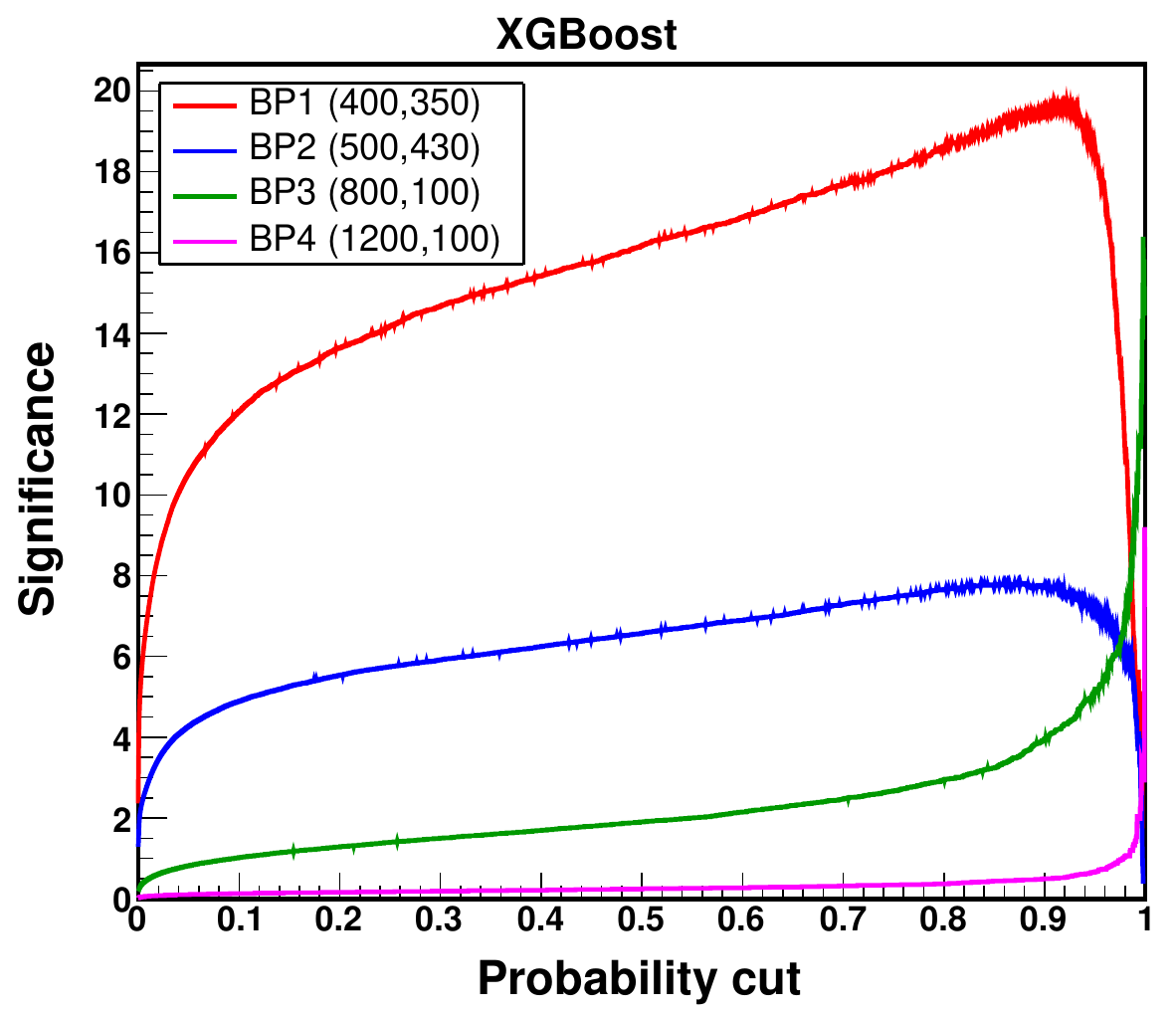}
\includegraphics[scale=0.35]{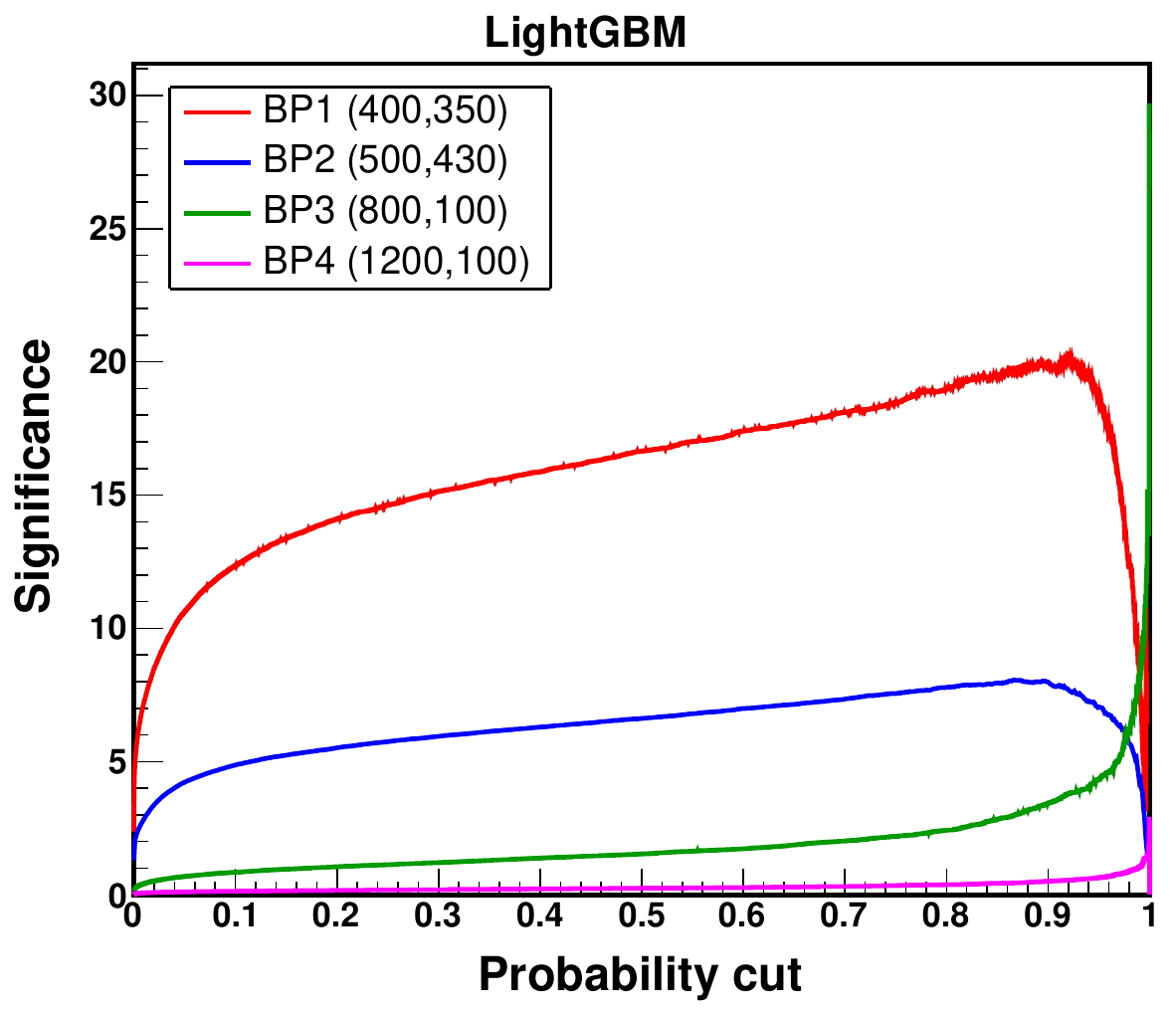}
\caption{The distributions of the signal significance with the probability cut are shown for all the benchmark points corresponding to each different algorithm.}
\label{fig:ams_dist}
\end{center}
\end{figure}

\begin{table}[!htb]
\centering
\setlength{\tabcolsep}{0.01pt}
\renewcommand{\arraystretch}{1.4}
\small
\begin{tabular}{||c||c||c||c||}
\hline\hline
\multicolumn{4}{||c||}{Optimized hyperparameters}\\
\hline\hline
\texttt{Random Forest} & \texttt{AdaBoost} & \texttt{XGBoost} & \texttt{LightGBM} \\
\hline\hline
\textit{n\_estimators} = 300 & \textit{n\_estimators} = 30 & \textit{num\_trees} = 500 & \textit{boosting\_type} = \textit{`gbdt'}\\
\textit{max\_depth} = 5 & \textit{max\_depth} = 6 & \textit{max\_depth} = 8& \textit{num\_leaves} = 70 \\
\textit{max\_leaf\_nodes} = 150 & \textit{learning\_rate} = 0.03 & \textit{learning\_rate} = 0.07 & \textit{max\_depth} = 8\\ 
\textit{max\_features} = \textit{`sqrt'} & \textit{algorithm=``SAMME.R"} & \textit{subsample} = 0.5 & \textit{learning\_rate} = 0.1\\
\textit{criterion} = \textit{`gini'} & & \textit{min\_child\_weight} = 2.5 & \textit{n\_estimators} = 200\\
& & \textit{alpha} = 0.21 & \textit{subsample} = 0.5 \\
& & \textit{gamma} = 0.68 & \textit{min\_child\_weight} = 1.5 \\
& & & \textit{colsample\_bytree} = 0.8 \\
\hline\hline
\end{tabular}
\caption{The values of optimized hyperparameters that lead to the best performance of the model corresponding to different algorithms are mentioned in this table.}
\label{tab:param_set}
\end{table}

\end{document}